\documentclass[]{aa}
\usepackage[T1]{fontenc}
\usepackage[utf8]{inputenc}

\usepackage{graphicx}
\usepackage{natbib}
\usepackage{ulem}
\usepackage{textcomp}
\usepackage{gensymb}
\usepackage{longtable}
\usepackage{threeparttable}
\usepackage{multicol}
\usepackage{multirow}
\usepackage{float}
\usepackage{todonotes}
\usepackage[Symbol]{upgreek}
\usepackage{amsmath}
\usepackage{etoolbox}
\usepackage{txfonts}
\usepackage{url}
\usepackage{xcolor}

\usepackage[most]{tcolorbox}
\usepackage{hyperref}
\usepackage{xcolor}
\newcommand{\enzo}{{\it {\small ENZO}}}
\newcommand{\CRaTer}{{\it {\small CRaTer}}}

\begin{document}
 
\title{Life cycle of cosmic-ray electrons in the intracluster medium}

\author{F. Vazza\inst{1,2,3}, D. Wittor\inst{2,1},  L. Di Federico\inst{1}, M. Br\"{u}ggen\inst{2}, M. Brienza \inst{1,3}, G. Brunetti\inst{3}, F. Brighenti\inst{1}, T. Pasini\inst{2}}

\offprints{%
 E-mail: franco.vazza2@unibo.it}
\institute{Dipartimento di Fisica e Astronomia, Universit\'{a} di Bologna, Via Gobetti 93/2, 40122, Bologna, Italy
\and  Hamburger Sternwarte, University of Hamburg, Gojenbergsweg 112, 21029 Hamburg, Germany
\and Istituto di Radioastronomia, INAF, Via Gobetti 101, 40122, Bologna, Italy}

\authorrunning{F. Vazza, D. Wittor, L. Di Federico, et al.}
\titlerunning{Cosmic-ray electrons in the ICM}

\date{Accepted ???. Received ???; in original form ???}

\abstract{We simulate the evolution of relativistic electrons injected into the medium of a small galaxy cluster by a central radio galaxy, studying how the initial jet power affects the dispersal and the emission properties of radio plasma. 
By coupling  passive tracer particles to adaptive-mesh cosmological MHD simulations, we study how cosmic-ray electrons are dispersed as a function of the input jet power. We also investigate how the latter affects the thermal and non-thermal properties of the intracluster medium, with differences discernible up to $\sim$ Gyr after the start of the jet. We evolved the energy spectra of cosmic-ray electrons, subject to energy losses that are dominated by synchrotron and inverse Compton emission as well as energy gains via re-acceleration by shock waves and turbulence. We find that in the absence of major mergers the amount of re-acceleration experienced by cosmic-ray electrons is not enough to produce long-lived detectable radio emissions. However, for all simulations the role of re-acceleration processes is crucial to maintain a significant and volume-filling reservoir of fossil electrons ($\gamma \sim 10^3$) for several Gyrs after the first injection by jets.
This is important to possibly explain recent discoveries of cluster-wide emission and  other radio phenomena in galaxy clusters.}
\maketitle

\label{firstpage}

 \section{Introduction}\label{sec::intro}

Jets from radio galaxies are among the most spectacular examples of how radically different temporal and spatial scales are connected in astrophysics: their accretion power originates from  supermassive black hole discs ($\leq 10^{-3} \rm ~pc$), and their energy is fed back into intergalactic space (also as non-thermal plasma components) remaining visible for a few $\sim 10^2 \rm ~Myr$ after their injection and out to $\sim~\rm Mpc$ scales. 

Radio galaxies are important sources of non-thermal energy in the intracluster medium (ICM). In particular, radio-bright, bipolar plasma outflows from active galactic nuclei (AGN)
 are commonly found in clusters of galaxies, and are routinely studied by radio observations \citep[e.g.][]{hardcastlecroston}.
 
A large diversity of radio galaxy morphologies, powers and
duty cycles exists. Their relations with the property of host galaxies, of the mass accretion rate onto their SMBH, or of the surrounding environment are still unclear \cite[e.g.][]{2008A&ARv..15...67M,mingo19,vard19,mingo22}. 

Numerical simulations over the last years have investigated the dynamics of relativistic jets as they expand into the ambient medium, and the key role of magneto-hydrodynamical instabilities in determining the jet morphology and stability 
\citep[e.g.][]{2007MNRAS.382..526P,2014MNRAS.443.1482H,2016A&A...596A..12M,2018MNRAS.481.2878E}. 
Supported by strong observational evidence from X-ray and radio observations of diffuse radio emissions \citep[e.g.][]{2004ApJ...607..800B,2007ARA&A..45..117M,2012ARA&A..50..455F}, cosmological simulations predict that the radio mode of  feedback (i.e. mediated by kinetic jets dissipating their energy via heating of the gas reservoir undergoing cooling) is crucial in shaping the thermodynamic structure of galaxy groups and clusters at low redshifts \citep[e.g.][]{2008ApJ...687L..53P,2010MNRAS.401.1670F,2017MNRAS.470.1121T,2017MNRAS.470.4530W}. However, no cosmological simulation has ever produced (or even idealised) radio galaxies, including their magnetic and cosmic ray content, possibly stressing the fact that feedback recipes in cosmological simulations may be highly idealised. 
Tailed radio galaxies are also often found to mix with diffuse radio emission from the ICM \citep[e.g.][]{2017PhPl...24d1402J,nolting19a}, both in the form of radio halos or radio relics \citep[e.g.][and references therein]{2019SSRv..215...16V}, and recent low-frequency observations are detecting complex morphologies of remnant plasma, injected by radio galaxies and in different stages of mixing their non-thermal content with the ICM \citep[e.g.][]{wilber18,2020ApJ...897...93B,2020A&A...634A...4M,brienza21}. 

A  volume-filling distribution of fossil relativistic electrons is also often required in order to explain the observed radio power of radio relics, under the hypothesis that weak merger shocks re-accelerate a mildly relativistic electrons, rather than injecting them from the thermal pool \citep[e.g.][]{ka12, 2013MNRAS.435.1061P,2020A&A...634A..64B,2021Natur.593...47C}.  However, cosmological simulations so far could not assess whether the normal activity by radio galaxies is sufficient to enrich the ICM with the necessary amount of fossil electrons. The overall role of radio galaxies in magnetising the Universe is also uncertain \citep[e.g.][]{Volk&Atoyan..ApJ.2000,xu09,va17cqg} and the modelling of recent 
radio observations leaves the room open to, both, primordial seed fields (with amplitude $\leq 0.5 \rm ~nG$) and the magnetisation by active galaxies \citep[][]{os20,vern21,va21magcow}. 

In our first paper \citep[][ hereafter Paper I]{va21jets} we studied the energy evolution of the relativistic electrons advected into the ICM by cluster-wide motions, for jets injected prior ($z=1$) or after ($z=0.5$) the formation of the host group. The spectral energy evolution of relativistic electrons was modelled considering ageing and acceleration processes.  This allowed us to track the dispersal of fossil electrons in the cluster as a function of time, and their detectability despite the expected energy losses, for different plausible re-acceleration mechanisms in the ICM. 
Our model thus offers an unprecedented view of the large-scale circulation and advection of fossil electrons on Megaparsec scales, in a regime scarcely affected by radiative cooling and other processes neglected in this work. First applications of our simulations to the modelling of real radio sources have been recently presented in \citet{Hodgson21a}, \citet{2021Galax...9...93V} and \citet{brienza22}.

In this new paper, we study how the dynamical and energy evolution of electrons injected by radio jets scale with the total jet power, which we vary 5 times and always refer to the same epoch ($z=0.5$), i.e. after the cluster as assembled $\geq 50\%$ of its final mass.

Our paper is structured as follows:
in Section~2, we present the cosmological simulations and all numerical methods employed in this paper; in Section~3, we give our main results, focusing on the effects of jets on the gas dynamics (Section~3.1) and on the spectra energy evolution of electron spectra (Section~3.2). The radio properties of our simulated jets are given in Section~4, while the discussion on the impact of jet power in the evolution of radio emission is presented in Section~5. Our conclusions are given in Section~6, and additional numerical tests on our algorithms are presented in the Appendix.

Throughout this paper, we use the following cosmological parameters: $h = 0.678$, $\Omega_{\Lambda} = 0.692$, $\Omega_{\mathrm{M}} = 0.308$ and $\Omega_{\mathrm{b}} = 0.0478$, based on the results from the \citet{2016A&A...594A..13P}. 

\section{Methods}\label{sec::simu}
 
Most of the relevant methods have already been extensively discussed in \citet{va21jets}, so we only summarise them here. 

\subsection{\enzo-simulations}

We used cosmological, adaptive mesh refinement \enzo-MHD \citep{enzo14} simulations using the MHD solver with a Lax-Friedrichs (LLF) Riemann solver to compute the fluxes in the Piece-wise Linear Method (PLM), combined with the Dedner cleaning method \citep[][]{ded02} implemented by \citet{wa09}. We start from a simple uniform magnetic field at $z=50$ with a value of $B_0=0.1  \ \mathrm{nG}$ in each direction. 

The total simulated volume is of  (50 Mpc$/h$)$^3$ and is sampled with a root grid of $128^3$ cells and dark matter particles. We further use four additional nested regions with increasing spatial resolution until the (4 Mpc/$h$)$^3$ volume where a  $M_{100} \approx 1.5 \cdot 10^{14} M_\odot$ cluster forms is sampled down to a $\Delta x=24.4$~kpc$/h$ uniform resolution. At run-time, two additional level of mesh refinement are added using a local gas/DM overdensity criterion ($\Delta \rho/\rho \geq 3$), allowing the simulation to reach a maximum resolution of $\approx 6 \rm ~kpc/h= 8.86$~kpc.  As a result of our nested grid approach, the mass resolution for dark matter in our cluster formation region is of  $m_{\rm DM}=2.82 \cdot 10^{6} ~M_{\odot}$ per dark matter particle, for the highest resolution particles that are used to fill the innermost AMR level since the start of the simulation.

\begin{table}
\begin{center}
\caption{Main jet parameters for the runs as a function of the assumed $\alpha_B$ boost factor in the Bondi accretion model (Equation~\ref{eq:bondi}). All quantities are measured 
$32 \rm ~Myr$ after the injection of jets , and the average, total or maximum quantities are taken from the distribution of tracer particles injected in jets.}
\footnotesize
\centering \tabcolsep 2pt
\begin{tabular}{c|c|c|c|c|c|c}
  Parameter&RunB&RunC&RunD&RunE&RunF&Run 0  \\ 
   $\alpha_B$ &1&3&10&30&50&0  \\\hline
      $P_{\rm j}$  [erg/s] & $3 \cdot 10^{43}$ & $9 \cdot 10^{43}$ & $3 \cdot 10^{44}$& $9\cdot 10^{44}$ & $1.5 \cdot 10^{45}$&0\\
        $\dot{M}_{\rm BH}/\dot{M}_{\rm Edd}$  & $0.002$ & $0.006$ & $0.02$& $0.06$ & $0.1$&0\\
      $B_{\rm av,j}$  [$\rm \mu G$] & $28.6$ & $70.1$ & $66.4$ &  $70.4$ & $41.5$ &0\\
      $B_{\rm max,j}$  [$\rm \mu G$] & $49.8$ &  $106.9$ & $91.2$ & $109.0$ & $97.6$ &0\\
      $T_{\rm av,j}$  [$\rm K$] & $1.5 \cdot 10^7$ &  $3.6 \cdot 10^7$ & $1.8 \cdot 10^8$ & $4.3 \cdot 10^8$ & $5.8 \cdot 10^8$ &0\\
      $v_{\rm av,j}$  [$\rm km/s$] & $260$ &  $1754$ & $2750$ & $3888$ & $4215$&0 \\
      $v_{\rm max,j}$ [$\rm km/s$] & $424$ & $2567$ & $4355$ & $10286$ & $7943$ &0\\
      $N_{\rm CRe}$ & $2.7 \cdot 10^{64}$ & $3.9 \cdot 10^{64}$ & $6.7 \cdot 10^{64}$ & $9.9 \cdot 10^{64}$ & $1.2 \cdot 10^{65}$ &0\\
  \end{tabular}
  \end{center}
\label{table:tab1}
\end{table}

\subsubsection{Active galactic nuclei and radio jets}
\label{jet}
All simulations presented here are non-radiative and only differ in the feedback power of AGN bursts. The jets and relativistic electrons in this set of simulations are injected at $z=0.5$, after which minor mergers occur across the entire lifetime of the group. A more prominent second major merger occurs between $z=0.3$ and $z=0.1$, following the accretion of a second massive companion. We thus neglect radiative cooling, and cannot have a self-regulating mechanism to switch the feedback cycle on and off \citep[e.g.][for a few examples]{gaspari11b,2015ApJ...813L..17R,2019MNRAS.489..802R,2021MNRAS.504.3619T}. 

The simulation of supermassive black holes (SMBH) is built on the numerical implementations by \citet[][]{2011ApJ...738...54K} on the public {\enzo} 2.6 version, and on a number of modifications by our group (as in Paper I). In particular, we used combination of thermal and magnetic feedback around the SMBH, which we place at $z=0.5$ at the centre of mass of a  $\sim 10^{14} M_{\odot}$ group of galaxies.
We assume that the SMBH accretes matter from the surrounding cells based on the Bondi–Hoyle formalism: 

\begin{equation}
    \dot{M}_{\rm BH}=\frac{4 \pi \alpha_B G^2 M_{\rm BH}^2\rho}{c_s^3} ,
    \label{eq:bondi}
\end{equation}
where $M_{\rm BH}$ is the SMBH's mass, $c_s$ is the sound speed of the gas at the SMBH's location (which we assume to be relative to a fixed $10^6 ~\rm K$ temperature of the accretion disc) and $\rho$ is the local gas density. 

 Given our resolution and the lack of radiative cooling, we cannot resolve the Bondi radius or the multi-phase interstellar medium around the host galaxy. Hence, as often done in the literature \citep[e.g.][]{2009MNRAS.398...53B,gaspari12,2016Natur.534..218T}, we introduce a boost factor, $\alpha_B$, which is meant to account for clumpy accretions (leading to a larger $\rho$ in Equation~\ref{eq:bondi}) which cannot be resolved by the finite spatial resolution of our method. 
We tested five different variations of $\alpha_B$ (=$1$,$3$,$10$,$30$,$50$)  
meant to explore a reasonable range of plausible variation in $\alpha_B$, which can follow from a randomly different episode of clumpy accretion onto the central SMBH at a given time.

All jets simulated in this work initially contain only thermal energy and magnetic energy, with absolute values depending on the assumed SMBH accretion rate, as explained below. Starting immediately after the short injection phase of thermal and magnetic energy (i.e. $\approx 32 \rm ~Myr$ after the injection), the added thermal and magnetic pressure drive powerful outflows around the SMBH region, which are then focused onto a bipolar jet-like structure by the assumed magnetic field topology, similar to previous works \citep[][]{2006ApJ...643...92L}, as we explain in more detail below.  On the other hand,  the relativistic electron component is only tracked in post-processing using tracer particles (see next two Sections), and thus it does not affect the jet dynamics or its interaction with the surrounding ICM. 

The bolometric luminosity of the black hole is defined as  $L_{\rm BH} = \epsilon_r \dot{M}_{\rm BH} c^2$, where $\epsilon_r$ is the radiative efficiency of the SMBH. 
Following \citet{2011ApJ...738...54K}, the SMBH particle releases thermal feedback on the surrounding  27 gas cells, in the form of an extra thermal energy output from each black hole particle, and that $\epsilon_{\rm BH} is $ the factor that converts the bolometric luminosity of the SMBH into the thermal feedback energy. 
Therefore the total feedback efficiency between the accreted mass, $\Delta M$, and the energy delivered by jets is $\epsilon_r \epsilon_{\rm BH} = E_{\rm jet}/(\Delta M \Delta t c^2)$, for each $\Delta t$ timestep. 
In this work we used   $\epsilon_r=0.1$ and $\epsilon_{\mathrm{BH}}=0.05$ \citep[e.g.][]{2011ApJ...738...54K,enzo14}, which also yielded a good match with observed galaxy clusters scaling relations as found in our previous work \citep[][]{va13feedback,va17cqg}. 

With our choices of $\alpha_B$, the average accretion rate onto the SMBH in the five runs have values  $\approx$ 0.2\% (run B), 0.6\% (run C),2 \% (run D), 6 \% (run E) and  10\% (run F) of the corresponding Eddington limited accretion rate ($ \dot{M}_{\rm Edd} =\frac{4 \pi G M_{\rm BH} m_p}{\epsilon_r \sigma_T c}$, with the usual meaning of symbols).  

In this work, we only allowed the SMBH feedback to last one root-grid timestep, which is $\Delta t \approx 32 \rm ~Myr$ at the epoch of activation ($z=0.5$).

The corresponding values of total power released by both jets, $P_j=\epsilon_r \epsilon_{\rm BH}  \dot{M}_{\rm BH} c^2$, are given in Table 1. Given our explored variations of $\alpha_B$, they range from $3 \cdot 10^{43} \rm erg/s$ to $1.5 \cdot 10^{45} \rm erg/s$. These values are compatible with the range of values typically inferred from the modelling of FRI radio galaxies in clusters and groups (typically through the analysis of X-ray cavities, e.g. \citealt{2004ApJ...607..800B,2008MNRAS.386.1709C,2013ApJ...767...12G}) and are also within the range of power that single sources in clusters can produce in multiple events \citep[e.g.][]{2021A&A...650A.170B}.
We shall also see in Section 4 that the radio power generated by the electrons injected by our jets, at least during their initial, visible stage, is compatible with the distribution of radio vs X-ray power reported by recent observations \citep[][]{pasini20,pasini21}, further confirming that the range of jet power simulated in this group of galaxies is realistic.

Simulated SMBH particles are also allowed to release additional magnetic field energy during each feedback event. The magnetic fields is injected in the form of magnetic loops ($2 \times 2$ cells at the highest resolution level), located at $\pm 1$ cell along the $z$-direction from the SMBH).
Such a simple topology is required by the limited available spatial resolution in the SMBH region ($\approx 8 \rm ~kpc$), while  more sophisticated choices are possible at  a higher spatial resolution \citep[e.g.][]{2020ApJ...896...86C}.  Of course, imposing a fixed jet alignment along the $z$ axis is not very physical, yet for the sake of our analysis the jets are launched only once and, in the lack of any physical prescription to link gas accretions to the SMBH axis in this case, any launching direction seems equally likely. 
We notice that with an additional set of runs (run DY and DZ, discussed in the Appendix) we tested different choices in  the initial direction of jets (i.e. we released jets along the other two possible perpendicular directions, in two alternative runs), meant to explore whether the large scale circulation of injected electrons can show different properties, depending on the different sectors in the ICM the jets expand into, and in the possible different amount of frustration that jets can experience depending on the ICM flow they encounter.
To a zero order, our tests report no significant differences in the statistics of the thermodynamical properties of the ICM. Therefore, while we only consider runs having the same jet orientation, we refer the reader to the Appendix for a more detailed discussion about the additional role of jet orientation. Dependencies on the radio jets morphologies and the initial jet orientation will be instead explored in future work. An additional run with radiative gas cooling (cun coolD) is also discussed in the Appendix. 

The injected magnetic energy is normalised to a fixed fraction of the total feedback energy $E_{\rm B,jet} = \epsilon_{\rm B,jet} E_{\rm jet}$, with $\epsilon_{\rm B,jet}=0.1$ as tested in  previous works \citep[e.g.][]{va17cqg}. Since the thermal energy is isotropically spread over a larger volume (i.e. 27 cells), the typical magnetic energy of our jets (see e.g. Tab. 1) is a factor $\sim 10$ larger than the kinetic and thermal energy assigned effectively  magnetically dominated soon after their creation. As in Paper I (e.g. Table 1), in all runs we measure that after the injection phase (lasting $\approx 32 \rm ~Myr$) the kinetic, thermal and magnetic energy within the jet region (marked by our tracers) are very close to equipartition, $E_{\rm th} \sim E_{\rm B} \sim E_{\rm kin} \sim E_{\rm jet}/3$. In reality, we expect that also the cosmic-ray component should be important for the jet internal pressure of radio jets embedded in clusters \citep[e.g.][]{2018MNRAS.476.1614C}, yet in this simulation we can only track cosmic-ray electrons as a passive fluid, with no contribution to the dynamical pressure (see Section 6 for a discussion on the physical limitations of our model).

Our approach is in many ways similar to early work by \citet[][]{2006ApJ...643...92L} and \citet[][]{xu09}, who initialised  toroidal magnetic fields on opposite sides of the SMBHs. Their setup evolved into self-collimating outflows, leading to a supersonic expansion in the cluster core region. This could lead up to a $\sim 90\%$ conversion of the initial magnetic energy into thermal and kinetic energy. 
In addition to the prescription by \citet[][]{2006ApJ...643...92L}, we initially inject thermal energy at each feedback event. This is motivated  by the fact that even our highest resolution is typically not high enough to properly resolve the injection of kinetic energy by jets, and its fast thermalisation at a few tens of kiloparsecs from the injection site. 

About $30 \rm ~Myr$ after the burst, bipolar large scale outflows emerge as a natural byproduct of the vertical current, associated with the toroidal magnetic field. Next to the injection site, the outflows accelerate gas particles up to several $v_j \sim 10^3\rm km/s$ (see Table 1), depending on the output AGN power, and the kinetic energy is mostly thermalised within a few $\sim 10^2 ~\rm kpc$ from the SMBH. 

Table 1 gives more details on the list of parameters describing the jet launching. We also added a simple control run (run 0) without jet feedback. 

 \subsection{\CRaTer-simulations of Lagrangian tracer particles}

As in Paper I, we are only concerned in the fate of relativistic electrons injected by radio jets. Therefore, the passive Lagrangian tracers used to simulate electron spectra were injected only once in each run - i.e. at the epoch when jets from our radio galaxies were released into the computational domain.
 We used the Lagrangian code \CRaTer \ to follow the spatial evolution of the cosmic-ray electrons in our runs, as detailed in our Paper I. In post-processing, we injected  $\sim 3 \cdot 10^4$ particles in the jet region at  $z=0.5$ with a mass resolution $m_{\rm trac}=5\cdot 10^5 ~M_{\odot}$, and evolved them using all snapshots of the simulation for $\sim 100$ timesteps.  The various grid quantities, e.g. density, velocity, are assigned to the tracers using a cloud-in-cell (CIC) interpolation method. Further details on the full procedure implemented to advect tracers in our simulations can be found in \citet{wi16} and in Paper I.
Shocks are measured on tracers based on the temperature jump they experience from one timestep to the following.  Our tracer particles also keep track of the local fluid divergence, $\nabla \cdot \vec{v}$, and of the fluid vorticity, $\nabla \times \vec{v}$, which serve as proxies for the local turbulence.

\begin{figure*}
\begin{center}
\includegraphics[width=0.95\textwidth]{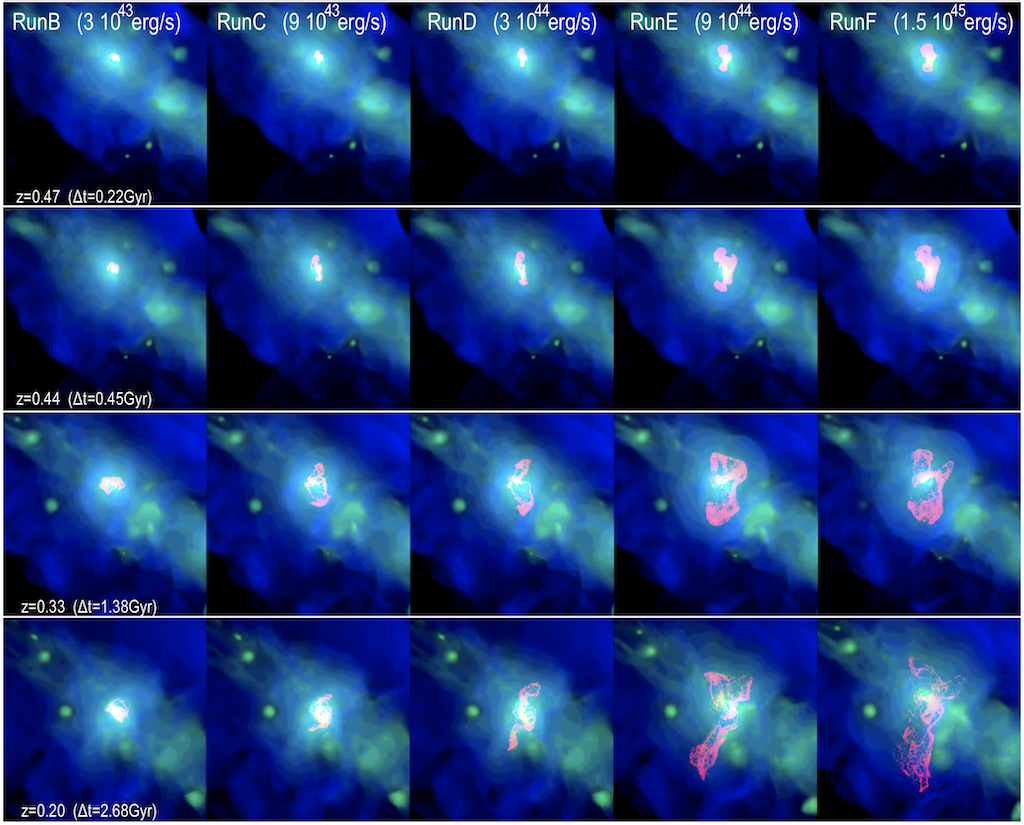}
\end{center}
\caption{RGB rendering of the evolution of thermal gas and passive tracers (marking relativistic electrons injected by jets) in our runs. The red colour marks the density of passive tracers, the blue the gas temperature and the green the X-ray emission. Each panel has a side of  5.5 Mpc (comoving).}
\label{fig:movie2}
\end{figure*}

\subsection{Injection of relativistic electrons by radio galaxies}
\label{subsec:brats}

In all runs considered in this work, tracers were only released into the ICM by the feedback event of the central active galaxy (and its central SMBH).
Unlike in our Paper I, we do not initialise electrons with a simple power law energy distribution. Instead, we generate a more realistic initial distribution of electron energies, capable of reproducing the observed radio spectra of real radio galaxies.
To do so, we rely on the "CIOFF" (Continuous Injection-Off, e.g. \citealt{1994A&A...285...27K}) model, which typically gives a good fit of observed radio spectra of remnant/old radio plasma  \citep[e.g.][]{2011A&A...526A.148M,2016A&A...585A..29B,2020MNRAS.496.3381R}. 
This model assumes a first phase during which new particles are continuously injected by the source (e.g. a continuous injection phase), with a power-law energy spectrum  $p~N(p)\propto p^{-\delta}$, which translates into a power-law radiation spectrum with spectral index $\alpha_{\rm inj} = (\delta-1)/2$. During this phase the radio spectrum steepens above a break frequency $\nu_1$ as $\alpha_{\rm inj}+0.5$. The acceleration of new particles stops after a time  $t_{\rm CI}$, and a new break appears in the spectrum, $\nu_2$, beyond which an exponential drop develops. In our simulations we assumed an energy injection index of  $2.2$, leading to a radio spectral injection index equal to $0.6$.

Our initial spectra are referred to an age of $t_{\rm on}\approx 32 \rm ~Myr$ (one root grid time step, corresponding to the entire elapsed time since the SMBH feedback began), followed by a short "off" phase, lasting $t_{\rm off}$ (typically $1-10 \rm Myr$ in our tests).  

The comparison of our simulated spectra (see Appendix) with the BRATS predictions for sources with the same age shows that, to a very good level, the input spectra of our simulated radio galaxies are compatible with reality. This aspect is not particularly crucial for this work, in which we are mostly focused on the long time evolution of fossil electrons, but allows us also to use our simulated radio galaxies for a comparative study of X-ray and radio estimates of the age and energetics of cavities (subject of forthcoming work).

Exactly as in Paper I, each injected tracer particle evolves a full spectrum of electrons in momentum space, $N(p)$, exclusively injected by radio jets, including loss and gain terms as detailed in the next Section. 

The normalisation of the relativistic electron spectra is a free parameter in our model. For the sake of simplicity, we imposed the number density of relativistic electrons to be a fixed fraction of the proton density in the jet region,  similar to other numerical works in the literature \citep[e.g.][]{2012ApJ...750..166M,2021MNRAS.505.2267M}. In all runs we adopted $X_e=10^{-2}$ in all runs, which is within the plausible range of uncertainties in the literature ( $\sim 10^{-3}-0.1$), and it also produces a realistic level of radio emission from our jets (Section \ref{subsec:radio}).

The total number of relativistic electrons (see the following Section) simulated by our tracers is given in the last row of Table \ref{table:tab1}: it ranges from $N_{\rm CRe} \approx 1.2 \cdot 10^{64}$ to $2.7 \cdot 10^{65}$ (for reference, the typical number of thermal electrons within the entire virial volume of the group is in the $\sim 5 \cdot 10^{70}$ ballpark), meaning that each of our simulated tracers evolved $\sim 10^{60}-10^{61}$ electrons.
It shall be noticed that the exact number of relativistic electrons injected in each model is non constant, as it depends on the local proton number density within jets, as well as on the volume covered by jets when tracers are first initialised. Tracers are injected at the first root grid time step since the feedback from SMBH particle is activated, i.e. after $t_{\rm on}=32 \rm ~Myr$. Since different jets expand in a slightly different amount during $t_{\rm on}$, as well as they carve more rarefied cavities as function of their power, the total number of injected relativistic electrons in each run depends on the combination of the two. 

It shall be noticed that only the spectra of our models where only energy losses are included can be rigidly renormalised, in the following,  for any other assumed value of $X_e$.
Instead in models where the additional acceleration by shocks is included, we shall see that the total number of radio emitting electrons gets typically dominated by the injection of new electrons accelerated via diffusive shock acceleration.
Therefore, the late time evolution of our electron spectra, in models with shocks and turbulence, is rather independent on the assumed value of $X_e$. 

\subsection{Evolution of electron spectra}
\label{subsec:fokker}

We solve the time-dependent diffusion-loss equation of relativistic electrons represented by tracer particles with an updated version of the relativistic electron solver \footnote{We shrare the public version of our ROGER solver at the url   \url{https://github.com/FrancoVazza/JULIA/tree/master/ROGER}}
method already given in Paper I. It is based on the standard \citet{1970JCoPh...6....1C} finite-difference scheme, written in 
Julia (https://julialang.org) and is typically run in parallel on 8 cores on a laptop.
We used $N_b=100$ momentum bins equally spaced in $\rm log (p)$, in the $p_{\rm min} \leq p \leq p_{\rm max}$ momentum range (with $P=\gamma m_e v $ and $p=P/(m_e c)$ is the normalised momentum of electrons. In all production runs we used $p_{\rm min}=10$ and $p_{\rm max}=10^6$ (hence $d\log (p)=0.05$).

We compute the evolution of the number density of relativistic electrons  as a function of  momentum $N(p)$ for each tracer: 

\begin{equation}
    {{\partial N}\over{\partial t}}
    =
    {{\partial}\over{\partial p}} \left[
    N \left( \left|{{p}\over{\tau_{\rm rad}}}\right| + \left|{{p}\over{\tau_{\rm c}}}\right| +
    {{p}\over{\tau_{\rm adv}}} - \left|{{p}\over{\tau_{\rm acc}}}\right| \right) \right].
    \label{eq11}
\end{equation}
In the absence of diffusion, injection and escape terms the Equation \ref{eq11} essentially is Liouville's equation, for which a numerical solution is obtained as: 

\begin{equation}
N(p,t+dt)=\frac{{{N(p,t)}/{dt}} + N(p+dp,t+dt){{\dot{p}}}} 
{1/dt + {\dot{p}}/dp} + Q_{\rm inj} ,
    \label{eq13}
\end{equation}
where in the splitting-scheme for finite differences we assumed $N(p +dp/2)=N(p+dp)$ and $N(p-dp/2)=N(p)$, and $Q_{\rm inj}$ accounts for the injection by radio galaxies or shocks. Whenever present, injection is treated as an instantaneous process owing to the time scales much shorter than the time step of our integration.

Compared to Paper I, a relevant modification of our solver is that we employ sub-cycling between the time steps used for the advection of tracers in order to resolve the fast cooling of electrons at high frequency ($\geq 1 \rm ~GHz$). 
Our tests have confirmed that the steepening of the emission beyond the synchrotron break frequency is typically well matched only if $dt'=10 \rm ~Myr$ is used in our RES, which therefore requires a subcycle with additional four steps in between the advection timesteps.
With this approach, if a particle is shocked between two advection timesteps, we adopt the shock injection only at the last of the solver's sub-cycling timesteps, so that the prompt radio signal by freshly injected electrons is not artificially dampened by a too long solver integration time.  

The gain/loss processes described by our model are:

\begin{itemize}
\item Energy losses from radiative, Coulomb and expansion (compression) processes are treated as in Paper I. We neglected bremsstrahlung losses since their time scale is significantly larger for all other losses for the ICM physical condition considered here.

\item Acceleration from diffusive shock acceleration (DSA). The shock kinetic energy flux that we assumed to be converted into the acceleration of cosmic rays is: $\Psi_{\rm CR} = \xi_e ~\eta(\mathcal{M}) \rho_u (V_s^3 dx_t^2)/2$ ,
in which $\rho_u$ is the pre-shock gas density,  $V_s$ is the shock velocity and the combination $\xi_e ~\eta({\mathcal{M}})$ gives the cosmic-ray acceleration efficiency. The latter comprises a prescription for the energy going into cosmic rays, $\eta(\mathcal{M})$, and the electron-to-proton acceleration rate, $\xi_e$.  As in Paper I, we use the polynomial approximation by \citet{kj07} for $\eta(\mathcal{M})$, which includes the effects of finite Alfv\'{e}n wave drift and wave energy dissipation in the shock precursor. The working surface associated with each tracer, $dx_t^2$, is adjusted at run-time as  $dx_t^3 = dx^3/n_{\rm tracers}$, where $dx^3$ is the volume initially associated with every tracer and ($n_{\rm tracer}$ is the number of tracers in every cell.

The value of $\xi_e$ is very uncertain for the weak ($\mathcal{M} \leq 5$) shocks in the ICM \citep[e.g.][]{2011ApJ...733...63R,guo14a,guo14b,2020ApJ...897L..41X}. As in Paper I, we link the injection of fresh electrons to the injection of  protons in the DSA scenario,  by requiring an equal number density of cosmic-ray electrons and protons above a fixed injection momentum,  $\xi_e=(m_p/m_e)^{(1-\delta_{\rm inj})/2}$, following \citet{2013MNRAS.435.1061P}{\footnote{We notice that our group has also recently explored more self-consistent dependencies between the minimum momentum of electrons and the shock acceleration efficiency \citep[][]{inc22}, following \citet{kang21}, yet here we prefer to keep the same formalism of Paper I, for the sake of comparison.}.}

Here we also neglect the dependence of the shock acceleration efficiency on the local magnetic field topology. In previous work, we showed that shocks in the ICM are mostly quasi-perpendicular and, hence, suitable for the acceleration of electrons by DSA \citep[e.g.][]{wittor20,Banfi20}.  The injection momentum  of electrons $P_{\rm inj}$ is linked to the thermal momentum of particles, i.e. $P_{\rm inj}= \xi P_{\rm th}$ ($P_{\rm th}=\sqrt{2 k_b T_d m_p}$). 

Relativistic electrons are injected by shocks with a power-law momentum distribution \citep[e.g.][]{sa99}:
\begin{equation}
    Q_{\rm inj}(p) = K_{\rm inj} ~p^{-\delta_{\rm inj}} \left(1-\frac{p}{p_{\rm cut}}\right)^{\delta_{\rm inj}-2} ,
\end{equation}
where the initial slope of the input momentum spectrum, $\delta_{\rm inj}$, follows from the standard DSA prediction, $\delta_{\rm inj} = 2 (\mathcal{M}^2+1)/(\mathcal{M}^2-1)$.  
$p_{\rm cut}$ is the cut-off momentum, which we set for every shocked tracer as the maximum momentum, beyond which the radiative cooling time scale gets shorter than the acceleration time scale, $\tau_{\rm DSA}$. 

For all plausible choices for the diffusion description, the acceleration time is governed by the energy dependent diffusion coefficient of electrons, giving acceleration time scales that are many orders of magnitude smaller than the typical cooling time of radio emitting electrons. Hence, we simply assume the injected distribution of electrons to be a power law within our momentum range of interest.

This approach allows us to model the shock injection by DSA by adding the newly created population of particles at the end of each timestep (see Equation~\ref{eq13} above), without integrating a source term, which is instead needed for the much slower re-acceleration by turbulence (see below).  

The normalisation factor, $K$, follows by equating the cosmic ray energy flux crossing each tracer volume element, and the product between the total energy of cosmic rays advected with a post-shock velocity  ($v_d$): 
$\Psi_{\rm CR} ~dx_t = v_d E_{\rm CR}$ ($v_d$ is the post-shock velocity) and thus

\begin{equation}
E_{\rm CR} = \int_{p_{\rm inj}}^{p_{\rm cut}} Q_{\rm inj}(p) T(p) dp ,
\end{equation}
with $Q_{\rm inj}(p)$ defined as above and $T(p) = (\sqrt{1+p^2}-1)m_e c^2$. The integration yields 

\begin{equation}
E_{\rm CR} = \frac{K_{\rm inj} m_e c^2}{\delta_{\rm inj}-1} \left[\frac{B_x}{2} \left( \frac{\delta_{\rm inj}-2}{2},\frac{3-\delta_{\rm inj}}{2}\right) + p_{\rm cut}^{1-\delta_{\rm inj}} \left(\sqrt{1+p_{\rm cut}^2}-1  \right)  \right] , 
\label{eq:ECR}
\end{equation}
where $B_x(a,b)$ is the incomplete Bessel function and $\rm x=1/(1+p_{\rm cut}^2)$, as in \citet[][]{2013MNRAS.435.1061P}.

\item The process of diffusive shock {\it re}-acceleration by shocks waves is modelled following \citep[e.g.][]{2005ApJ...627..733M,kr11,ka12}, and the particle spectrum until the new shock, $N_0(x)$, becomes

\begin{equation}
N(p)=(\delta_{\rm inj}+2) \cdot p^{-\delta_{\rm inj}} \int_{p_{min}}^p N_0(x) x^{\delta_{\rm inj}+1} dx ,
\end{equation}
where $\delta_{\rm inj}$ is the local slope within each energy bin.

\item The Fermi II re-acceleration via stochastic interaction with diffusing magnetic field lines in super-Alfvenic turbulence is computed following \citet{bv20}, and it depends on the amplitude of the local turbulent velocity, $\delta V_{\rm turb}$, measured within the scale $L$.

Following Paper I, we used the gas vorticity measured by tracers to estimate the local level of solenoidal turbulence experienced by particles, $\delta V_{\rm turb}= |\nabla \times \vec{v}| L$. The gas vorticity reasonably prescribes the local turbulent velocity responsible for the stochastic acceleration, which is only related to the solenoidal component of turbulence \citep{bv20}. For simplicity, we used the same fixed reference scale of $L \approx 27$~kpc to compute vorticity via finite differences (i.e. 3 cells on the high-resolution mesh). The vorticity is used to compute the turbulent re-acceleration model outlined below, and thus the (solenoidal) turbulent kinetic energy flux
\begin{equation}
F_{\rm turb} = \frac{\rho \delta V^3_{\rm turb}}{2L}.
\label{eq:Fturb}
\end{equation}
The Fermi II re-acceleration mechanism adopted here closely follows what we presented in \citet{bv20} and it operates in large-scale super-Alfv\'enic solenoidal turbulence, allowing electrons to be reaccelerated stochastically while diffusing across regions of magnetic reconnection and dynamo. If solenoidal turbulent modes are dominant and the turbulence is 
super-Alfv\'enic ($\mathcal{M}_A^2=(\delta V_{\rm turb} /v_A)^2 \sim \mathcal{M}_{\rm turb}^2 \beta_{pl} \gg 1$, where $v_A$ is the Alf\'en velocity, $\mathcal{M}_{\rm turb}$ is the turbulent Mach number, $\mathcal{M}_{A}$ is the Alfv\'enic Mach number and $\beta_{pl}$ is the plasma beta), then this acceleration mechanism is expected to become faster than transit-time damping acceleration with compressive turbulence \citep[][]{bl07}. Following  \citet[][]{2016MNRAS.458.2584B}, we used a  diffusion
coefficient in the particle momentum space :
\begin{equation}
    D_{pp} \sim \left({{l_A}\over{\lambda_{\rm mfp}}} \right)^2 
    {{v_A^2}\over{D}} p^2
    \label{dpp1}
\end{equation}

where $\lambda_{\rm mfp}$ is the effective mean free path of electrons, $l_A$ is the Alfv\'en scale, defined as as the scale where the velocity of turbulent eddies equals the Alfven velocity $v_A$  \citep[e.g.][]{bl07}, and $D \sim 1/3 c \lambda_{\rm mfp}$ is the  spatial diffusion coefficient along magnetic field lines. The underlying assumption is that, in super-Alfv\'enic turbulence, the hydro motions set $\lambda_{\rm mfp} \leq l_A$ because electrons travelling in tangled magnetic fields change directions on this scale preserving the adiabatic invariant.

Following \citet[][]{2016MNRAS.458.2584B} and \citet{bv20}, we adopt a value of the effective $\lambda_{\rm mpf}$ that is a fraction of (similar to) $l_A$, specifically $\lambda_{\rm mfp} \approx 1/2 l_A$, under the assumption that  in super-Alfv\'enic turbulence the interaction is
driven by the largest moving mirrors on scales $\sim l_A$; indeed  this limits the
effective mean free path to $\lambda_{\rm mfp} \leq l_A$
 \citep[][]{2016MNRAS.458.2584B}. 
 Since in the Kolmogorov theory the energy flux is scale independent, while the Alfven scale (defined above) is not, we can equate $dV_{\rm turb}^3/L = v_A^3/l_A$, which allow us to rewrite the diffusion coefficient in momentum space of Equation \ref{dpp1} as:

\begin{equation}
D_{pp} \simeq
{{48}\over c} {{F_{\rm turb}}\over{\rho v_A}}
p^2 
\label{dpp2}
\end{equation}

\noindent
which is what we used also in this work, and it yields a re-acceleration timescale of order:

\begin{equation}
    \tau_{\rm ASA} = \frac{p^2}{4~D_{\rm pp}  } = 125  ~Myr \rm  \frac{L/(500) ~ B}{\sqrt{n/10^{-3}} ({\delta V_{\rm turb}/10^8})^3} ,
    \label{eq:ASA}
\end{equation}

in which the outerscale $L$ is measured in kpc, the magnetic field $B$ is measured in $\rm \mu G$, the density $n$ is measured in particle per cubic $\rm cm$ and the turbulent velocity is measured in $\rm cm/s$.

 Equation \ref{eq:ASA} gets incorporated in Equation~\ref{eq13} by adding a positive $dp/dt \approx p/\tau_{\rm ASA}$ term. 

It shall be noticed that the level of turbulent motions measured across the $L=27 \rm ~kpc$ scale used to compute vorticity is only a fraction of the the entire range of turbulent velocity dispersion developed in the ICM, which is always found to have an outer scale of the order of $\sim 100-500 \rm ~kpc$. However, it is important to stress here that the relevant quantity used in  Equation \ref{eq:ASA} is the (solenoidal) turbulent kinetic energy flux, $F_{\rm turb}$ (Eq. \ref{eq:Fturb}),  which is a constant if measured within the entire range of scales and if the case the turbulent spectrum is scale invariant, i.e. constant from the outer scale of turbulence to the dissipation scale. Therefore, given that our previous analysis of similar simulations \citep[e.g.][]{va09turbo,va11turbo,va17turb,2020MNRAS.495..864A,2022A&A...658A.149S} as well as many other independent works \citep[e.g.][]{ga13, 2018MNRAS.481.1075S,2019ApJ...874...42V,2021MNRAS.504..510V} have established that turbulence in the simulated ICM develops quite close to the standard Kolmogorov model, our estimated local $\delta V_{\rm turb}$ is ensured to give a reliable proxy for the turbulent kinetic energy flux across the cascade, which is used both to estimate the Fermi II turbulent reacceleration in Equation\ref{eq:ASA}, as well as to estimate the (unresolved) amount of magnetic field amplification in our runs (see Section\ref{subsec:Bfield}).

At variance with Paper I, in this work we do not include the effect of Fermi II re-acceleration during the same timestep in which a tracer is found to be shocked, because in this case the our estimate of turbulence is likely to be overestimated because of the vorticity generated close to shocks \citep[e.g.][]{wi17b}, is not part of a truly turbulent cascade. 

\end{itemize}

\subsection{Synchrotron emission}
\label{subsec:sync}

In order to speed up the computation of the synchrotron emission for our simulated distribution of relativistic electrons (which we wish to compute for $\sim 10^2$ timesteps, $\sim 10^4$ tracers, $3$ models and $5$ cluster resimulations),  we resort to fitting functions for the synchrotron emissivity, following \citet{2014MNRAS.442..979F}. 
 The instantaneous integrated synchrotron power for a distribution of relativistic electrons $N(\gamma)$, between $\gamma_1$ ad $\gamma_2$ and with an isotropic pitch angle distribution, is given by:
\begin{equation}
    P({\nu})=\sqrt{3} \, e^2 \,2 \, \pi \, \, \frac{\nu_{\rm L}}{c}\int^{\gamma_2}_{\gamma_1}\,N(\gamma) \, F\left(\frac{\nu}{\nu_{\rm c}}\right ) d\gamma .
    \label{eq:Psync}
\end{equation}{}
Here, $\nu_{\rm L}=eB/(2\pi m_e c)$ is the Larmor gyration frequency and $\nu_{\rm c}=(3/2)\gamma^2\nu_{\rm L}$ is the characteristic frequency of synchrotron radiation.
The synchrotron function $F(\nu/\nu_{\rm c})$ is,
\begin{equation}
    F(y)=y\int^{\infty}_{y}K_{5/3}(y') \, dy' ,
\label{eq:kernel}
\end{equation}{}
where $y=\nu/\nu_{\rm c}$ and $K_{5/3}(y')$ is the modified Bessel-function.

Following \citet{2014MNRAS.442..979F} we can parameterise the radio spectral power as $P({\nu})=P_1F_{\rm \delta}(x,\eta)$, i.e.  in terms of a dimensionless frequency $x=\nu/\nu_1$ with $\nu_1=(3/2)\gamma_1^2\nu_{\rm L}$ and of the Lorentz factor ratio $\eta\equiv \gamma_2/\gamma_1$.
\noindent The parametric function  $F_{\rm \delta}(x,\eta)$ is described by two fitting functions, depending on the ratio between $\nu$ and $\nu_1$ : 

\begin{equation}
\begin{centering}
\label{eqn:Fp}
F_{\rm \delta}(x,\eta)= \begin{cases}
    F_{\rm \delta}(x) \, - \, \eta^{-\delta+1} \, F_{\rm\delta}\left(x/\eta^2\right), & \text{for $x<x_{\rm c}$}, \\
    \sqrt{\frac{\pi}{2}} \, \eta^{-\delta+2} \, x^{-1/2} \, \exp{(-x/\eta^2)}\left[ 1 +  a_{\rm \delta} \, \frac{\eta^2}{x} \right], & \text{for $x\geq x_{\rm c}$}.
    \end{cases}
\end{centering}
\end{equation}
\noindent Here, $\rm \delta$ is the slope of the energy spectrum between to contiguous energy bins in the CR spectrum.
\noindent  The fitting formula for $F_{\rm \delta}$ is described by 
\begin{align*}
    F_{\rm \delta}\approx & \, \kappa_{\rm \delta} \, x^{1/3} \, \exp{(a_1 \, x^2 \, + \, a_2 \, x \, + \, a_3 \, x^{2/3})}  \\
    &+ \, C_{\rm \delta} \, x^{-(\delta-1)/2}[1 \, - \, \exp{(b_1 \, x^2)}]^{\delta/5+1/2} ,
\end{align*}{}
\noindent applicable for $1<\rm \delta<6$. Here, $\kappa_{\rm p}$ and $C_{\rm \delta}$ depend on the Gamma function as
\begin{align*}
    \kappa_{\rm \delta} &= \frac{\pi \,  2^{8/3}}{\sqrt{3} \, (\delta-1/3) \, \Gamma(1/3)} \\
    C_{\rm \delta} &=\frac{2^{(\delta+1)/2}}{\delta+1} \, \Gamma\left(\frac{\delta}{4}+\frac{19}{12}\right) \, \Gamma \left(\frac{\delta}{4}-\frac{1}{12}\right).
\end{align*}
\noindent In the equations above, $x_{\rm c}=(2.028-1.187\delta+0.240\delta^2) \, \eta^2$ and $a_{\rm \delta}=-0.033-0.104\delta+0.115\delta^2$, while the coefficients $a_1$, $a_2$, $a_3$ and $b_1$ in terms of $\delta$ are
\begin{align*}
    a_1 = &-0.14602+3.62307\times 10^{-2}\delta-5.76507\times 10^{-3}\delta^2\\
    &+3.46926\times 10^{-4}\delta^3 \\
    a_2 = &-0.36648+0.18031\delta-7.30773\times 10^{-2}\delta^2 \\
    &+1.12484\times 10^{-2}\delta^3-6.17683\times 10^{-4}\delta^4  \\
    a_3 = &9.69376\times 10^{-2}-0.48892\delta+0.14024\delta^2 \\
    &-1.93678\times 10^{-2}\delta^3+1.01582\times 10^{-3}\delta^4  \\
    b_1 = &-0.20250+5.43462\times 10^{-2}\delta-8.44171\times 10^{-3}\delta^2 \\
    &+5.21281\times 10^{-4}\delta^3  .
\end{align*}{}

For CR spectra steeper than $\delta=6$, the radio emission is negligible and therefore we set $P(\nu)=0$. 

For a recent application of this approach to compute the radio emission from Sagittarius A, we refer the reader to  \citet[][whose helpful compact notation we also used in the above derivation]{2021MNRAS.507.5281C}. 

This approach is very efficient to speed up the computation of synchrotron emission, as it
yields a correct prediction within a $\leq 10-20\%$ error, with a remarkable $\sim 10-100$ speedup (depending on the chosen number of energy bins) compared to more standard approaches based on the solution of the integral in the synchrotron kernel function (Equation~\ref{eq:kernel}).

\subsection{Treatment of magnetic fields}
\label{subsec:Bfield}

\begin{figure*}
    \centering
    \includegraphics[width=0.495\textwidth]{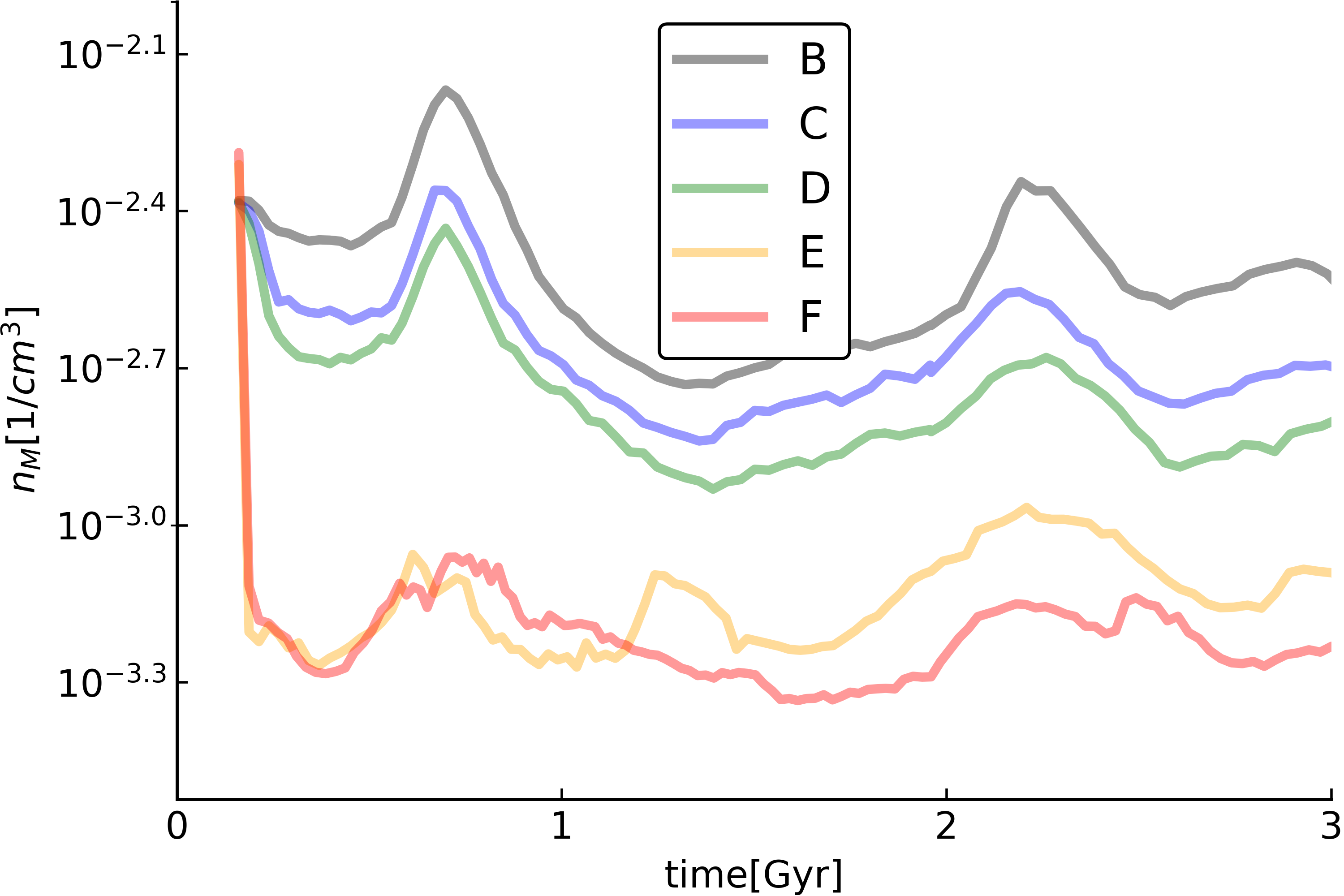}
    \includegraphics[width=0.495\textwidth]{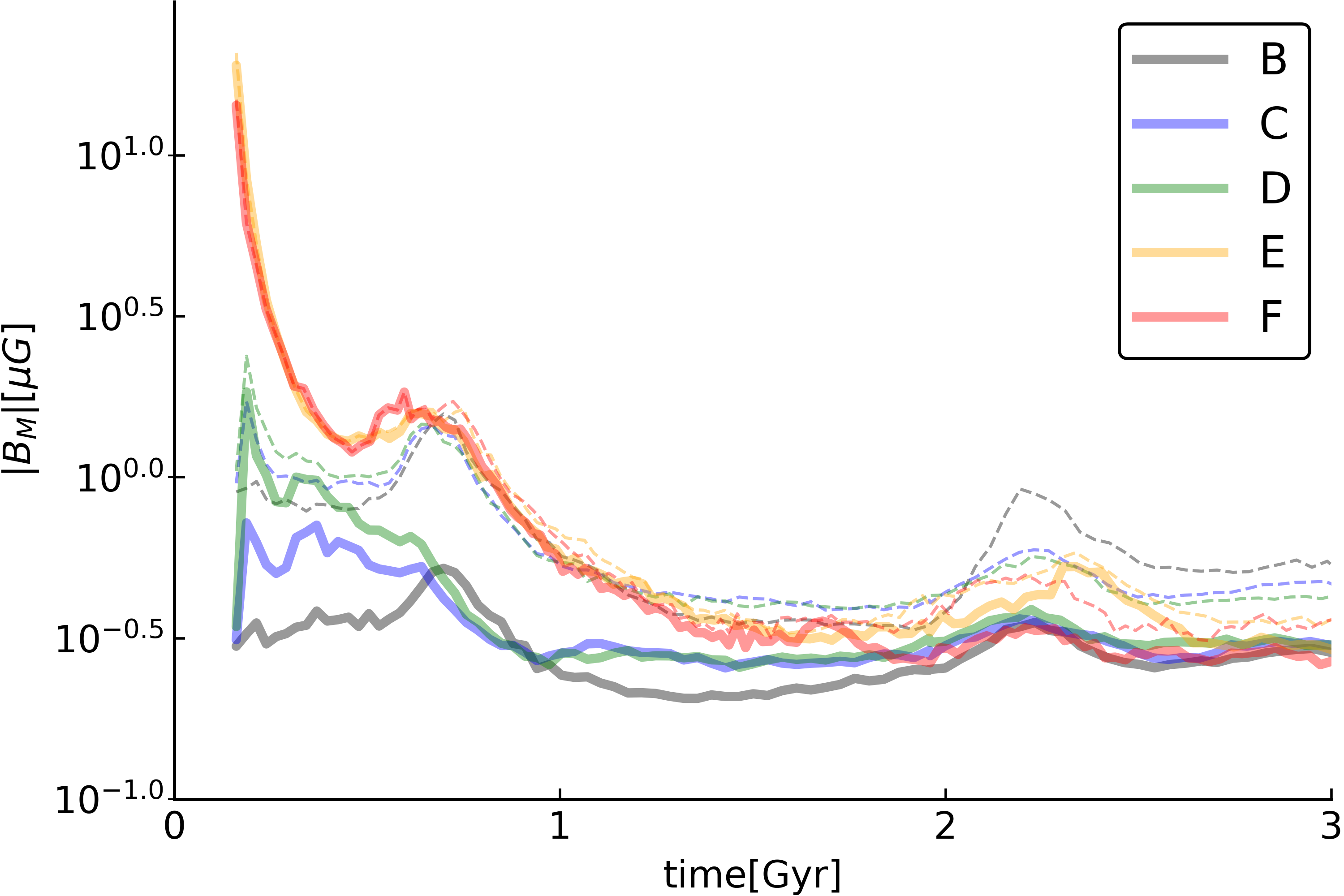}
     \includegraphics[width=0.495\textwidth]{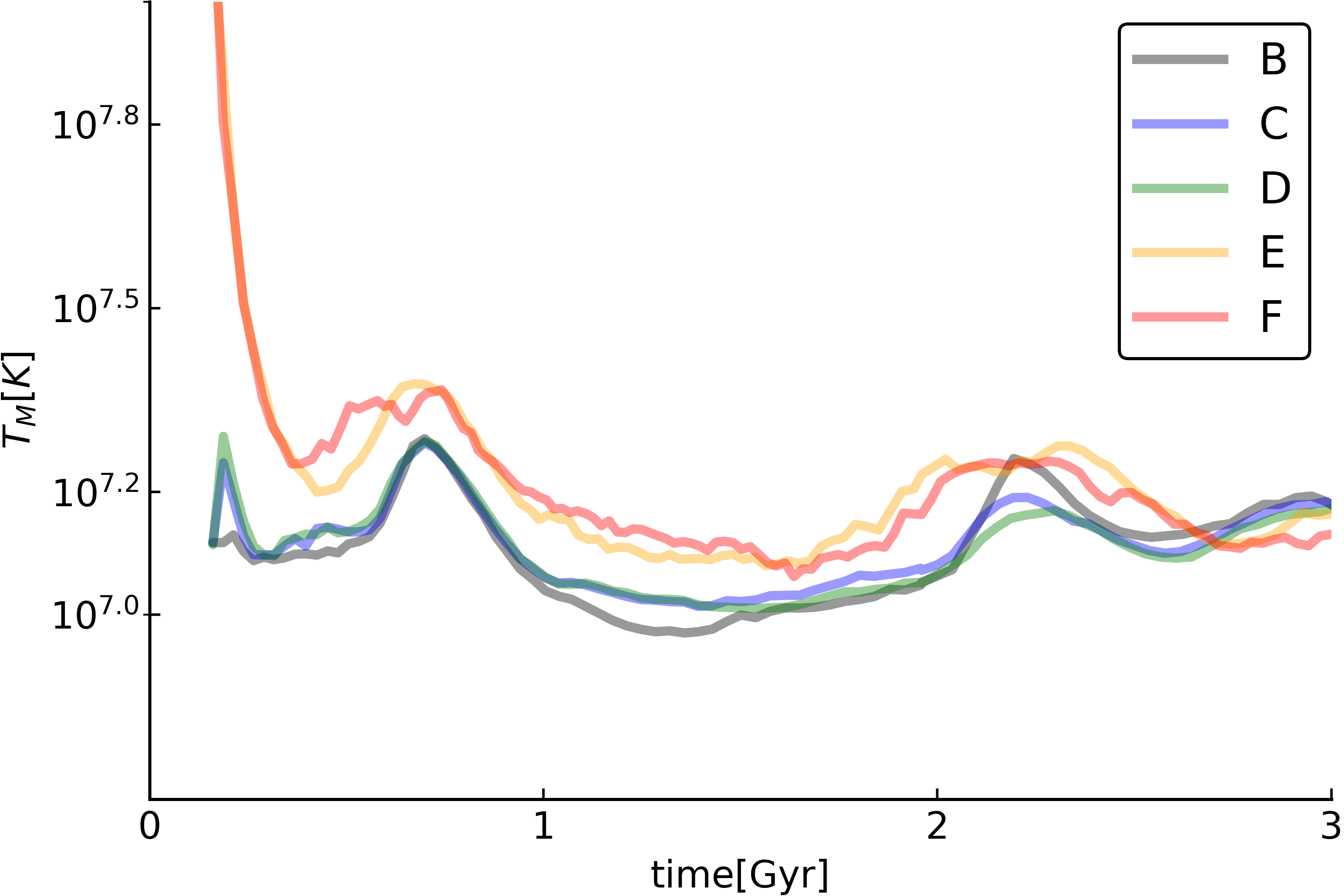}
     \includegraphics[width=0.495\textwidth]{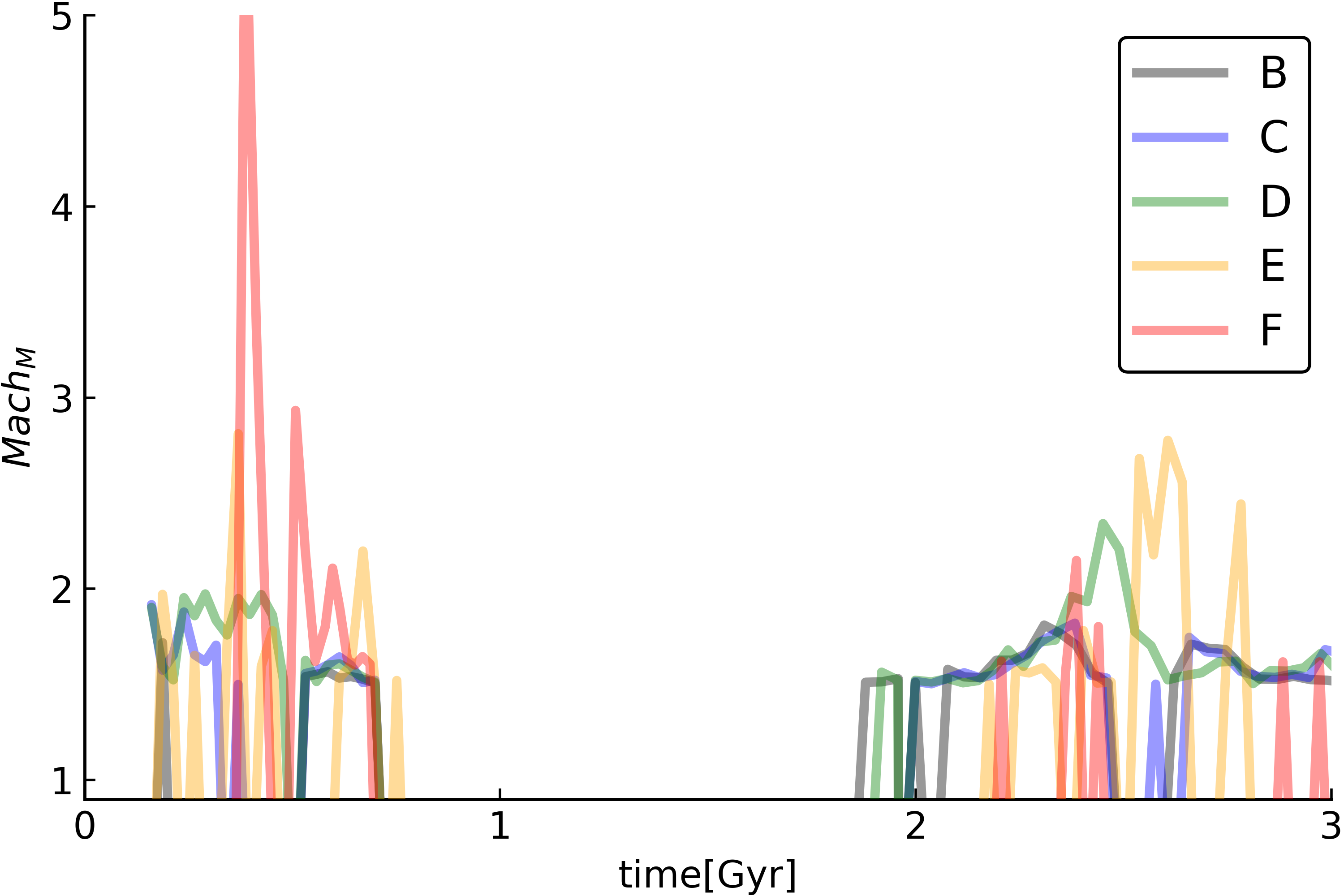}
      \includegraphics[width=0.495\textwidth]{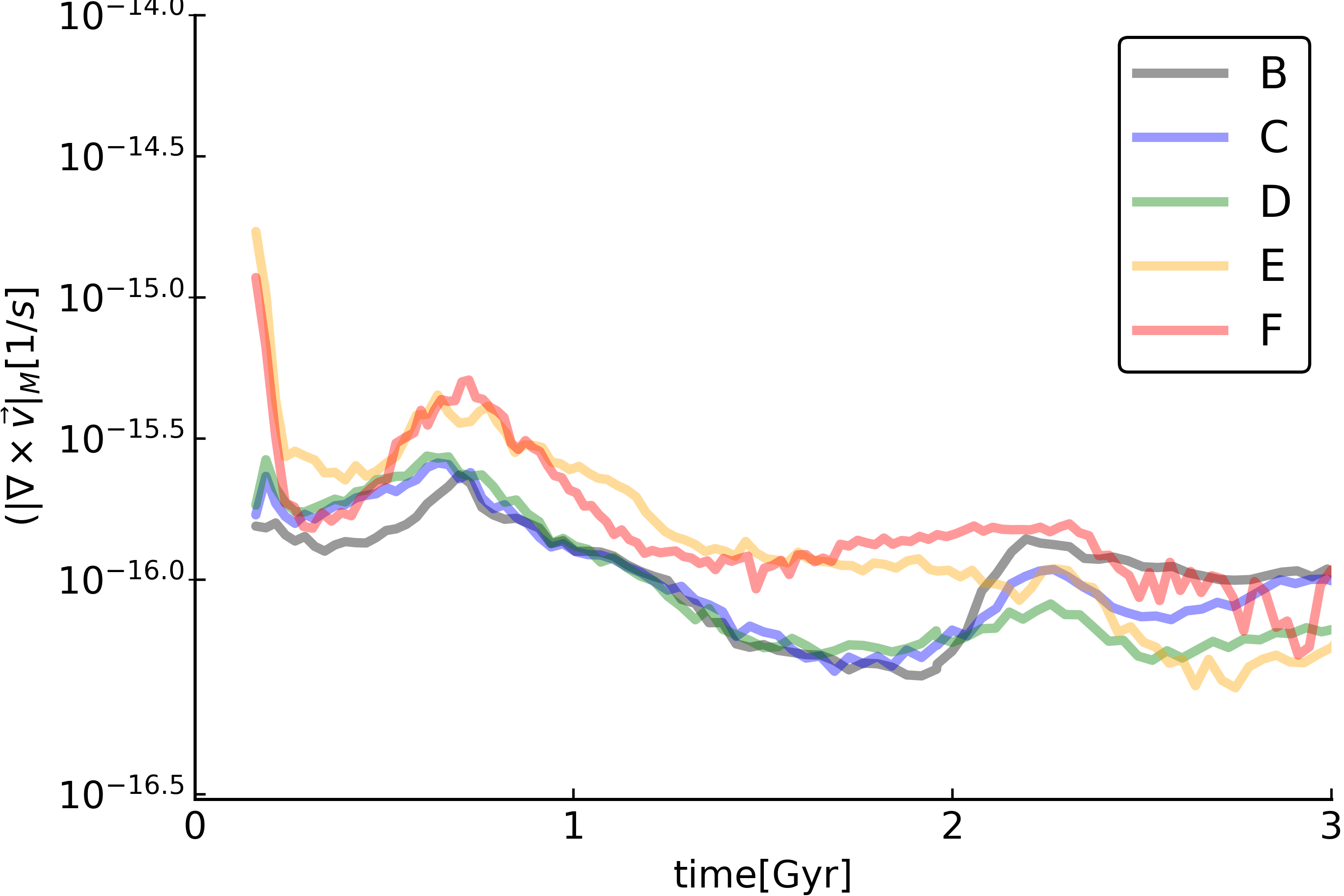}
      \includegraphics[width=0.495\textwidth]{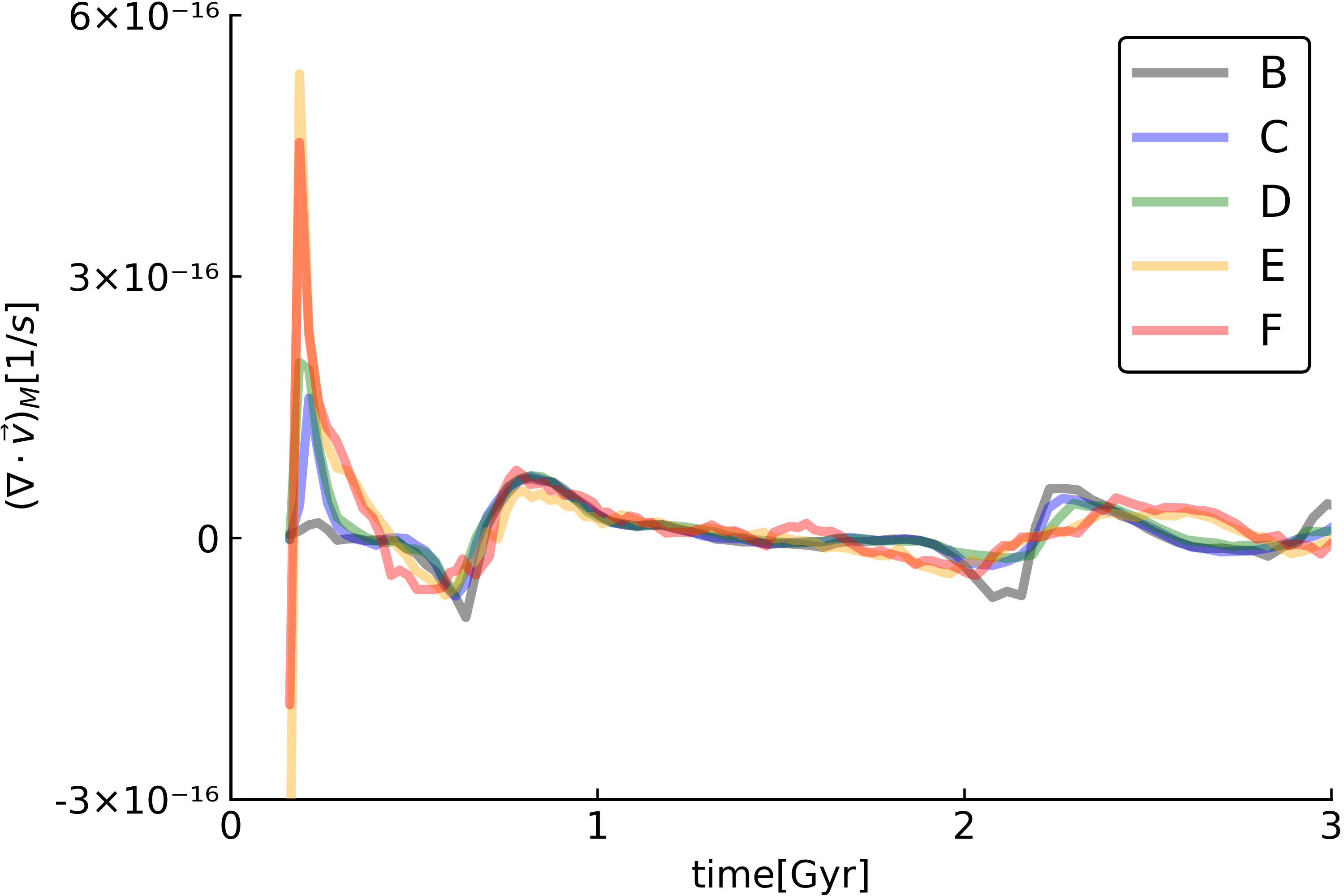}  
      
    \caption{Evolution of the median of comoving gas density, magnetic field strength, temperature, Mach number (only for shocked tracers), module of vorticity and divergence measured by tracers in all our runs with jets.  The time is measured since the start of the jets launching, i.e. from $z=0.5$. In the upper right plot, we show with solid lines the trend of the original magnetic field produced by our MHD simulation, while with dotted lines the rescaled magnetic field value, based on the local turbulent field, as estimated in Section\ref{subsec:Bfield}.}
    \label{fig:scatter}
\end{figure*}

\begin{figure*}
    \includegraphics[width=0.99\textwidth]{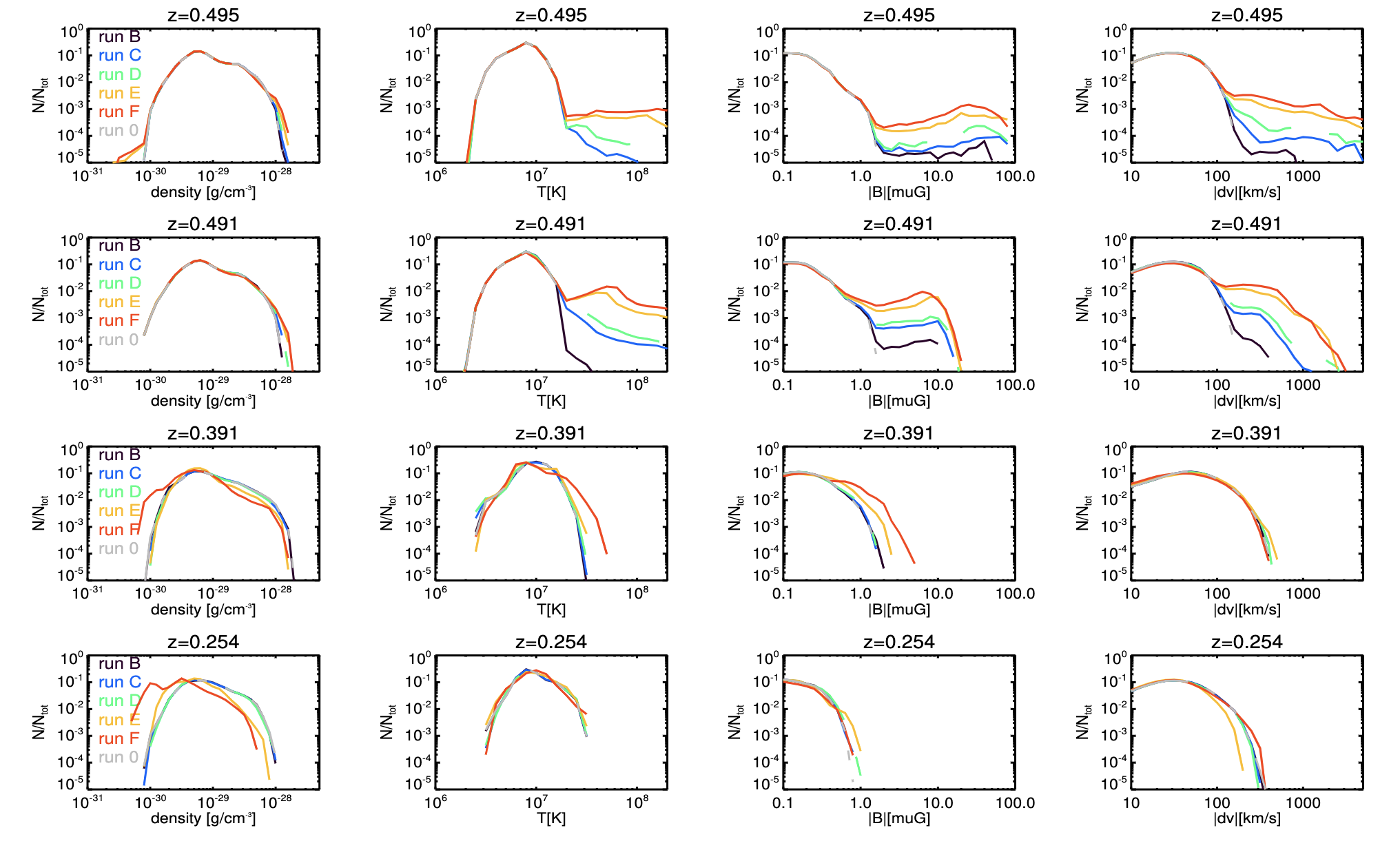}
    \caption{Distributions of gas density, gas temperature, velocity dispersion and magnetic field strength for four epochs, for a reference comoving $1^3 \rm ~Mpc^3$ around the (moving) cluster center of mass at four different redshifts. For comparison, we also report the distributions from the Run 0, without jets (grey lines).} 
    \label{fig:pdf_evol}
\end{figure*}

Magnetic fields evolve subject to compression and amplification of a uniform magnetic field with an initial value of $B_0=0.1  \ \mathrm{nG}$ in each direction. In addition, magnetised loops are injected by feedback events (Section~\ref{jet}). However, the mesh refinement and our MHD solver \citep[][]{ded02} do not ensure a large enough magnetic Reynolds number to prevent the amplification by a small-scale dynamo \citep[e.g.][]{review_dynamo}. 
As a result, the average magnetic field in our simulated cluster is in the $\langle B \rangle \sim 0.1 ~\rm \mu G $ range (see also Figure ~\ref{fig:pdf_evol} later), which is smaller than what is typically observed in galaxy groups of a similar size \citep[e.g.][]{2012MNRAS.423.1335G}, as well as predicted by numerical simulations at higher resolution \citep[e.g.][]{xu09,2014MNRAS.443.1482H}.
For this reason, unlike in Paper I, we decided to renormalize in post-processing the value of $B$ to be assigned to each tracer particle, in order to compute its energy evolution and synchrotron emission under a more realistic magnetisation level, expected in the plausible case of efficient small-scale dynamo amplification.
To do so, we followed the same approach of \citet{bv20} and assumed that after the turbulent kinetic energy cascade reaches dissipation scales, a fixed fraction ($\eta_B=2 \%$ in this case) of the energy flux of turbulence ($F_{\rm turb}$) is dissipated into the amplification of magnetic fields \citep[e.g.][]{bm16}. We thus estimated the magnetic field for each tracer from $B_{\rm turb}^2/8\pi \sim \eta_B F_{\rm turb} \tau \sim {1 \over 2} \eta_B \rho \delta V_{\rm turb}^2$, where $\tau= L/\delta V_{\rm turb}$ is the turnover time,  and $\delta V_{\rm turb}$ is the same of Equation~\ref{eq:ASA}. Whenever $B_{\rm turb} \geq B$ (where $B$ is the magnetic field recorded by a tracer), we replace the value of the MHD calculation with $B_{\rm turb}$ in the computation of radiative losses (Equation~\ref{eqn:Fp}), of the Fermi II re-acceleration (Equation~\ref{eq:ASA}) as well as of the synchrotron emission (Equation~\ref{eq:Psync}). We must notice that, since higher values of magnetic fields correspond to shorter radiative time scales, and longer re-acceleration time scales, the long term detectability in the radio band of our lobe remnant radio emission is considerably reduced compared to our estimates given in Paper I, where only the magnetic field directly produced by the simulation was considered. 
If we compare the median of the original magnetic field directly produced by the MHD simulation and the renormalised one, the difference between the average field strength across the entire sample of tracers is within a factor $\sim 2$ at most times (e.g. see Figure~\ref{fig:scatter}). In no cases, the rescaled magnetic field is comparable to the thermal gas pressure. This follows, by construction,  from the fact that the rescaled magnetic field energy is a fraction of the local kinetic gas energy, which is in turn only a fraction of the local thermal energy since the ICM motions are predominantly sub-sonic. 
However, the differences gets larger ($\sim 10$) for a sub-fraction of tracers in the most turbulent patches of the simulated ICM, where $\delta V_{\rm turb}$ is significant, but the effective resolution of the simulation is not enough to well resolve the "MHD scale" and produce dynamo amplification \citep[e.g.][]{review_dynamo}. Those tracers can dominate the detectable radio emission, and our post-processing renormalisation of the magnetic field is important in these cases.
While in the main paper we only discuss results obtained after renormalising the magnetic field of tracers to more realistic values, we also discuss in the Appendix the difference in the electron spectra simulated with or without our rescaling of the magnetic field.

\begin{figure*}
    \begin{centering}
    \includegraphics[width=0.45\textwidth]{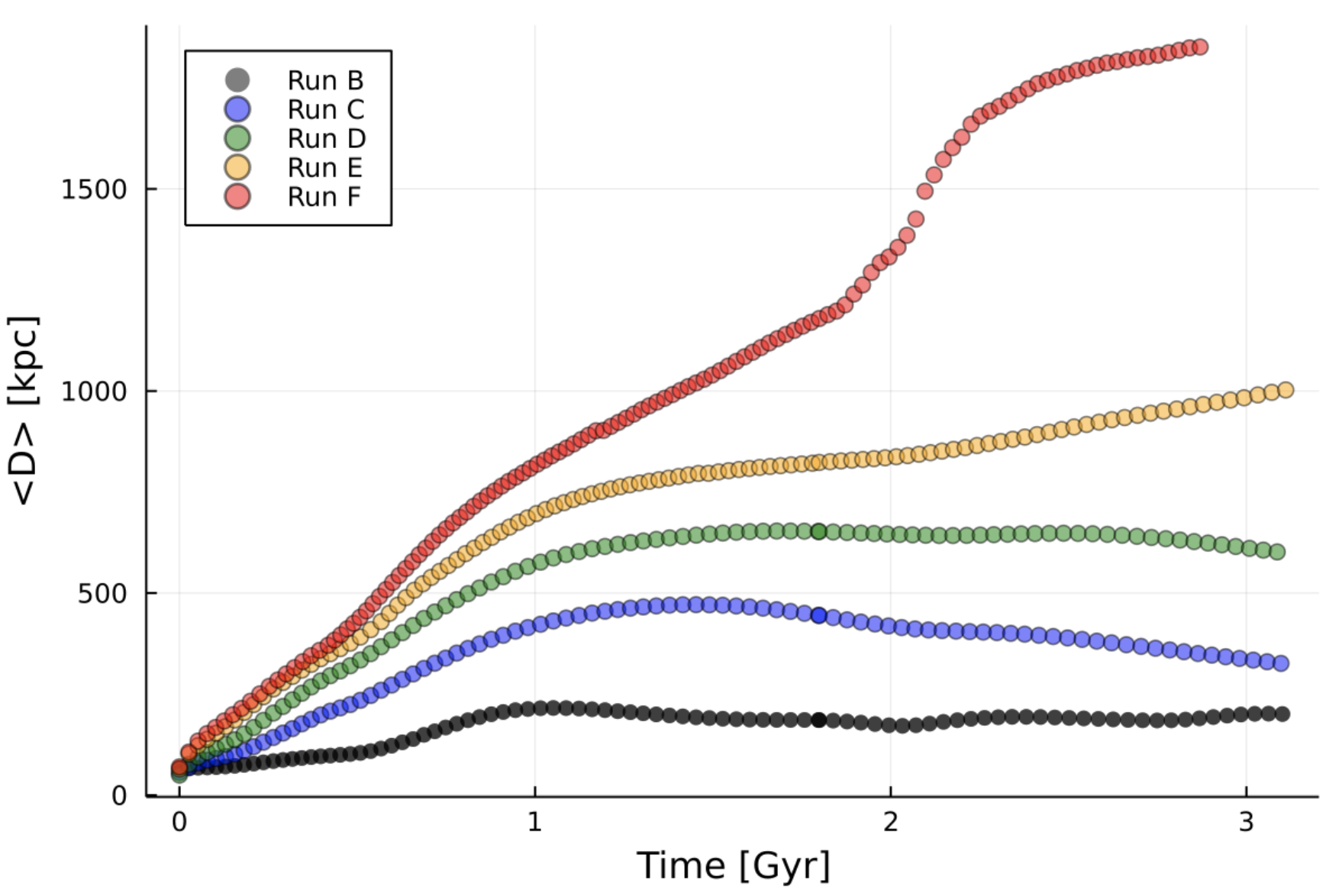}
    \includegraphics[width=0.45\textwidth]{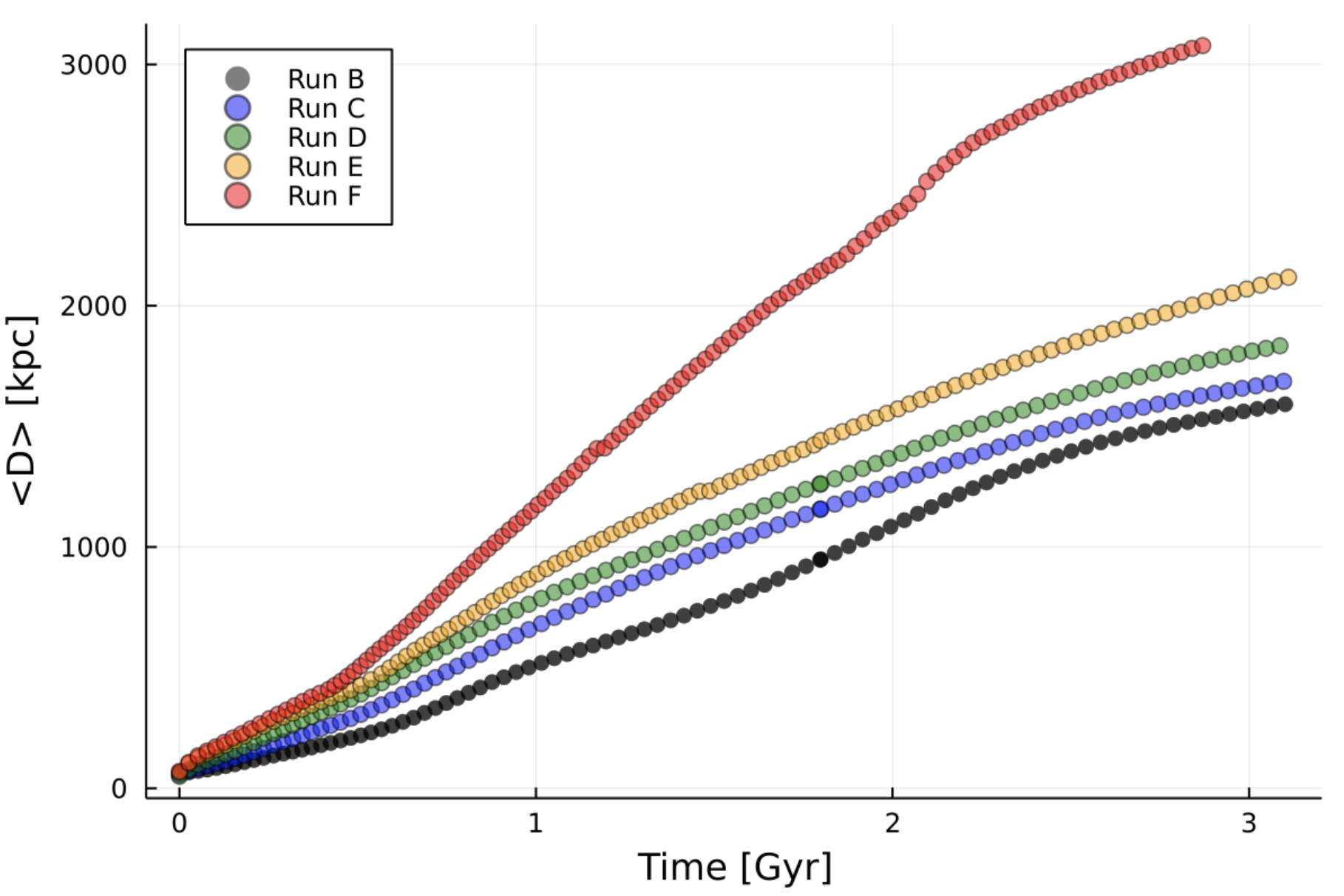}
    \caption{Evolution of the median distance travelled by tracers in all runs, either with respect to the particle center of mass at all epochs (left panel), or with respect to their initial injection site (right panel). The time is measured since the start of the jets launching, i.e. from $z=0.5$.}
    \end{centering}
    \label{fig:distance}
\end{figure*}

\begin{figure*}
\begin{centering}
\includegraphics[width=0.45\textwidth]{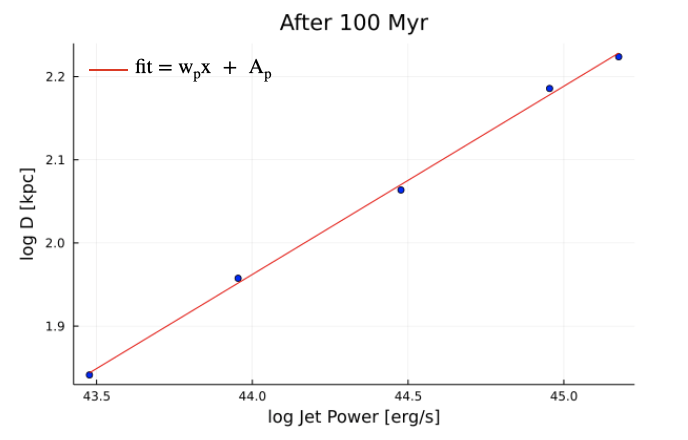} 
\includegraphics[width=0.42\textwidth]{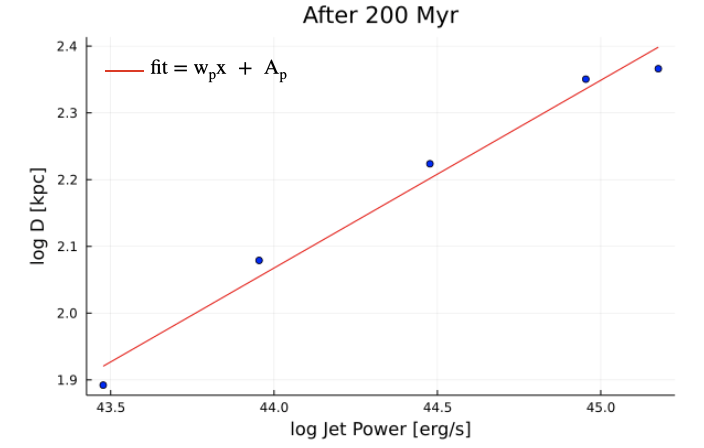}
\includegraphics[width=0.42\textwidth]{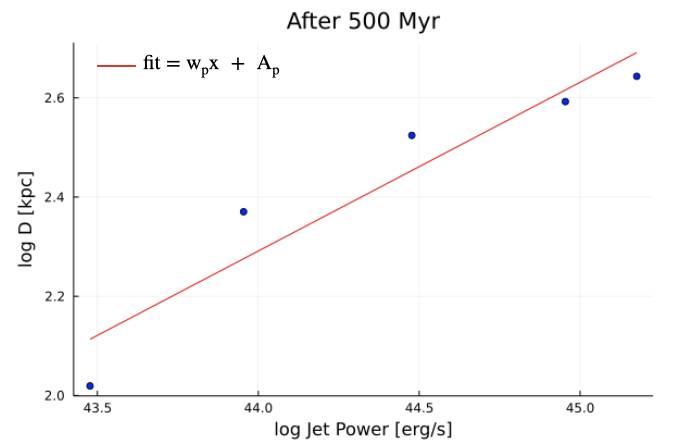}
\includegraphics[width=0.45\textwidth]{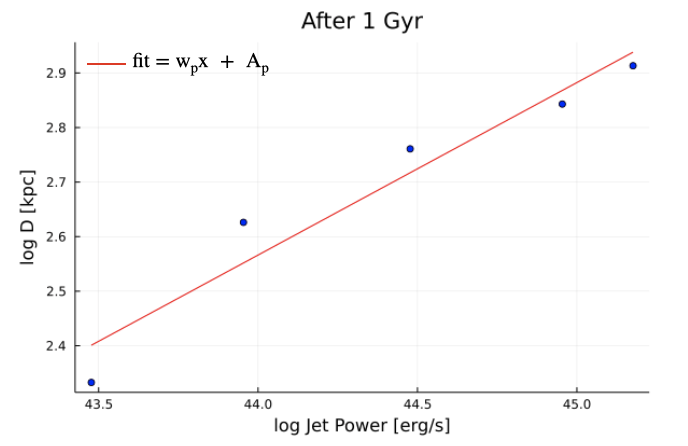} 
\caption{Relation between the average distance reached by tracers in four different epochs, as a function of the input jet power in the five runs.  The time is measured since the start of the jets launching, i.e. from $z=0.5$.} 
    \end{centering}
\label{fig:power_distance}
\end{figure*}

\section{Results}\label{sec::res}

\subsection{The effects of radio jets on gas dynamics} 
\label{kinematics}

Figure~\ref{fig:movie2} shows the evolution of the thermal gas and of our passive tracers at different epochs, from the injection of jets to  the end of our simulations. 

Since our runs are fully cosmological, jets soon interact with their surrounding environment, and they get increasingly affected and distorted by their relative motion with respect to the intracluster medium, as well a by the ram pressure of the jets' host galaxy as it travels relative to its environment, which we measure to be 
 $v \sim 300$ km/s.
Although simulated jets were released in a perfect alignment with the $z$-axis of the simulation (i.e. the vertical axis in Figure ~\ref{fig:movie2}), their content gets dispersed also transversally to it, in a few Gyr.

At the qualitative level, the morphology of the structures that are formed by the ejected material and the mixing with the ICM resemble the appearance of observed radio lobes. However, the detectability of such features depends on the re-acceleration mechanism (later in Section~\ref{subsec:radio} below), and in most cases only the structures formed within $\leq 100 \rm ~Myr$ are radio detectable in all models. 

The various panels in Figure ~\ref{fig:scatter} show the evolution of the median density, magnetic field (the one directly produced by the simulation), temperature, Mach number (only measured for shocked tracers), velocity curl ($\vec \nabla \times \vec v$) and velocity divergence ($\nabla \cdot \vec v$) for all tracers in the various runs, as a function of time, referred to the initial epoch of jet launching, i.e. $z=0.5$. In the case of the magnetic field, we show here the value of the original magnetic field directly produced by our MHD simulation, and the renormalised magnetic field produced based on the local solenoidal turbulence (and later used to evolve our electron spectra), as explained in Section\ref{subsec:Bfield}. 

In order not to be too biased by the fact that, at any given time, tracers on average travel to a larger distance for runs with a larger jet power, we measure all median quantities within a radius of $\leq 500 \rm ~kpc$. 
In all runs, particles are subject to the largest fluctuations of thermodynamic values in the first $\sim 0.5 \rm ~Gyr$ since their injection, which is a clear signature of the dynamics induced by jets with increasing power (which is the only different player in the different runs at this early stage).  
As found in Paper I, during the entire first billion years after the release of jets, the difference in the median magnetic field between runs remains always more significant than the difference in temperature. This strengthens the idea that the non-thermal content of jets delivers a long lasting memory of ancient AGN events into the ICM, while the violent mixing of gas phases, typically promoted by the AGN itself, erases the thermal unbalances between jets and the ICM more quickly. 

Since our tracer particle carry a significantly higher magnetic field in the highest power runs (a factor $\sim 3-4$ higher than in the lowest power runs, corresponding to a $\sim 10$ larger radio power), even $\sim 500-750 \rm ~Myr$ after the jet release, also their radio properties are expected to be a good marker of the input jet power, while at their X-ray signature (e.g. cavities) are invisible after such a long period of time  \citep[e.g.][]{2012A&A...547A..56D,2012MNRAS.427.3468B,2018MNRAS.476.1614C}. This makes the radio window a potentially very powerful probe of the past activity and energetics of AGN feedback in such sources - provided that the total energetics and age of the radio lobes can be reliably estimated during such advanced evolutionary stages. 

Figure~\ref{fig:pdf_evol} gives the complementary Eulerian view of of the evolving distributions of gas density, temperature, velocity dispersion (measured here after removing the bulk gas velocity on scales $\geq 100 \rm ~kpc$, with a simple high-pass filtering approach) and magnetic field strength in the innermost $r \leq 500 ~\rm kpc$ (comoving) sphere around the cluster centre. We have selected here four epochs that well  illustrate the different changes in visible regime  in the different models. In order to better highlight the (thermo-)dynamical impact of jets on the intracluster medium, in all panels we also show the distributions from the Run 0 model, in which no jets were included. 
Just after jets have been launched ($z=0.495$) and also after $\approx 60 \rm ~Myr$ ($z=0.491$) we measure increasingly more prominent tails of high gas temperature, magnetic fields and velocity dispersion going from run B to run F.
For reference, before the jets are activated, the gas in this region has a 
median gas temperature of $\sim 10^{7} \rm ~K$, a
median velocity dispersion $\sim 50-100 \rm ~km/s$ and a median magnetic field strength $\sim 0.1-0.3 \rm ~\mu G$.
As soon as the jets are launched, the region experiences tails of shocked hot gas (up to $T \sim 5 \cdot 10^8 \rm ~K$ in the E/F runs), high magnetic fields ($\sim 50-100 \rm ~\mu G$ in the E/F runs) and large transonic turbulent velocities ($\sim 5000 \rm ~km/s$, in the E/F runs). While the median values of all fields remain the same, the amplitude of the tails of temperature, magnetic field strength and velocity roughly  scale with the input jet power at $z=0.491$ (i.e. $\sim 100 \rm ~Myr$ since the jet injection)  and present a clear excess compared to the baseline, Run 0 model without jets. Based on the difference between the various velocity distributions, we can say that up to $\sim 100 \rm ~Myr$ since the injection of jets, the ICM can develop a $\sim 10$ times larger turbulent velocity compared to the baseline level induced by matter accretions in this system, for roughty at least for a $\sim 10\%$ of the central volume being considered here. 
For longer evolutionary times, the distribution of all fields in the central regions settle to very similar distributions, with the most significant exception in the distribution of magnetic fields, which remain larger even at $z=0.391$ (i.e. after $\approx 800 \rm ~Myr$ since the release of jets), consistently with our Lagrangian analysis above. 

Only after $\sim 2 \rm ~Gyr$ of evolution since the release of jets ($z=0.254$ in the Figure) all distributions have become similar, and similar to the baseline Run 0 model without jets, thus broadly marking the maximum time after which a single, powerful jet episode can leave significant dynamical difference in the host gas environment.
There is one noticeable exception, however: the distribution of gas density in run E/F is significantly lowered (by a factor $\sim 50 \%$), which we interpret as the integrated effect of the gas diffusion promoted by jets in these latter two runs.

\bigskip

In order to monitor the circulation of matter ejected by jets we used our Lagrangian tracers and a simple metric to assess their global dispersal with time. Unlike in Paper I, we explore two options to compute such distance: a) by comparing the median distance covered by tracers with respect to their initial location in the absolute reference frame of the simulation (i.e. the cells where jets were first injected), and b) by computing the distribution of distances at every snapshots, with respect to the position of the moving galaxy, which is reasonably marked by the center of mass of the distribution of tracers.
While the first choice (adopted in Paper I) is the standard approach to compute the pair dispersion statistics of Lagrangian particles in idealised turbulent simulations, the second is motivated by the fact that the center of mass of the tracer distributions, together with the galaxy hosting jets, travels across the ICM with a velocity of $\sim 300 \rm ~km/s$. 
The time evolution of the mean distance of tracers, measured in these two ways, is given in Figure ~\ref{fig:distance}.

Regardless of the adopted approach, tracers clearly travel on average out to a larger distance if the jet power is larger.  
However, there is no simple fit formula for the observed $\langle D(t) \rangle$ at all observed epochs. 
In the lowest power runs (B,C,D) jets are not powerful enough to break out of the cluster core ($\leq 100 \rm kpc$ from the centre), which has a total thermal energy  $E_{\rm core} \approx 7 \cdot 10^{58} \rm erg$, and the average separation of particles relative from the SMBH settles to a constant value, with a normalisation dependent on the initial jet power. This is understandable because, in the higher power runs, the initial launching phase propels a fraction of tracers to beyond the cluster core where mixing motions by cluster-wide turbulent motions  become dominant  \citep[e.g.][]{lau09,2020MNRAS.495..864A}, and continue to spread tracers to even larger radii, in a turbulent diffusion process \citep[][]{va10tracers}.

While we could not find a good, simple relation that can fit the entire measured trend of the distance $\langle D(t) \rangle$  across the time range of our simulated outputs, a simple $\langle D(t) \rangle \propto t^{w}$ relation is found to provide a good fit for the average tracer expansion in the first $\leq 400 \rm ~Myr$ after the jets injection, with slopes for the $\langle D(t) \rangle \propto P_j^w$ relation of $w=0.082 \pm 0.007$ (run B), $w \approx 0.351 \pm 0.021$ (run C), $w \approx 0.564 \pm 0.0194$ (run D), $w \approx 0.653 \pm 0.02$ (run E)  and $w \approx 0.690 \pm 0.028$ (run F), respectively.  

Predicting the exact expansion rate of jets into the surrounding halo atmosphere is a non-trivial task \citep[e.g.][]{1989ApJ...345L..21B}, even in the simplistic case of static power-law density profiles ($\rho(R) \propto R^{-\alpha}$) is non-trivial, due to a number of effects related to the later expansion of jets, their progressive entrainment of the surrounding halo gas, and the onset of fluid instabilities \citep[e.g.][]{2017MNRAS.472.4707B}. 
 The average expansion relation measured for our tracers, at least for the most powerful runs, is approximately linear with time, unlike the softer 
 $\propto t^{0.7}$ trend reported by \citet{2007MNRAS.382..526P} towards the end ($\sim 7 \ \rm Myr$) of their high-resolution simulation of a Fanaroff-Riley type I source, designed to model the propagation of a $P \approx 10^{45} \ \rm erg/s$ jet into an atmosphere resembling the real 3C31 radio galaxy. In principle, a number of important differences can account for such discrepant behaviour: unlike in previous work, the resistance of the surrounding ICM, swept up by the expanding lobes, becomes increasingly more important and, in addition, the interaction with pre-existing turbulent motions further affects the expansion dynamics of the jets, but widening the aperture of the initial velocity cone. Furthermore, from the sharp drop of  temperature and density in Figure \ref{fig:scatter}, one can clearly see that the pressure in the lobes is strongly affected by expansion (and further dissipation of thermal gas energy into the environment). For all these reasons, deriving an analytical estimate of the $\langle D \rangle (t)$ relation, also depending on the varying jet power, is non trivial (and beyond the scope of this work).
 
An interesting astrophysical application of our simulation is to investigate to which extend it would be possible to derive the total jet power in each model, provided  that the jet injection epoch can be robustly inferred from observations (for example through the modelling of the radio emission spectra of radio lobes, and/or from the modelling of X-ray cavities carved by jets into the hot surrounding atmosphere).
Figure \ref{fig:power_distance} gives the measured average distance of tracers after 100, 200, 500 and 1000 Myr since the start of jets, as function of the input jet power. 
At all epochs, the measured average distance is reasonably well fitted by a $\langle D(P_j,t) \rangle  = A_p P_j^{w_p}$ relation, with best fit parameters given in Tab.~\ref{tab:P_D}. 

Especially for the epochs 100 and 200 Myr (which are also are closer to the maximum age of observable remnant radio galaxies and X-ray cavities, \citealt[e.g.][]{wise07,2017A&A...600A..65S}) this simple fit relation can predict reasonably close the overall tracers dynamics, meaning that for $\leq 200 \rm ~Myr$ the jet power is the main parameter determining the propagation of tracers). This suggests that the combination of an analysis of the age of particles in lobes, and of the relatively easy (at least in simple geometry cases, in which projection effects can be kept under control) the measurement of their average distance from their host galaxy can lead to an accurate determination of the total jet power of jets  (typically with $\leq 5-10\%$ error). In reality, the average distance covered by jets is also a function of the environment, which in turn is known to be strongly linked to the jet power and morphology \citep[e.g.][ and references therein]{hardcastlecroston}. 

A systematic study of the reliability of age estimates of radio lobes based on the analysis of their observable radio spectrum will be the subject of forthcoming work (Di Federico et al., in prep.).

\begin{table}[h]
\centering
\begin{tabular}{|c|c|c|}
\hline
\textbf{Elapsed time} & \textbf{$w_p$} & \textbf{$A_p$} \\
\hline
\textbf{100 Myr} &  $ 0.226 \pm 0.016$ & $ -7.997 \pm 0.733$\\
\hline
\textbf{200 Myr} & $0.281 \pm 0.073 $&$ -10.310 \pm  3.254 $ \\
\hline
\textbf{500 Myr} &$ 0.340\pm 0.210$ &$ -12.675 \pm 9.3271 $\\
\hline
\textbf{1 Gyr} &$ 0.316 \pm 0.151 $ & $ -11.348 \pm 6.724 $\\
\hline
\textbf{2 Gyr} &$ 0.469 \pm 0.188 $ & $ -18.069 \pm 8.358$ \\
\hline
\end{tabular}
	\caption{Best fit parameters for the $\langle D(P_j,t) \rangle  = A_p P_j^{w_p}$ relation between the average distance covered by tracers and the input jet power, for 5 different epochs.}
	\label{tab:P_D}
\end{table}

\begin{figure*}
    \centering
    \includegraphics[width=0.98\textwidth]{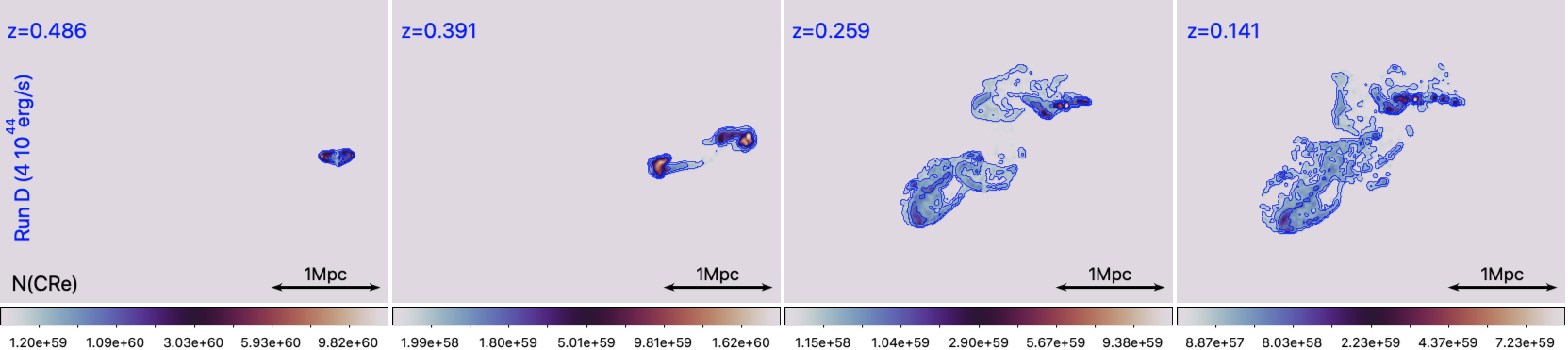}
    \includegraphics[width=0.98\textwidth]{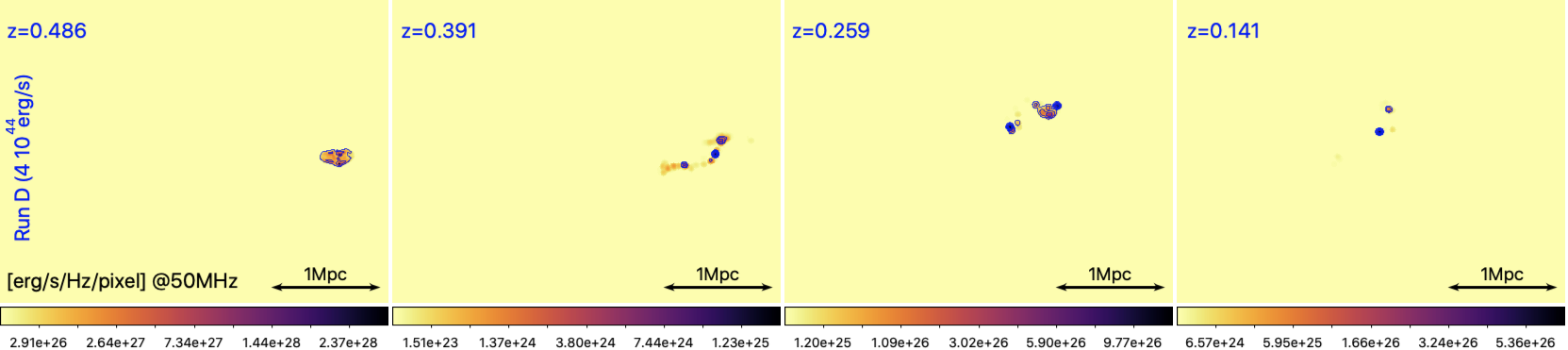}
   \caption{Projected number of relativistic electrons within the same column of cells along the line of sight in run D (top), and radio emission at $50$ MHz  at four different epochs, in the electron model including all loss and re-acceleration terms. The blue contours in the lower raw show the regions potentially detectable with LOFAR LBA.}
    \label{fig:mapD_evol}
\end{figure*}

\begin{figure}
    \centering
    \includegraphics[width=0.47\textwidth,height=0.18\textheight]{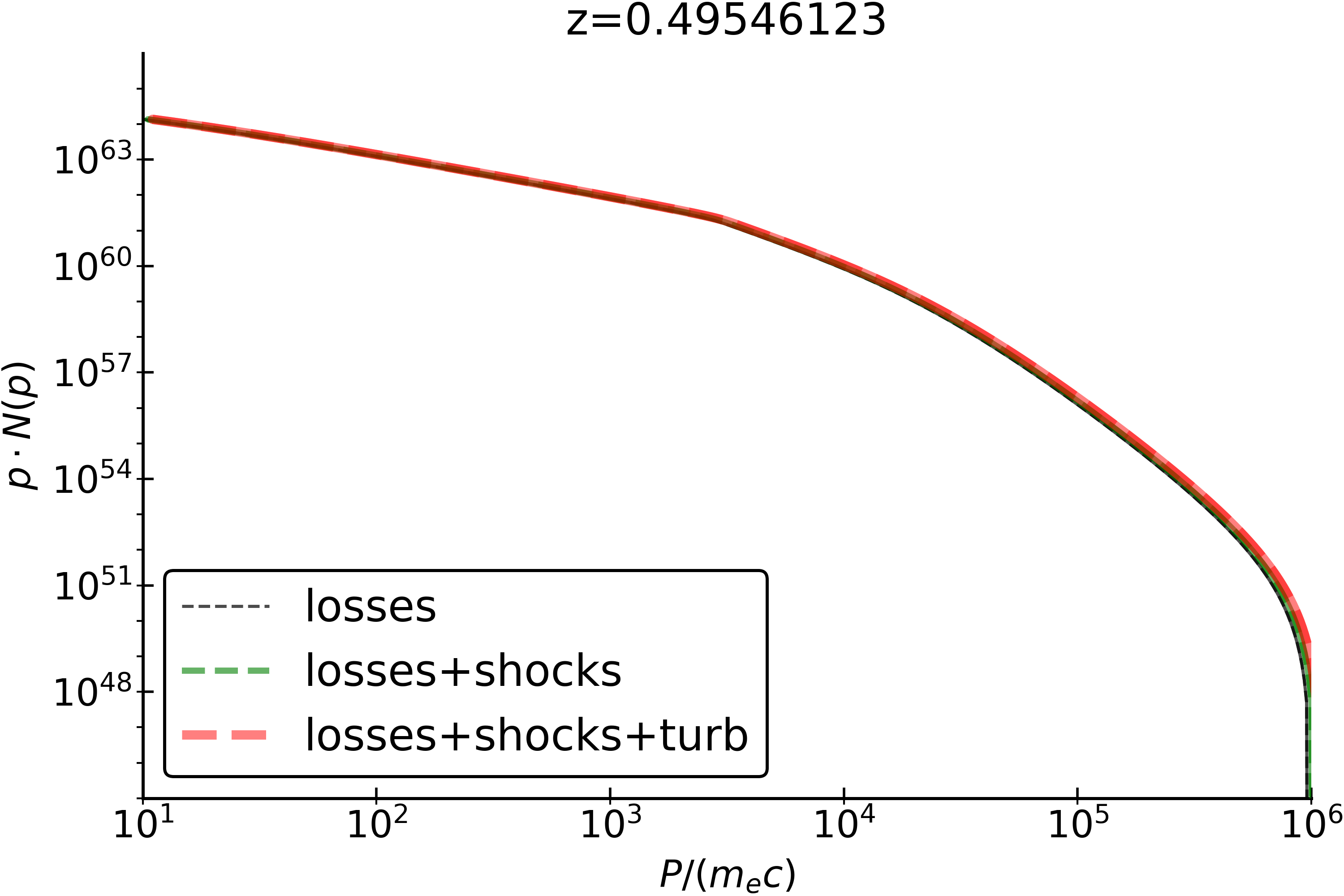}
        \includegraphics[width=0.47\textwidth,height=0.18\textheight]{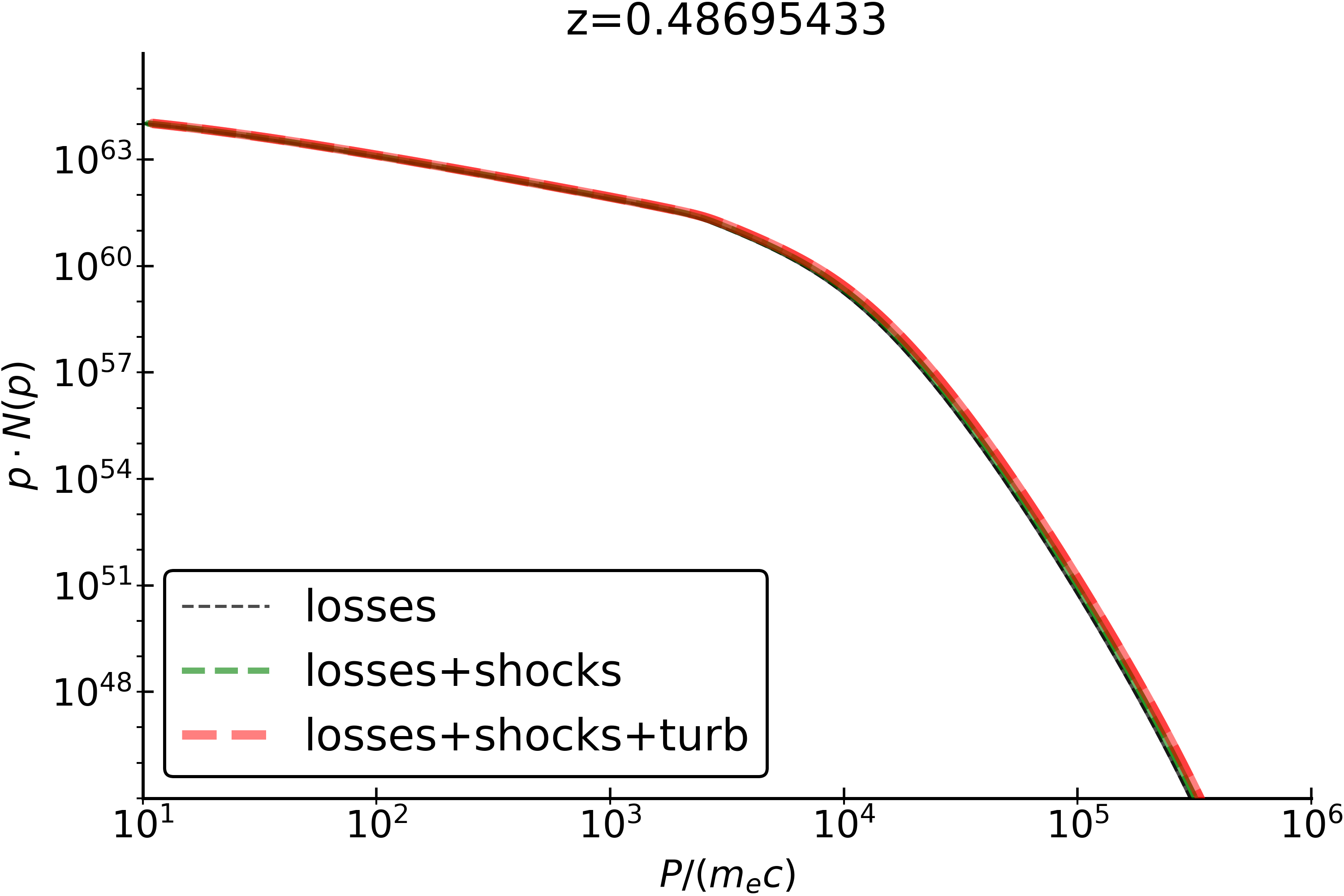}
            \includegraphics[width=0.47\textwidth,height=0.18\textheight]{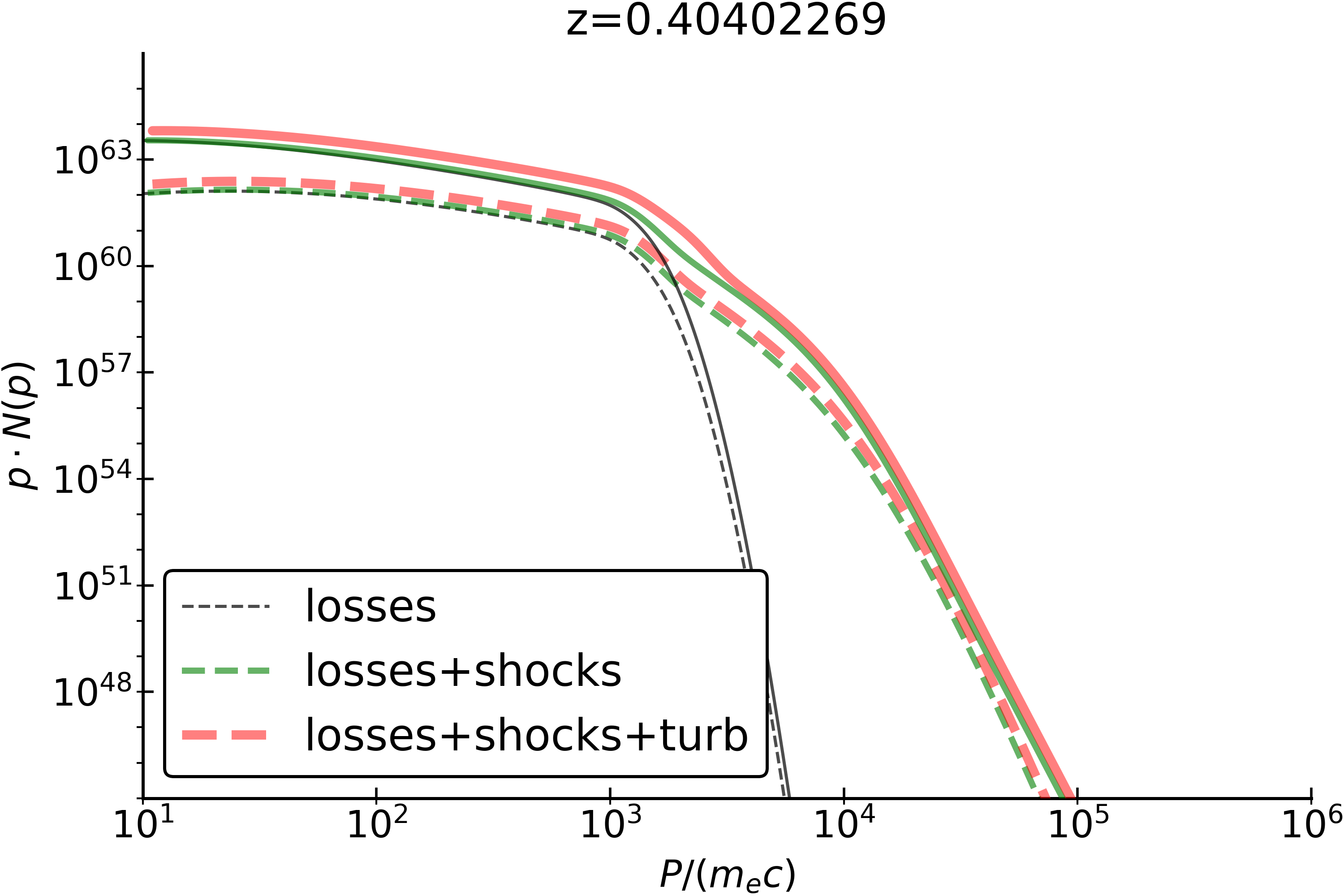}
                \includegraphics[width=0.47\textwidth,height=0.18\textheight]{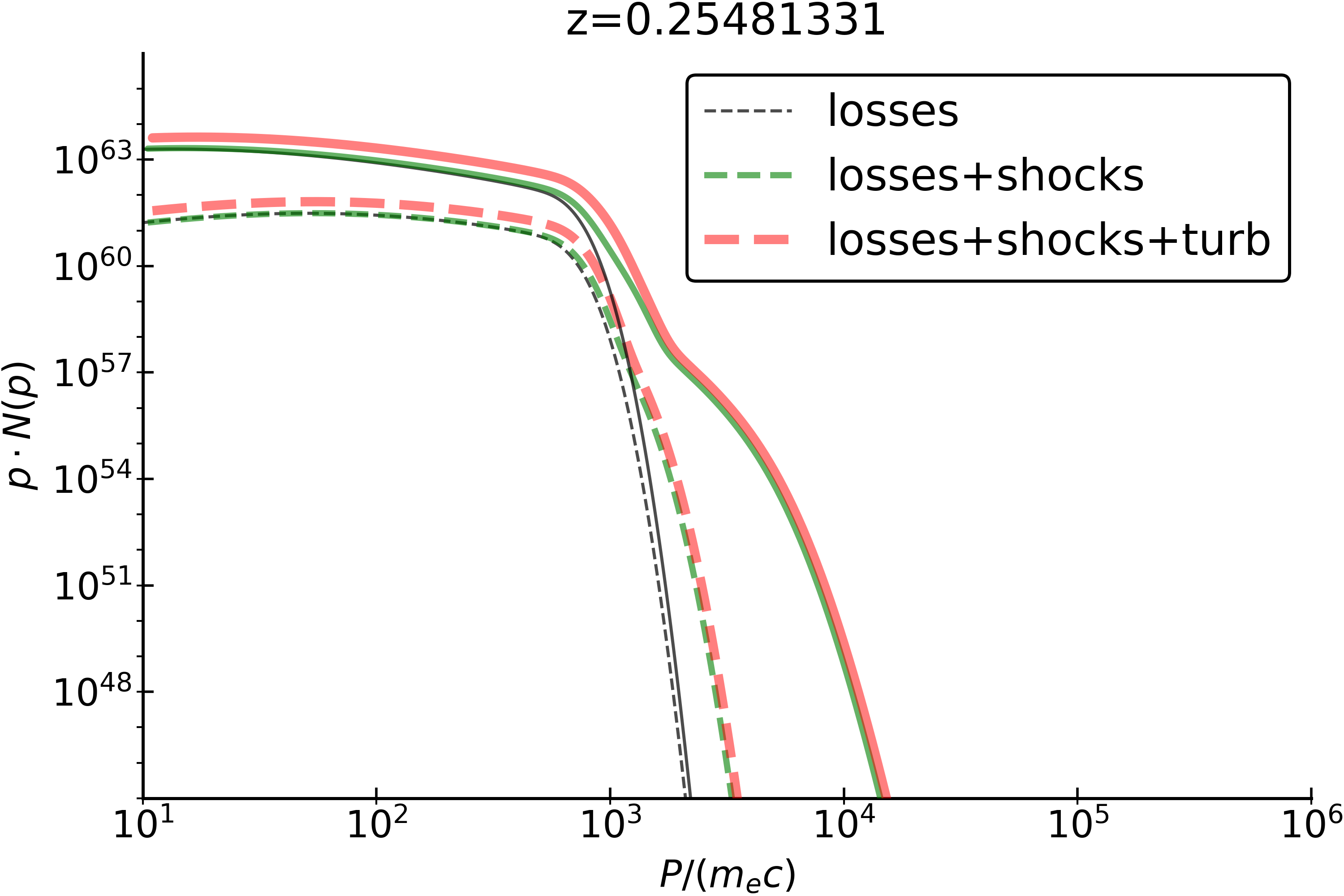}
                    \includegraphics[width=0.47\textwidth,height=0.18\textheight]{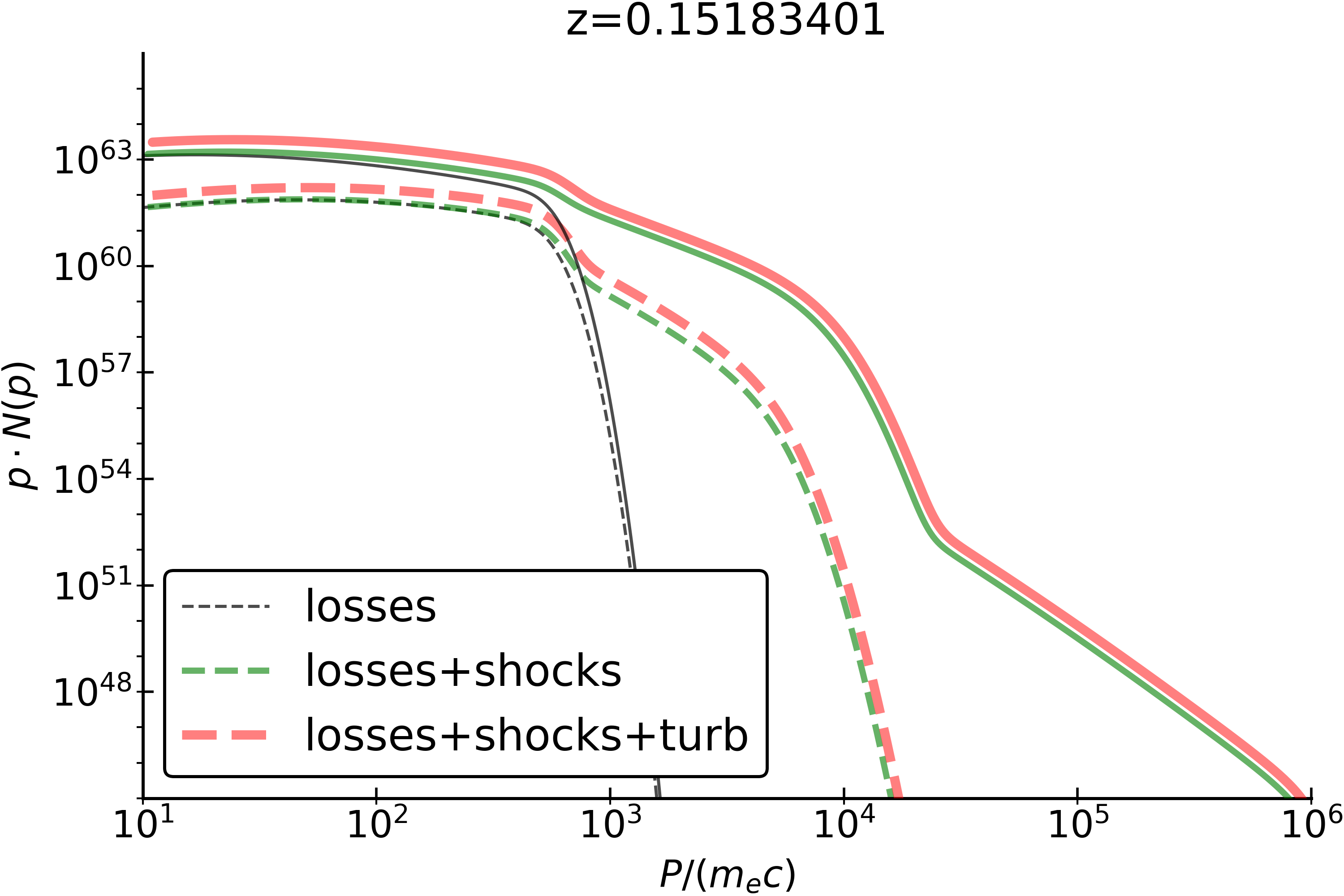}
    \caption{Evolution of the momentum distributions of electrons in the run D, for five different epochs and for the three different acceleration/cooling models (colours). The dashed lines are referred to the cluster region $r \leq 300$~kpc while the solid line are for electrons at any distance from the cluster centre.} 
    \label{fig:spectra_evol}
\end{figure}

\subsection{The spectral energy evolution of relativistic electrons} 
\label{subsec:spectra}

The energy evolution of electrons injected by radio galaxies is a complex combination of different mechanisms (e.g. adiabatic expansion, cooling, mixing with the ICM and re-acceleration following cluster-wide turbulence and shocks), with specific duration and features which are observed to change from run to run. 

First, we focus on the evolution of the  momentum spectrum for the medium power run of our suite, run D. 
We give in the top row of Figure~\ref{fig:mapD_evol} the spatial distribution of the projected number of cosmic ray electrons (computed by summing up all spectra from tracers located along the same column of $1 \times 1 \times 640$ cells along the line of sight, each cell being $8.86^3 \rm kpc^3$ in size) for four epochs.

The bottom row in the same Figure shows instead the evolution of the radio detectable parts of the electron distribution, which we shall comment later in Section\ref{subsec:radio}. 
This time, the maps shows the projection along a perpendicular line of sight compared to Figure~\ref{fig:movie2}. 
All re-acceleration and loss terms are considered in this case. These epochs corresponds to the emergence of very salient features in our simulated electron spectra, given in Figure~\ref{fig:spectra_evol}.  
 In all cases (with the exception of the very first stage, in the first column) the radio detectable fraction projected volume is only a tiny fraction of the total projected volume filled by electrons, as a combined effect adiabatic expansion and of Inverse Compton and synchrotron losses, which take a few $\sim 10^2 \rm ~Myr$ to devoid the radio emitting part ($p \geq 10^4$) of the momentum spectrum, for most of our simulated families of electrons.
After $\sim \rm 800 ~Myr$ since the jet injection ($z \leq 0.4$), a steep spectrum distribution of new relativistic electrons is injected by shocks crossing the region and related to the initial AGN burst . Particles gets transported to larger distances and outside of the core, and are overall subject to the integrated effect of cooling losses ($z>0.3$), which produces a classic boxy distribution in $p ~N(p)$, peaking at $p \sim 10^3$. The presence of re-acceleration by Fermi I and Fermi II mechanisms, however, makes  the electron distribution to have a significantly more extended tails of particles for $p \sim 10^3-10^4$, which in turns makes particle available for subsequent re-acceleration by merger-driven shocks even at later times, giving rise to a complex momentum distribution, with several bumps, as well as to a very patchy morphology of transported electron ($z=0.151$). On top of this, the more continuous re-acceleration by Fermi II significantly increases the total budget of fossil electrons at $p \sim 10^3-10^4$. 

Given the different circulation patterns of electrons, related with the jet power, that we outlined already in Section~\ref{kinematics}, we can expect increasingly different morphologies and spectral signature of the ICM dynamics when closely comparing our different runs.

\begin{figure*}
    \centering
    \includegraphics[width=0.98\textwidth]{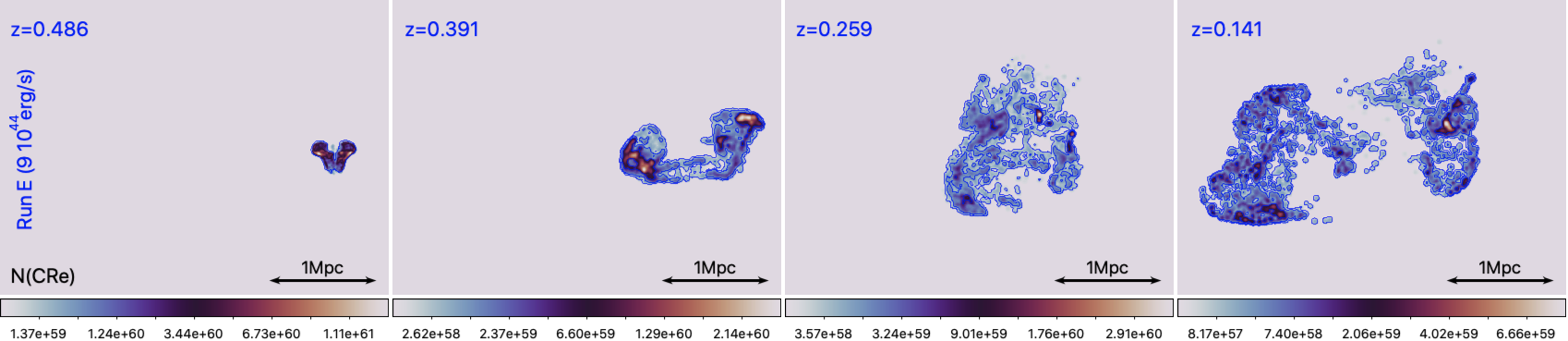}
    \includegraphics[width=0.98\textwidth]{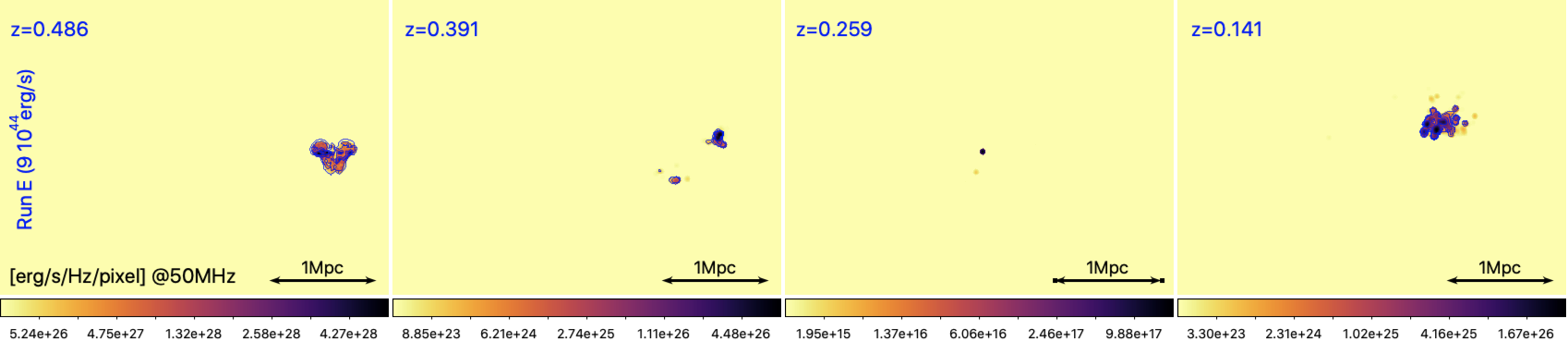}
    \caption{Projected number of relativistic electrons in run E (top) and radio emission at $50$ MHz  at four different epochs, in the electron model including all loss and re-acceleration terms. The blue contours in the lower raw show the regions potentially detectable with LOFAR LBA.}
    \label{fig:mapE_evol}
\end{figure*}

\begin{figure*}
    \centering
    \includegraphics[width=0.98\textwidth]{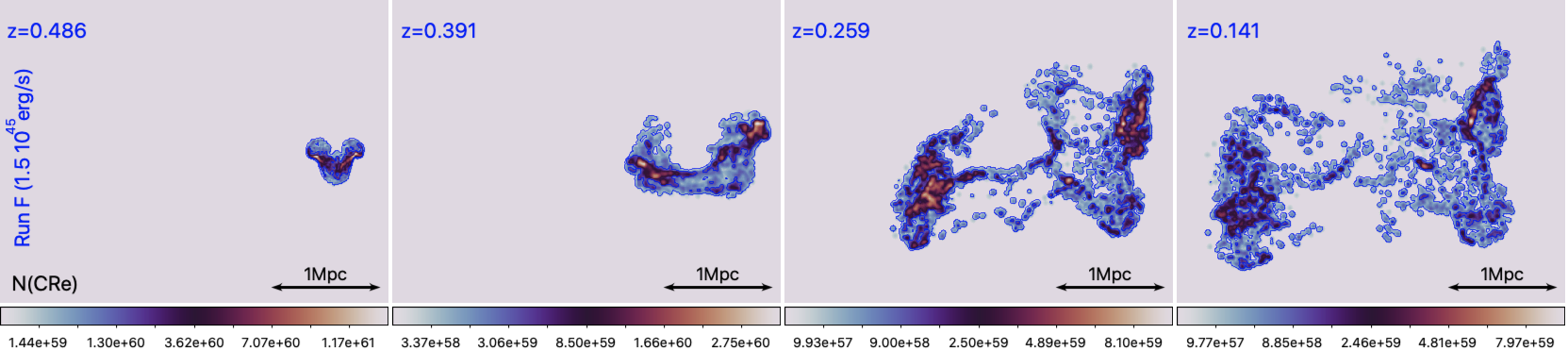}
    \includegraphics[width=0.98\textwidth]{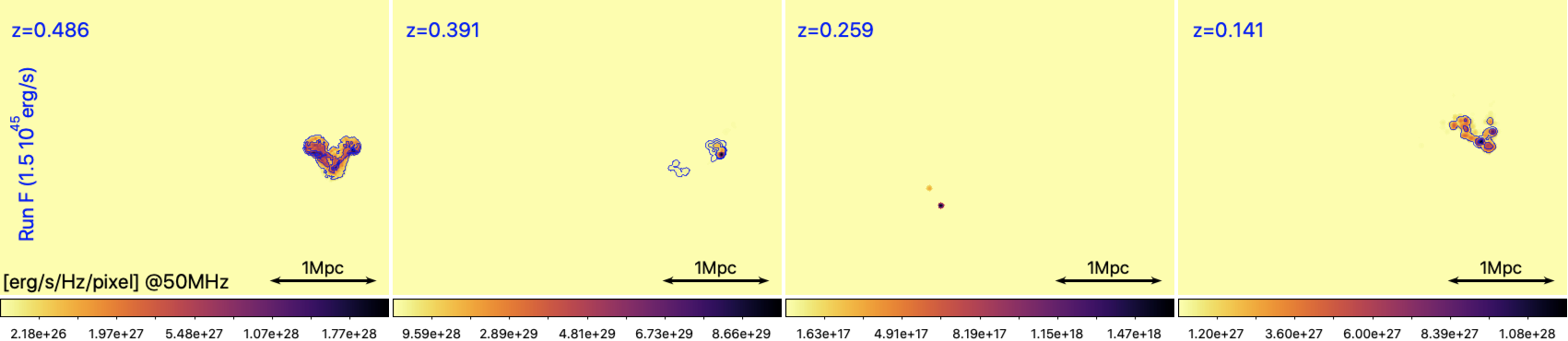}
   \caption{Projected number of relativistic electrons in run F (top) and radio emission at $50$ MHz  at four different epochs, in the electron model including all loss and re-acceleration terms. The blue contours in the lower raw show the regions potentially detectable with LOFAR LBA.}
    
    \label{fig:mapF_evol}
\end{figure*}

The top panels of Figure~\ref{fig:mapE_evol} and \ref{fig:mapF_evol} again show the spatial distribution of the projected number of cosmic ray electrons, for the same epochs of Figure~\ref{fig:mapD_evol}. Clearly, already at $z \approx 0.4$ the volume filling factor of electrons is larger than in the previous case, and more substructures are visible in the spatial distribution of the expanding lobes, which is understood by the increased ICM dynamics, also as a result of the shocks and turbulence injected by higher power AGN jets in run E and F. 
The spatial distribution of electrons injected in Runs B and C is considerably smaller (not shown, but see Figure~\ref{fig:movie2}) as the bulk of electrons is confined within $\leq 400 \rm kpc$ from the cluster centre, e.g. Section~\ref{kinematics}). 
Also in these cases, the radio detectable part of the electron distribution is just a tiny fraction of the underlying, undetectable and distribution of "fossil" electrons predicted by our models.

\begin{figure}
    \centering
    \includegraphics[width=0.47\textwidth,height=0.18\textheight]{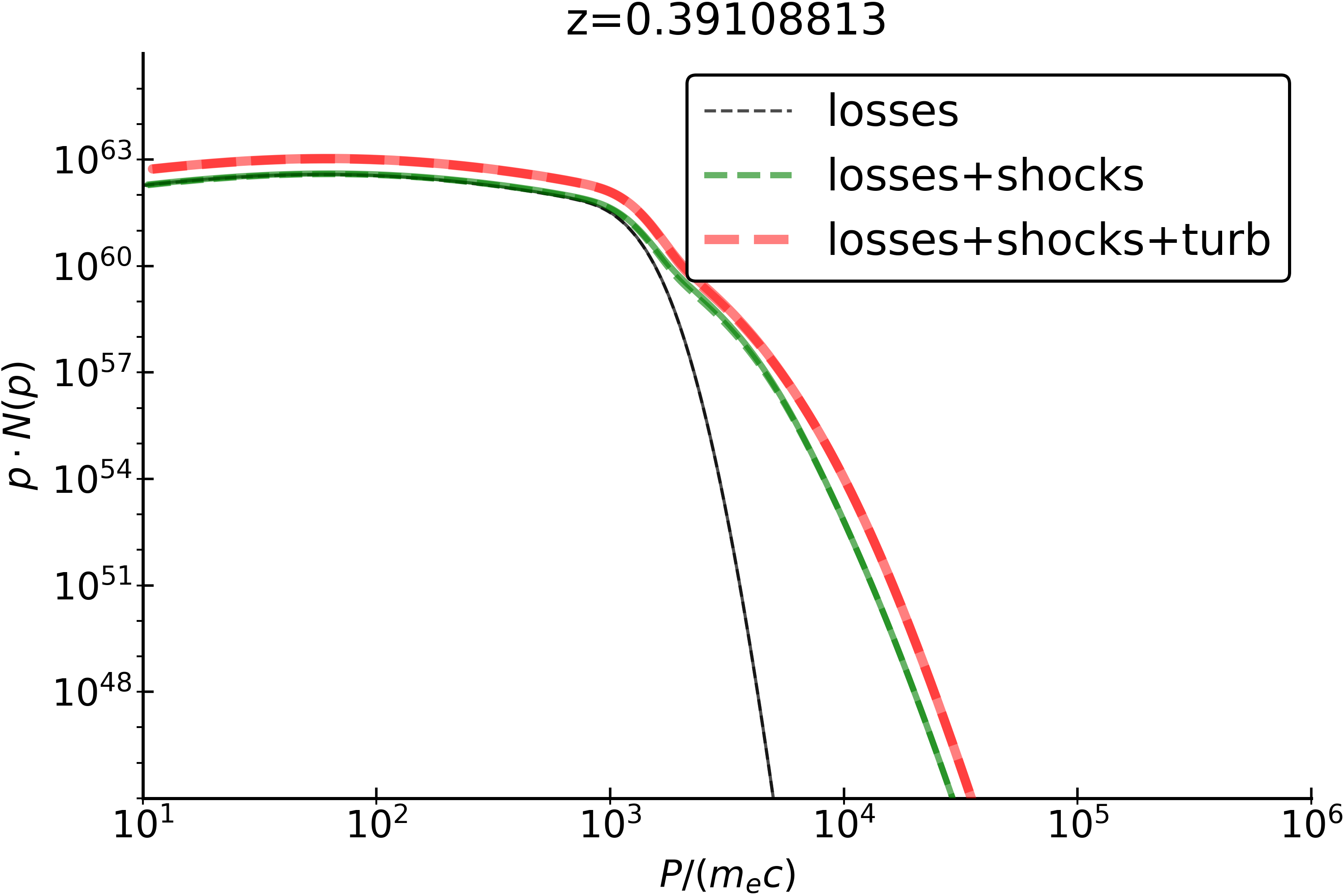}
    \includegraphics[width=0.47\textwidth,height=0.18\textheight]{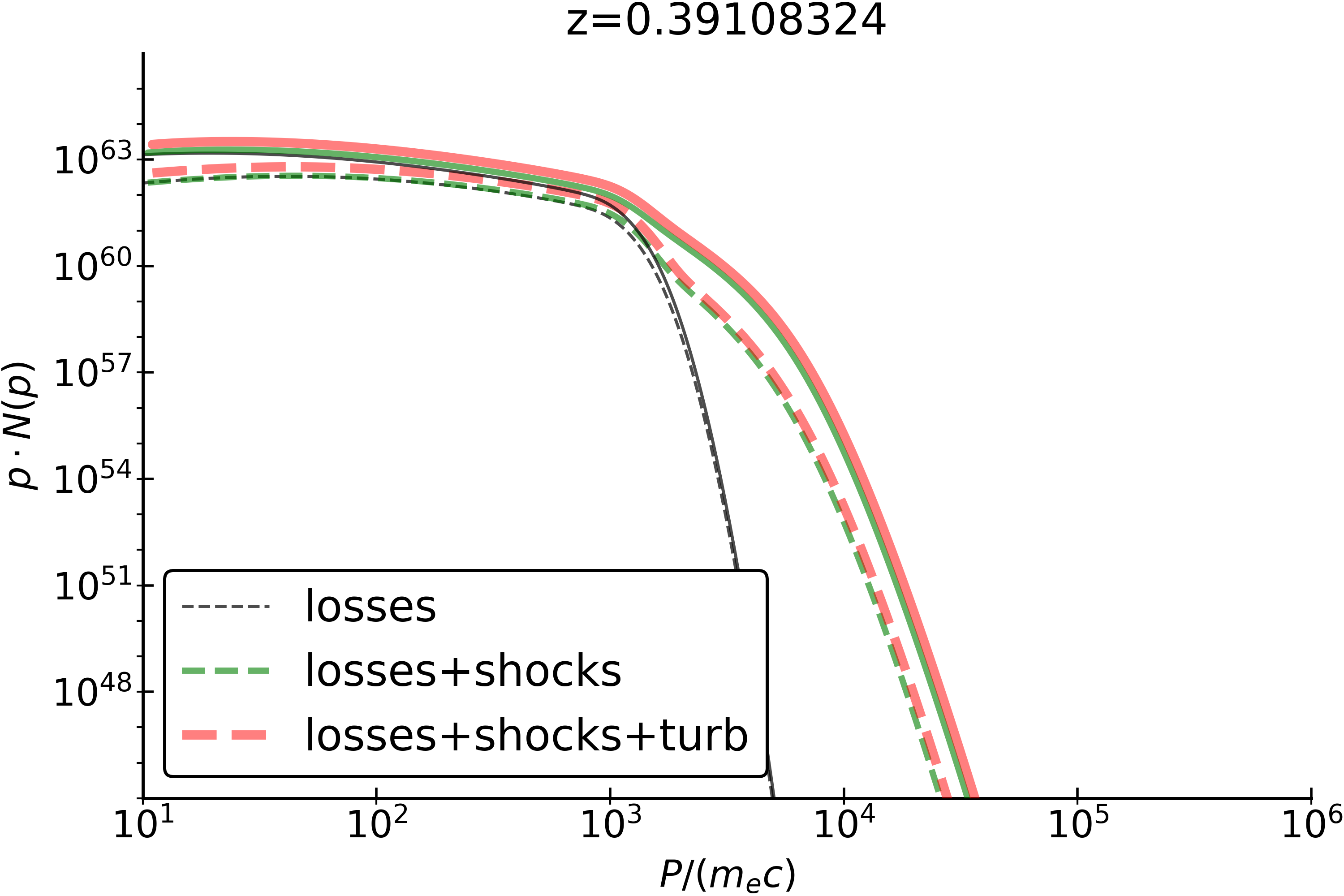}
     \includegraphics[width=0.47\textwidth,height=0.18\textheight]{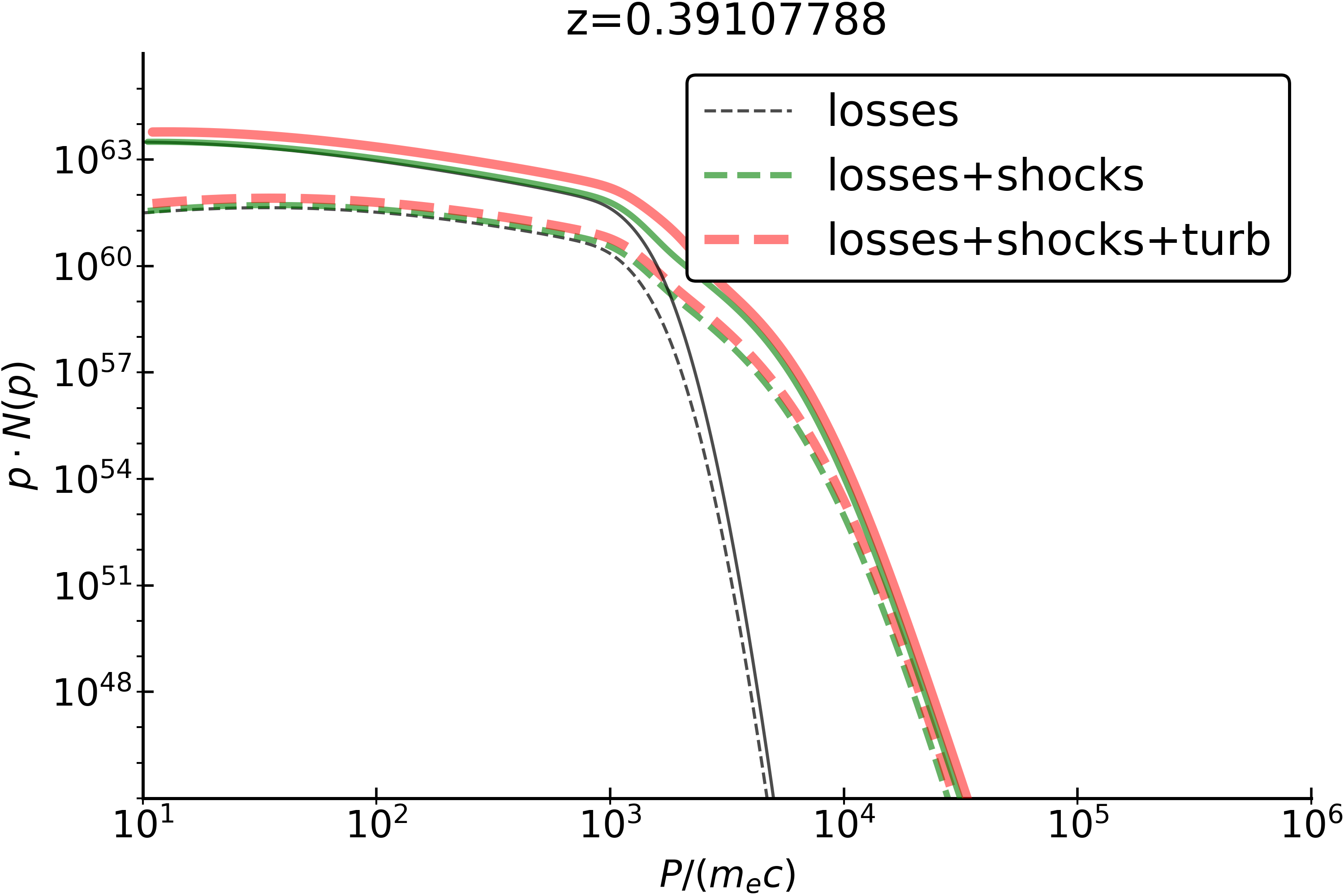}
     \includegraphics[width=0.47\textwidth,height=0.18\textheight]{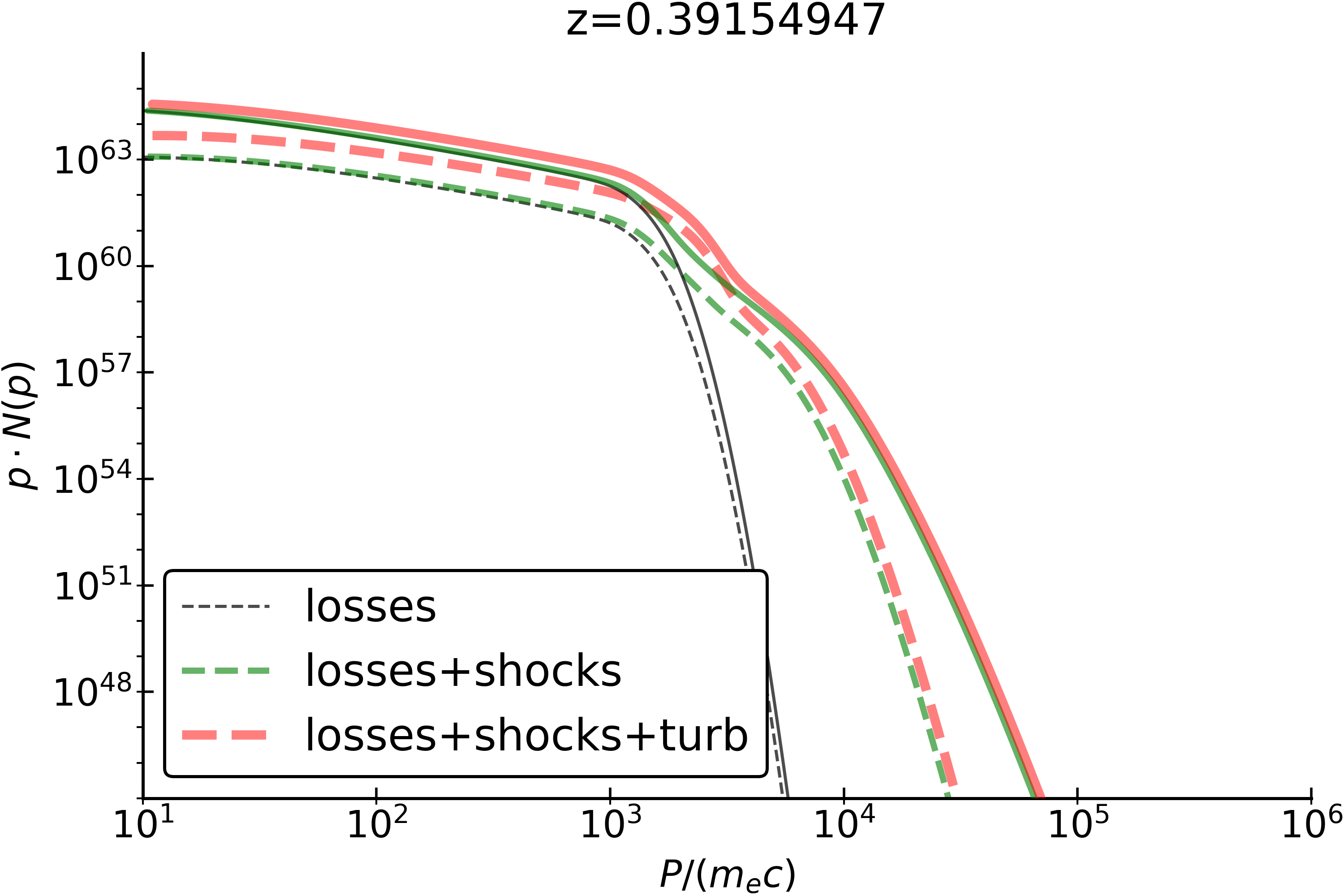}
          \includegraphics[width=0.47\textwidth,height=0.18\textheight]{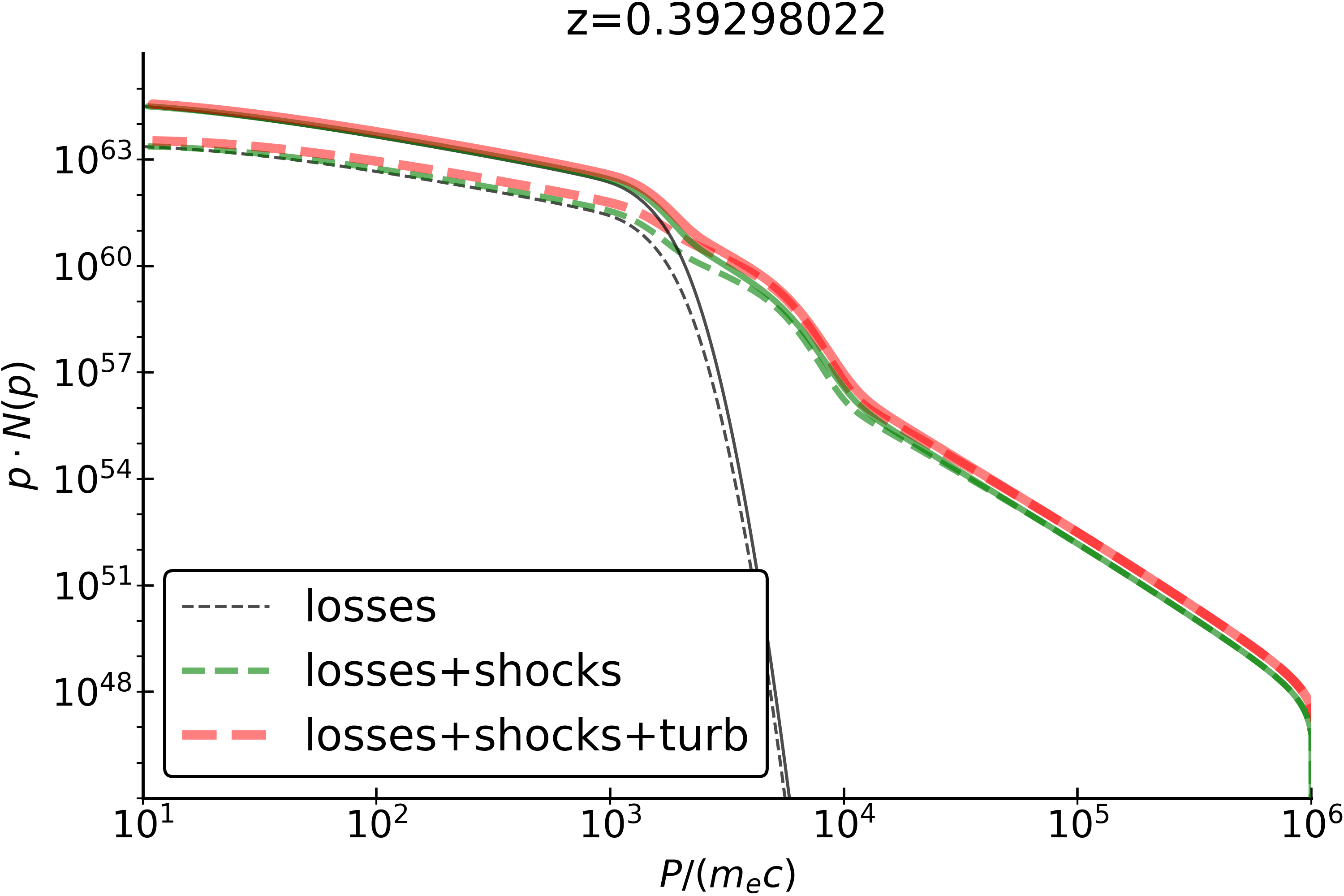}
    \caption{Momentum distributions of electrons at $z \approx 0.39$  in our five runs (i.e. $\approx 760 \rm ~Myr$ after the jet injection), for the three different acceleration/cooling models (colours). The dashed lines are referred to the cluster region $r \leq 300$~kpc while the solid line (here overlapping with the first) are for electrons at any distance from the cluster centre. The lowest power run B is at the top and the highest power run F is at the bottom.}
    \label{fig:spectra1}
\end{figure}

\begin{figure}
    \centering
        \includegraphics[width=0.47\textwidth,height=0.18\textheight]{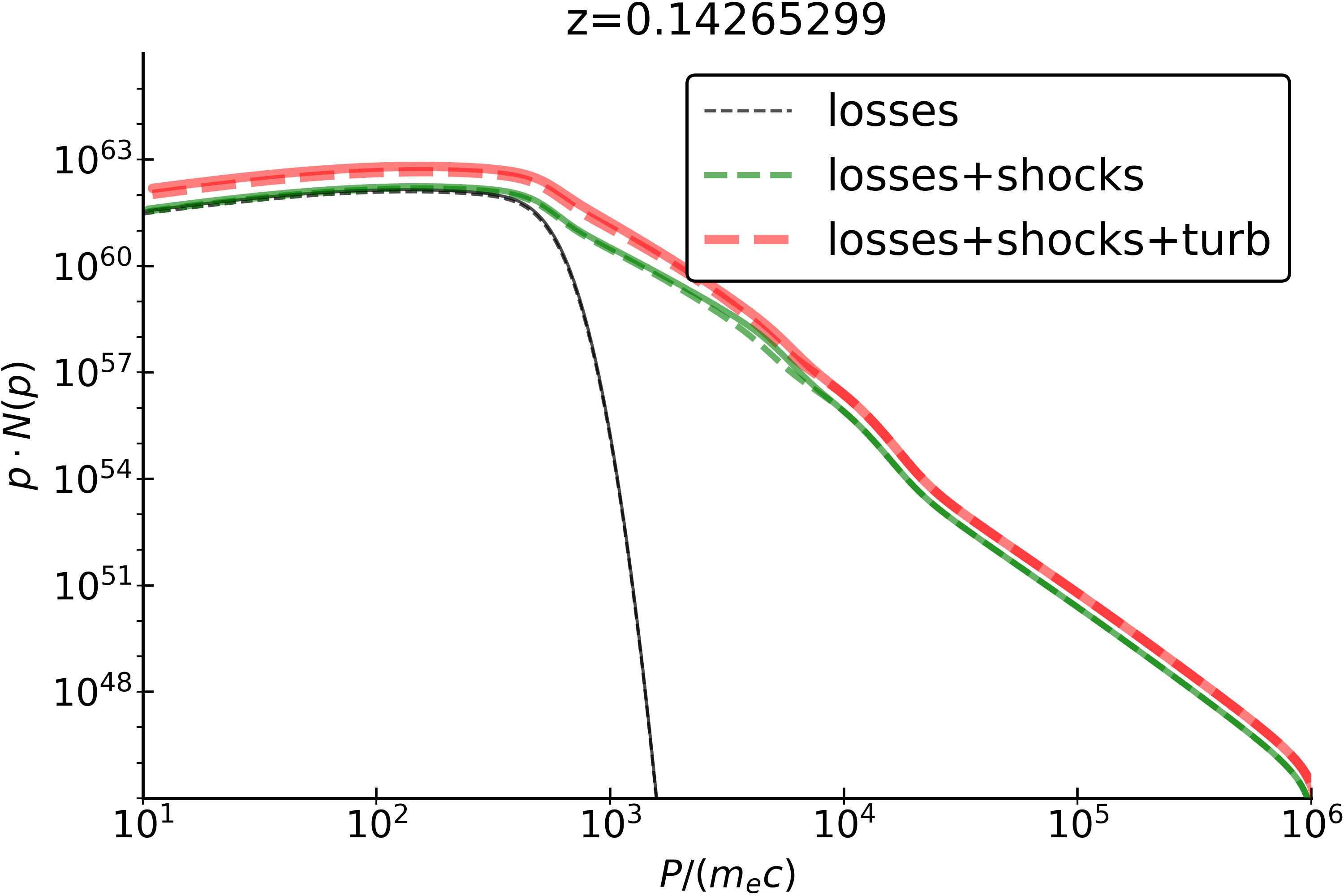}
    \includegraphics[width=0.47\textwidth,height=0.18\textheight]{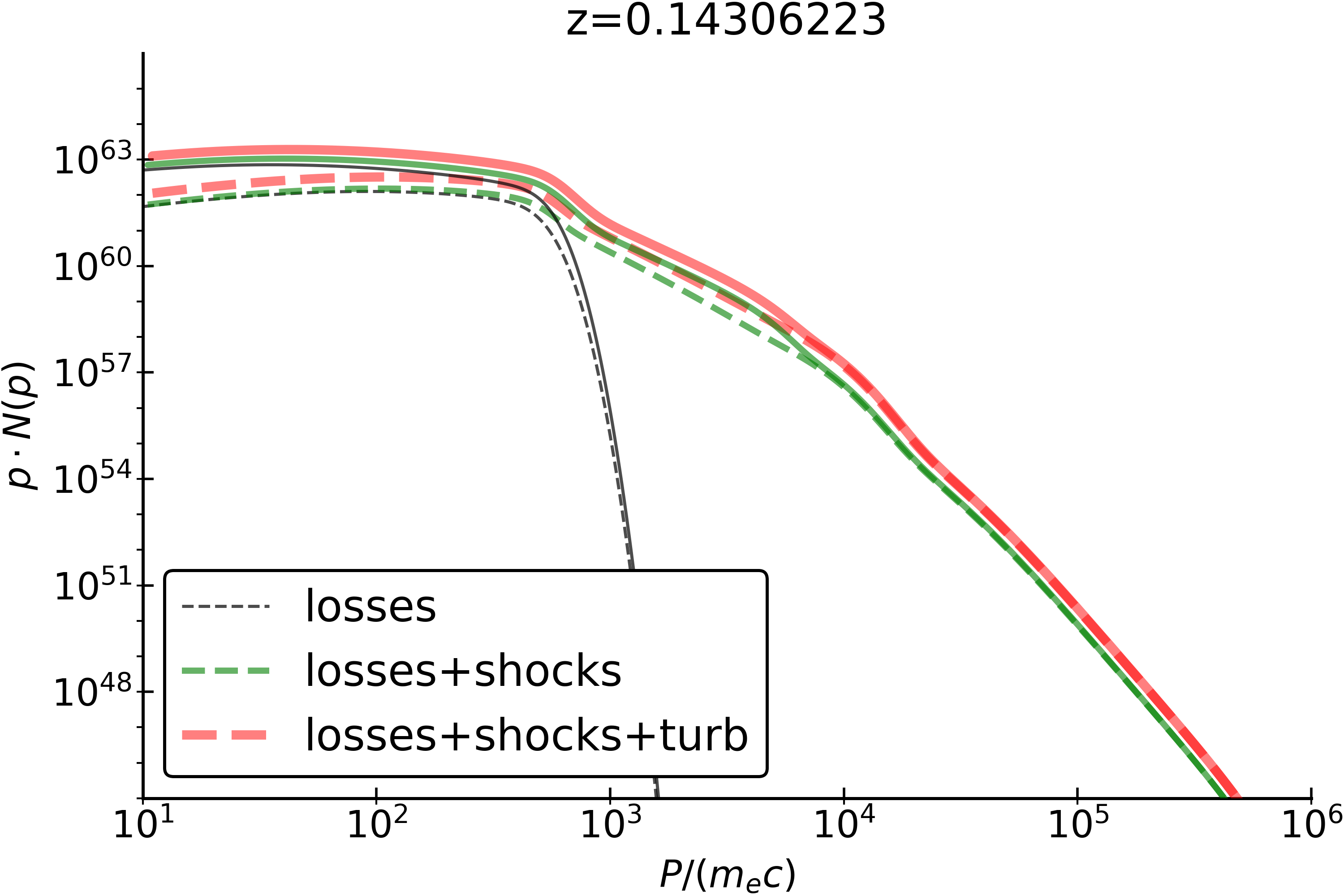}
     \includegraphics[width=0.47\textwidth,height=0.18\textheight]{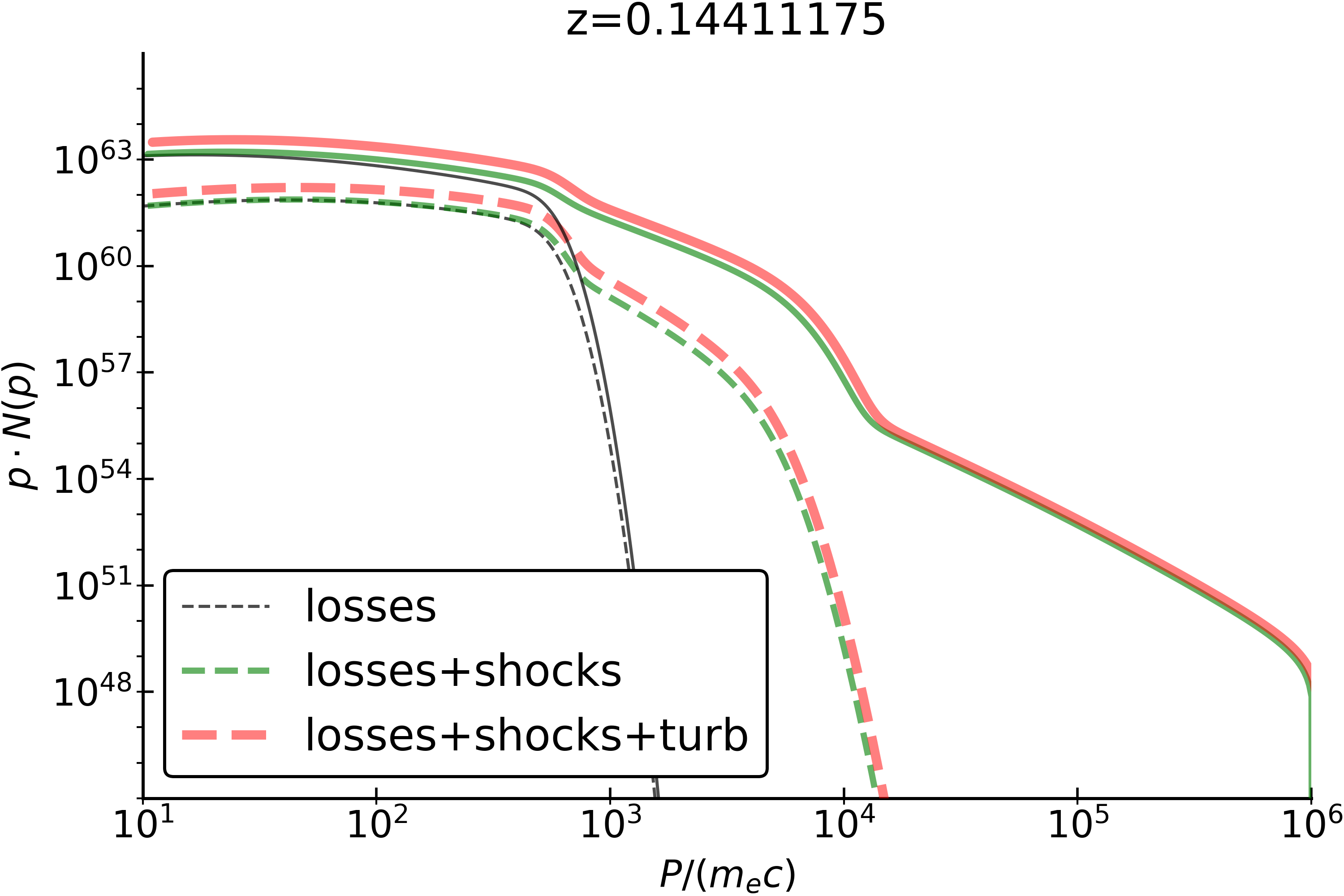}
     \includegraphics[width=0.47\textwidth,height=0.18\textheight]{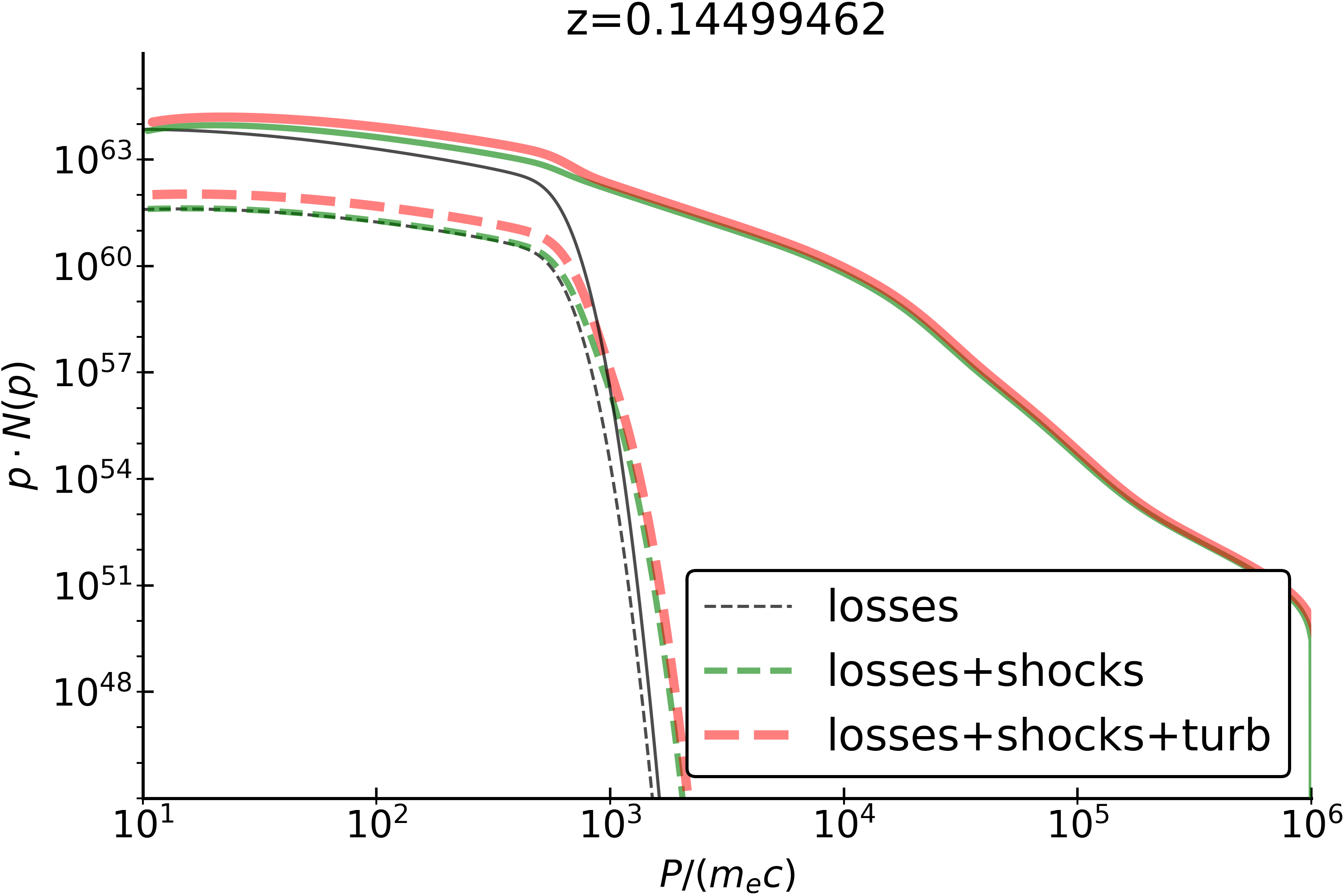}
          \includegraphics[width=0.47\textwidth,height=0.18\textheight]{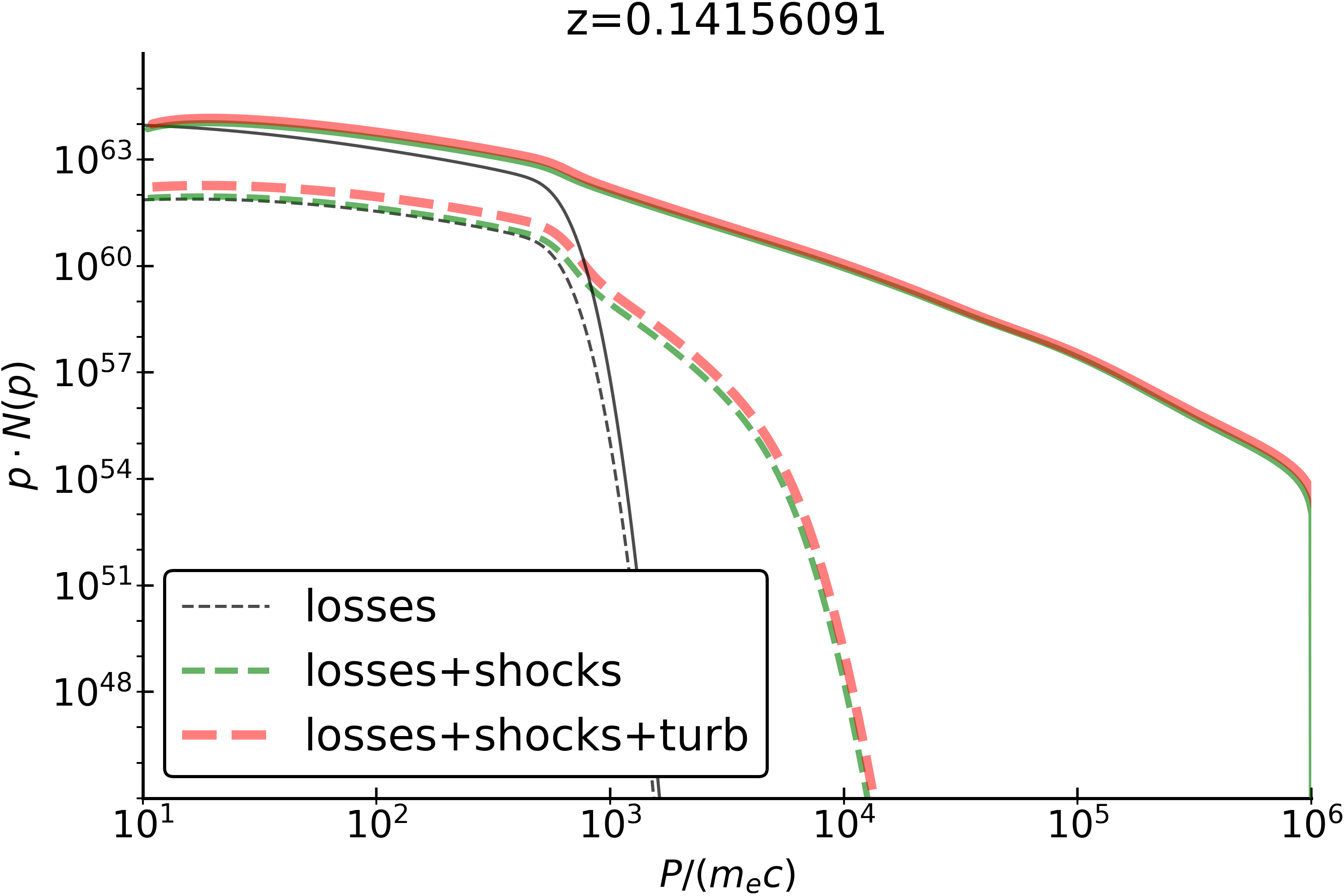}
     
    \caption{Momentum distributions of electrons at $z=0.13$ in our five runs (i.e. $\approx 3.2 \rm ~Gyr$ after the jet injection), for the three different acceleration/cooling models (colours) and marking the core cluster region $r \leq 300$~kpc (dashed) or the entire cluster volume (solid). The lowest power run B is at the top and the highest power run F is at the bottom. }
    \label{fig:spectra2}
\end{figure}

The reason for the quick disappearance from the radio band of such a large amount of electrons is best explained by the comparison of the particle spectra for all runs, for two of the epochs above, given in  Figure  \ref{fig:spectra1}-\ref{fig:spectra2}: at an epoch of $t \sim 800 \rm ~Myr$ since the injection ($z=0.39$, corresponding to the second row of plots in Figure~\ref{fig:movie2} and roughly marking the epoch of maximum distance from the source reached by particles in the lowest energy runs)  and $t \sim 3.4 ~\rm Gyr$ since the injection   ($z=0.14$, corresponding to the fourth row of plots in Figure~\ref{fig:movie2}, which roughly corresponds to a high dynamical activity in the ICM, following the last major accretion episode in the host group). 

At both epochs (and in general) the spectra show the tendency of producing an increasingly extended tails of high energy electrons ($\gamma \geq 10^4$) with the increase of the jet power. As we saw in the previous Section,  this stems from the combination of electrons being pushed to larger cluster radii (where they are subject to lower energy losses and to typically more efficient shock and turbulent re-acceleration) as well as to the increased ICM dynamical activity promoted by more powerful AGN, which in turn also enhances particle re-acceleration. 

\begin{figure*}
    \centering
    \includegraphics[width=0.9\textwidth]{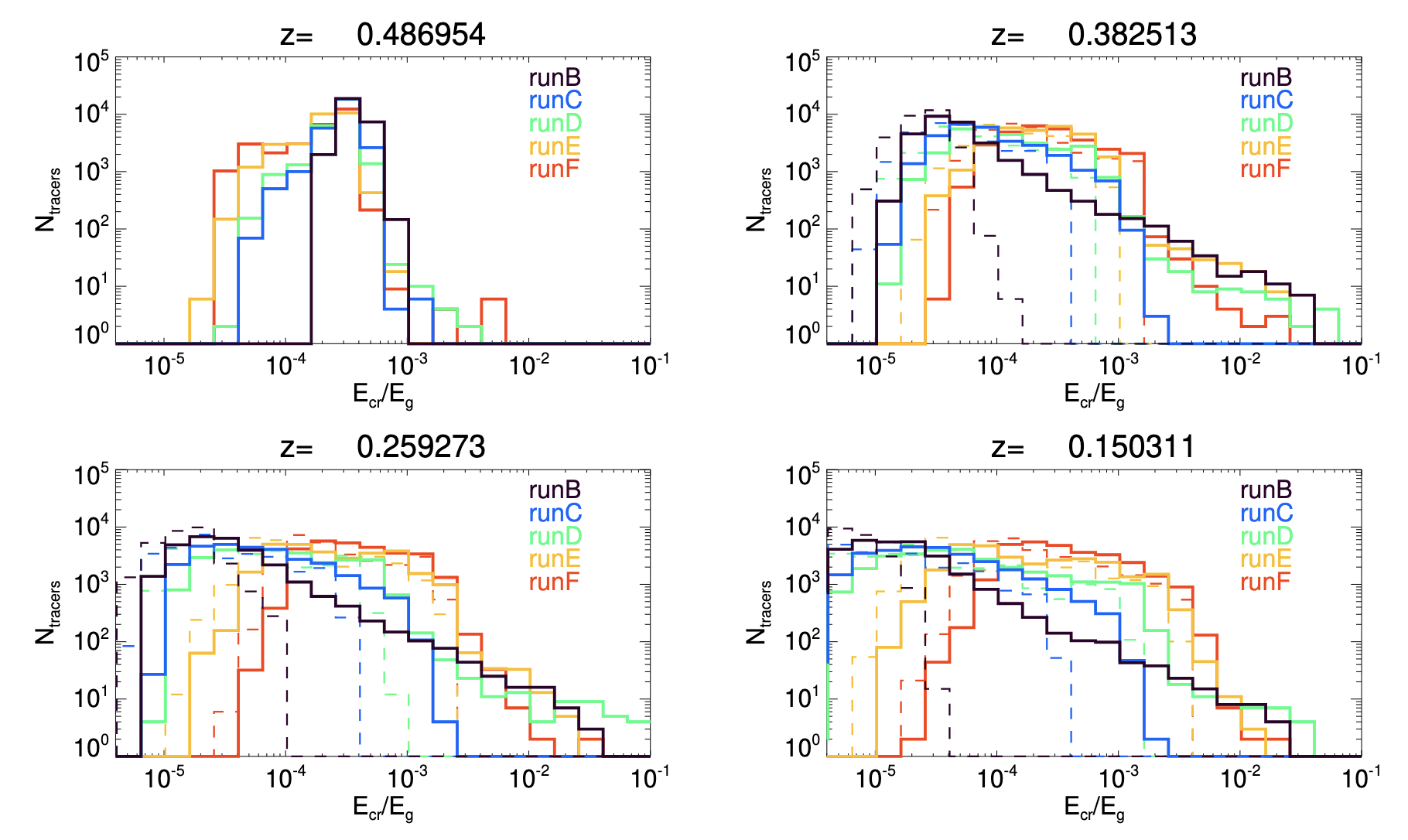}
    \caption{Distributions of the ratio between the total energy of all relativistic electrons we evolved in our runs at four different epochs, and the thermal gas energy associated with the same tracers. The solid lines give the distribution of $E_{\rm cr}/E_{\rm g}$ for the electron evolution model including shocks and turbulent re-acceleration, while the dashed lines are for the electron model only including radiative losses and adiabatic changes after the injection by jets.}
    \label{fig:pdf_Eratio}
\end{figure*}

With our simulations we can estimate the plausible budget of relativistic electrons that can be injected in the ICM by the activities of radio galaxies. 
To measure this, we computed the ratio between the total energy of relativistic electrons:
\begin{equation}
E_{\rm cr} \approx \int_{p_{min}}^{p_{max}} P c N(p) dp     
\end{equation}
and the total gas energy sampled by each tracers, $E_g = 3/2 k_B T (\rho /\mu m_p) ~dx_t^3$, as a function of time and for the different runs. Figure~\ref{fig:pdf_Eratio} gives the  distribution of $E_{\rm cr}/E_g$ for all tracers in the simulated volume, at the same four epochs considered above. 

Figure~\ref{fig:pdf_Eratio} gives the distribution of the energy ratio within the entire volume, for the same four epochs and runs considered above, and contrasting the prediction from the electron evolution model with shocks and turbulent re-acceleration (solid),with the model only including radiative losses and adiabatic changes (dashed). We do not divide these distributions as a function of the distance from the cluster centre as there is no significant dependence with radius (not shown). After $\sim 100 \rm ~Myr$ since the jet release the energy ratio in the lobes is $E_{\rm cr}/E_g \sim 10^{-3}$, but as the mixing with the ICM proceeds, the distribution widens and stretches to smaller values. 
At late epochs, we observe the progressive increase of the peak distribution of $E_{\rm cr}/E_g$ as a function of the initial jet power, with values ranging from $E_{\rm cr}/E_g \sim 10^{-5}$ in run B  to 
$E_{\rm cr}/E_g \sim 2 \cdot 10^{-4}$ (with tails stretching to larger values) in run E and F. Combined with the spectral evolution presented above, these trends can be understood considering that in higher power runs particle are typically spread to larger cluster radii, where the role of non-thermal energy components is larger, and particles suffer less of synchrotron and Coulomb losses. On the other hand, the role of re-acceleration processes is not dramatic here because the $E_{\rm cr}/E_g$ ratio is dominated by the low energy part of spectra: as it can seen by comparing the solid and dashed distributions of each colour in Figure ~\ref{fig:pdf_Eratio}, $E_{\rm cr}/E_g$ does not dramatically change with the inclusion of re-acceleration terms.

\begin{figure*}
    \centering
    \includegraphics[width=0.9\textwidth]{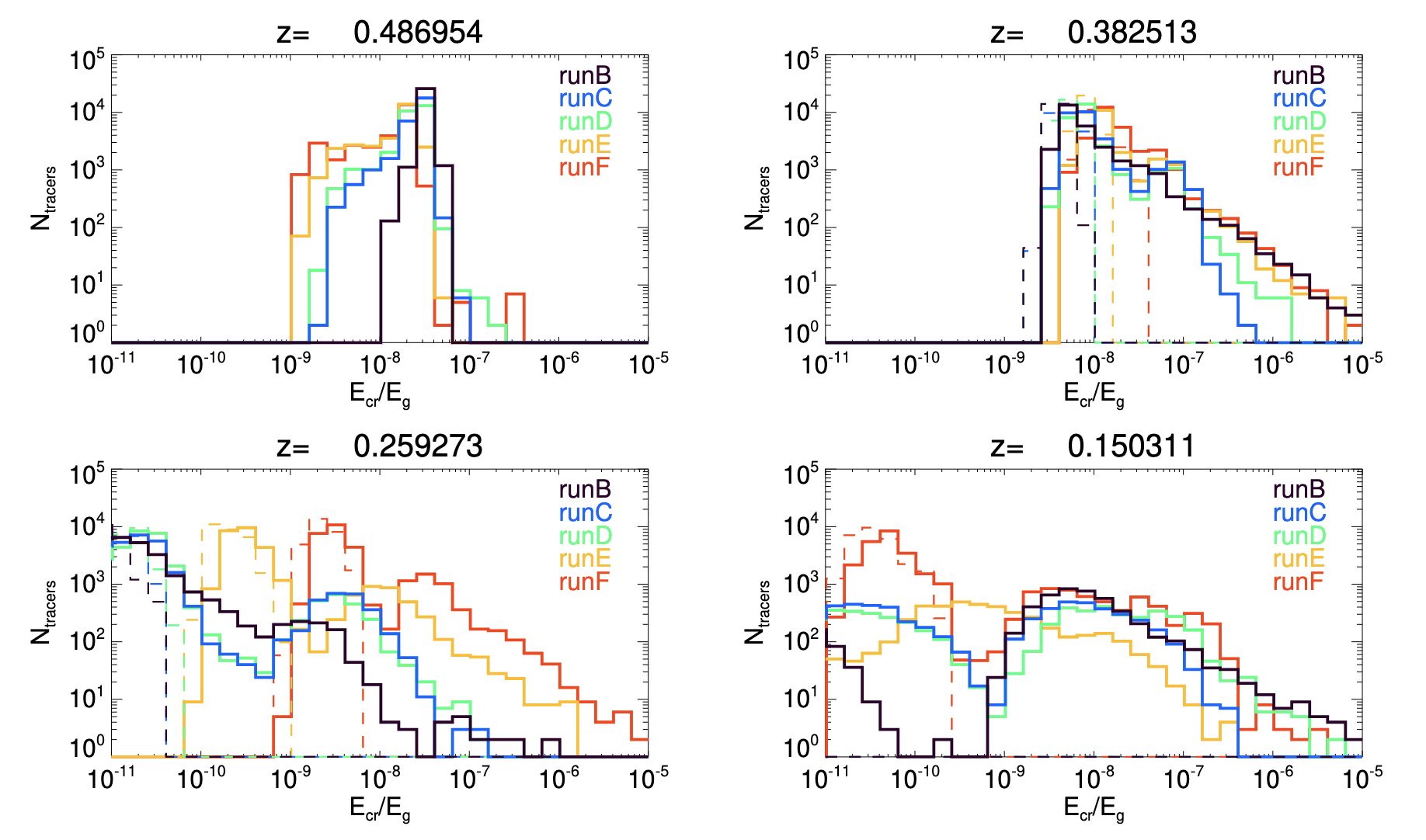}
    \caption{Similar to Figure ~\ref{fig:pdf_Eratio}, but only including $P/(m_e c) \geq 10^3$ electrons in the computation of $E_{\rm cr}$.}
    \label{fig:pdf_Eratio_1e3}
\end{figure*}

Finally, we show in Figure ~\ref{fig:pdf_Eratio_1e3} the same distribution of energy ratios, but only limited to the $p \geq 10^3$ part of the electron momentum distribution, which more closely tracks the budget of fossil electrons in the ICM. In this case, the role of re-acceleration terms is dominant, and at all epochs (but the very initial stage after the jet injection) and models the formation of reservoirs of fossil electrons with $E_{\rm cr}/E_g \geq 10^{-8}$ is possible only when re-acceleration is present (otherwise the energy ratio is seen to be a factor $\geq 10^2$ smaller).

\section{Observable radio properties}
\label{subsec:radio}

For each tracer in the simulation we compute its associated synchrotron radio emission for seven frequencies (50, 140, 650, 1400, 3000, 5000 and 10000 MHz)  using the formalism outlined in  Section ~\ref{subsec:sync}

The lower rows of Figure ~\ref{fig:mapD_evol}-\ref{fig:mapF_evol} show the radio synchrotron maps at 50 MHz for runs D, E and F at four epochs (always for the model with all loss and re-acceleration terms included), which gives an idea of the structures that might be in principle detectable with deep, LOFAR Low Band Antenna (LBA observations). To estimate which fraction of the emission would be detectable, we considered a fixed luminosity distance of $d_L=132 \rm ~Mpc$ for all snapshots, in which case our simulated pixel size corresponds to the resolution beam of LOFAR LBA ($\approx 8 \rm ~kpc$, considering a beam of $\theta=12.5"$), and a LOFAR LBA sensitivity of $\sigma = 5.7 \cdot 10^{-4} \rm ~Jy/beam$ (we considered here $\geq \sigma$ detections), which can be currently achieved with $\approx 72 \rm hr$ of integration (Botteon et al. submitted).

From the figures, it can be readily seen that the  emission detectable even with a deep LOFAR LBA observation is only a small fraction of the entire reservoir of electrons released by radio galaxies into the ICM. The detectable structures are often associated with weak shocks leading to small radio relic-kind of emissions, or with the most turbulent patches of the ICM, leading to more irregular patches of emission,
even on scales which are apparently not filled by non-thermal plasma associated with visible radio lobes. 

\begin{figure}
    \centering
    \includegraphics[width=0.47\textwidth]{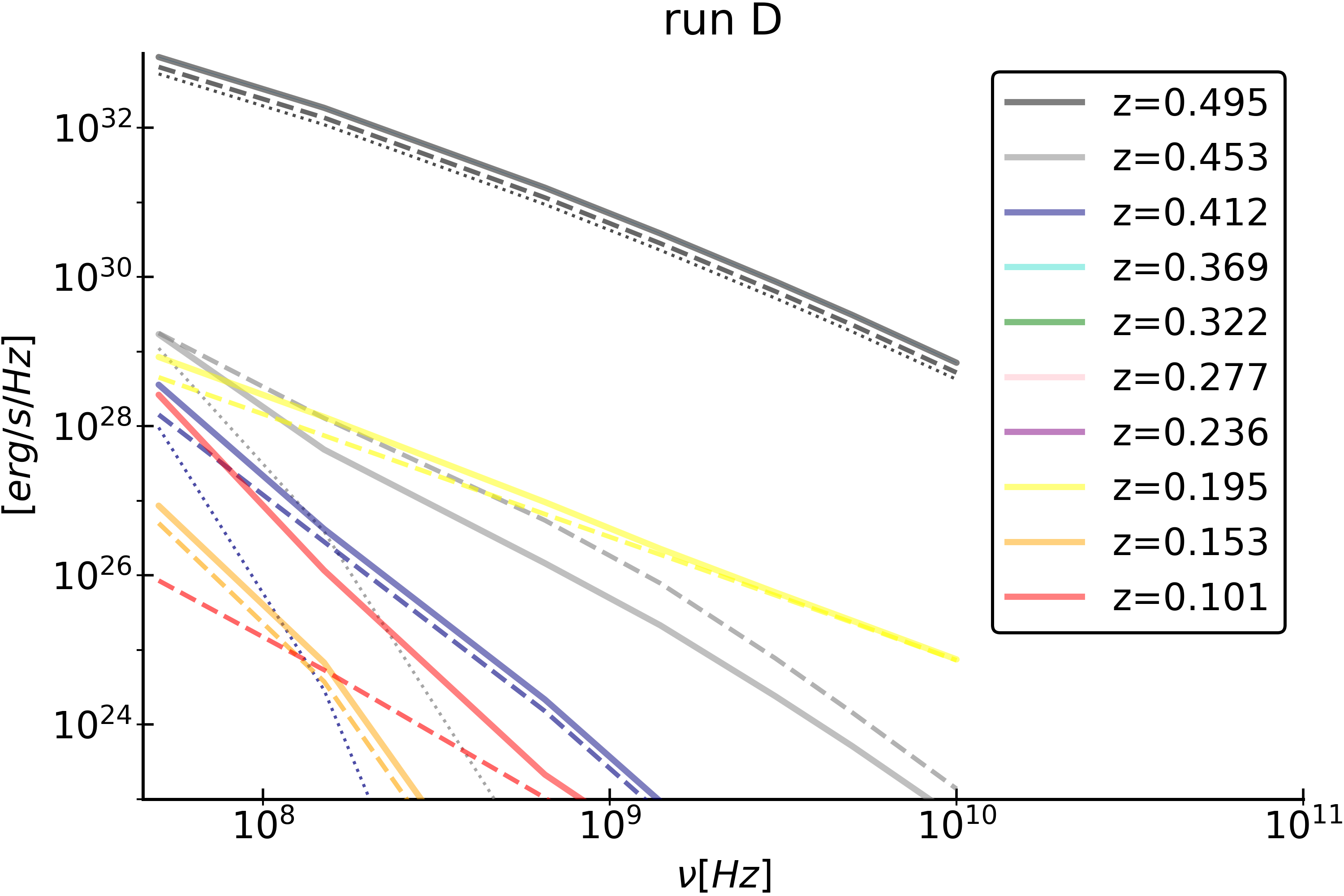}
    \caption{Evolution of the synchrotron radio spectra of all electrons in the run D, for ten different epochs.  The solid lines are the spectra including all loss and re-acceleration terms, the dashed line are for models including all loss terms and shock (re)acceleration, while the dotted lines are for models including only loss terms.} 
    \label{fig:radio_spectra_evol}
\end{figure}

\begin{figure}
    \centering
    \includegraphics[width=0.47\textwidth]{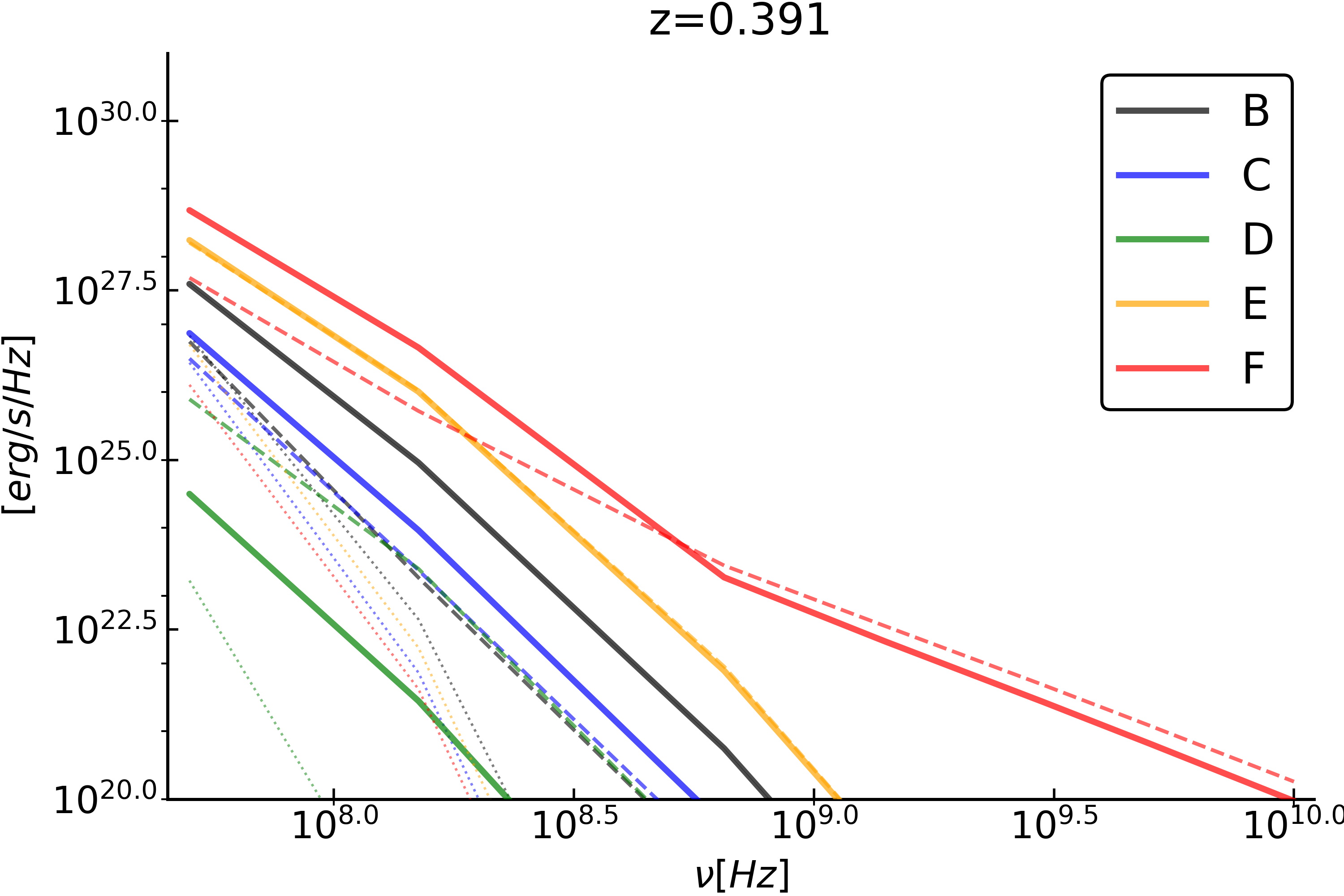}
        \includegraphics[width=0.47\textwidth]{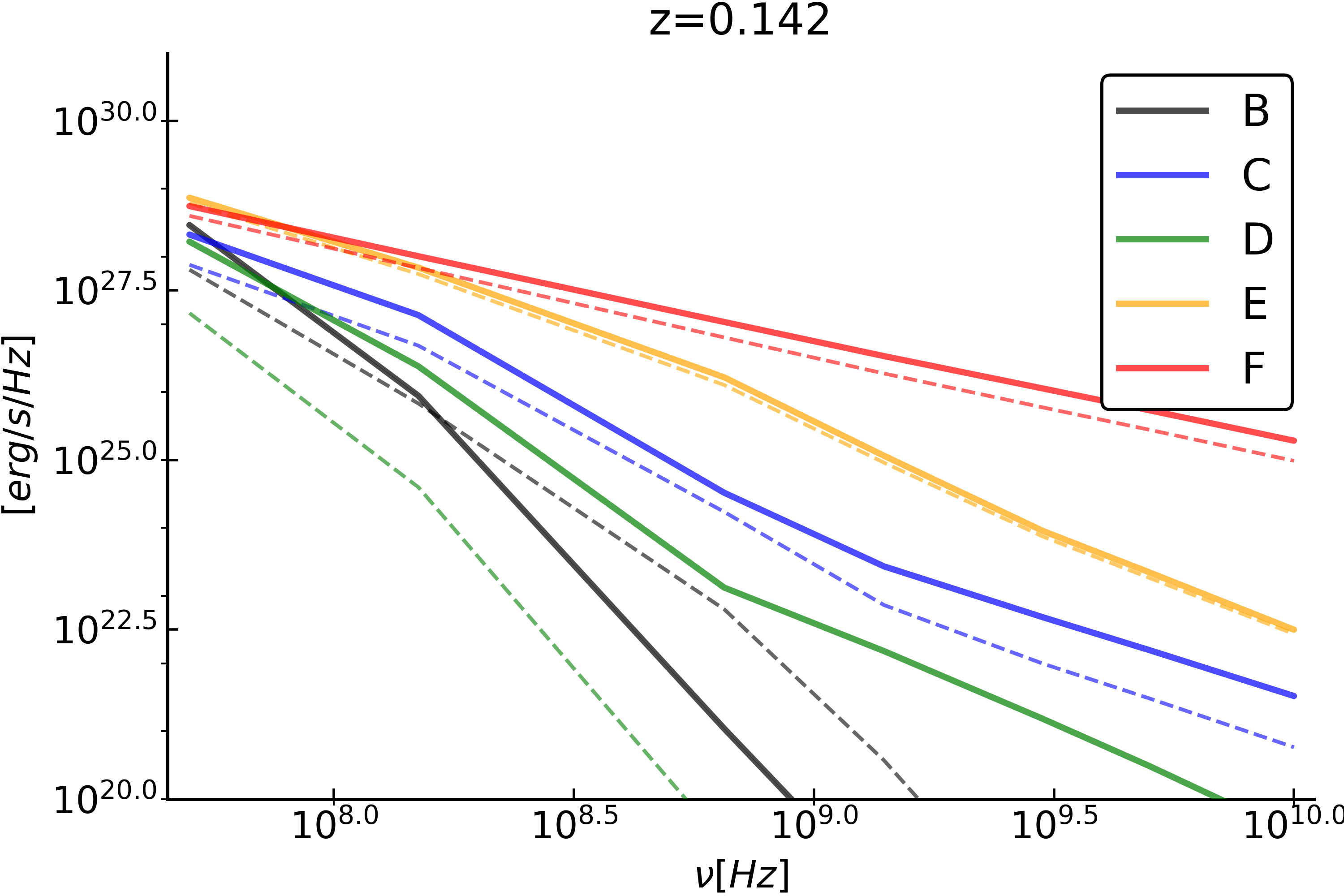}
    \caption{Synchrotron radio spectra in our five runs at $z=391$
    (i.e. $\approx 760 \rm ~Myr$ after the jet injection) and at $z=0.142$ ($\approx 3.2 \rm ~Gyr$ after the jet injection). The solid lines are the spectra including all loss and re-acceleration terms, the dashed line are for models including all loss terms and shock (re)acceleration, while the dotted lines are for models including only loss terms.  } 
    \label{fig:RADIO1}
\end{figure}

The integrated radio spectra, after the initial injection phase, are generally very steep, owing to the depletion of electrons with $\gamma \geq 5 \cdot 10^3$, which are mostly responsible for the emission at $\nu \geq 50 \rm ~MHz$.
The radio spectra for the evolving run D are shown in Figure ~\ref{fig:radio_spectra_evol}, while the comparison between the five runs at two different epochs (the same of Figure ~\ref{fig:spectra1}-\ref{fig:spectra2}) are shown in Figure ~\ref{fig:RADIO1}.

\begin{figure*}
    \centering
    \includegraphics[width=0.95\textwidth]{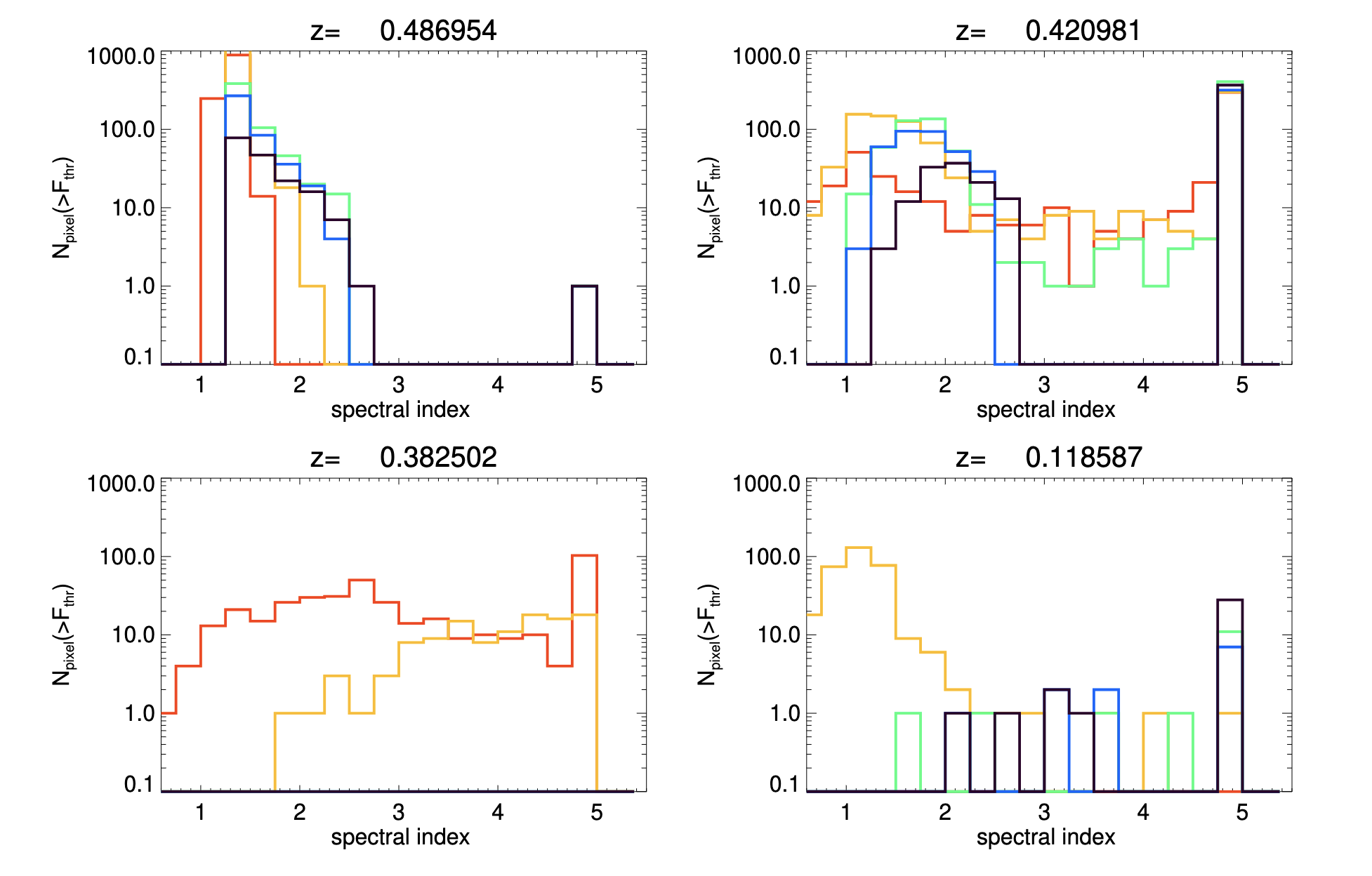}
    \caption{Distributions of radio spectral index between $140$ and $50$ MHz, only for $\geq 1 \sigma_{\rm rms}$ detectable pixels in radio maps of the run E at four different epochs.  The solid lines give the distribution of distribution of spectra for the electron evolution model including shocks and turbulent re-acceleration, the dashed lines are for the electron model only including shock re-acceleration and loss terms, and the dotted lines are for the electron model only  including radiative losses and adiabatic changes after the injection by jets.} 
    \label{fig:pdf_spectraE}
\end{figure*}

The distribution of radio spectral indices between 50 and 140 MHz, for $\geq \sigma$ detectable pixels, is shown in Figure ~\ref{fig:pdf_spectraE} for four different epochs in run E, showing the predictions for the three electron evolution scenarios: after $z \geq 0.48$, the only detectable patches are clearly related with the turbulent re-acceleration model, leading even to extremely steep ($\alpha \geq 2-3$) and small structures. The only contribution of shock re-acceleration can produce fewer detectable pixels, typically with $1 \leq \alpha \leq 2$ depending on the shock Mach number. No emission can be detected after $\approx 200 \rm ~Myr$ if the energy of relativistic electrons injected by radio galaxies  solely evolves under radiative losses and adiabatic changes. 

We shall remark that our assumed $\geq \sigma$ detection criterion is optimistic and is meant to highlight the tip of the iceberg of the potentially detectable structures in radio. State-of-the-art imagining pipelines typically detect diffuse emission from $\geq  2 \sigma$ at most \citep[e.g.][]{2018MNRAS.478..885B}. However, in recent works, we have shown that new signal analysis strategies based on Machine Learning \citep[][]{gh18deep} or Convolutional Deep Denoising Autoencoders \citep[][]{gh22denoise} can potentially detect diffuse correlated emission even at the $\sim 0.1 \sigma$ level, when opportunely trained with large training sets and under ideal observing conditions. While quantitatively estimating the actual gain in detection level compared to more standard techniques under realistic noise condition is still difficult,  such techniques are being massively explored as a solution to the challenges that the Square Kilometer Array will pose to the community \citep[e.g.][]{2022MNRAS.511.4305Y}, including the one of detecting and classifying the flurry of data from radio galaxies that future survey will deliver \citep[e.g.][for recent works on this topic]{2021MNRAS.503.1828B,2022MNRAS.510.4504T}. All this considered, the adopted $\geq  \sigma$ detection criterion considered here is a reasonable intermediate estimate between what is currently possible, and what should soon be achieved by standard detection algorithms to exploit the vastness of modern radio surveys.

\begin{figure*}
    \centering
    \includegraphics[width=0.95\textwidth]{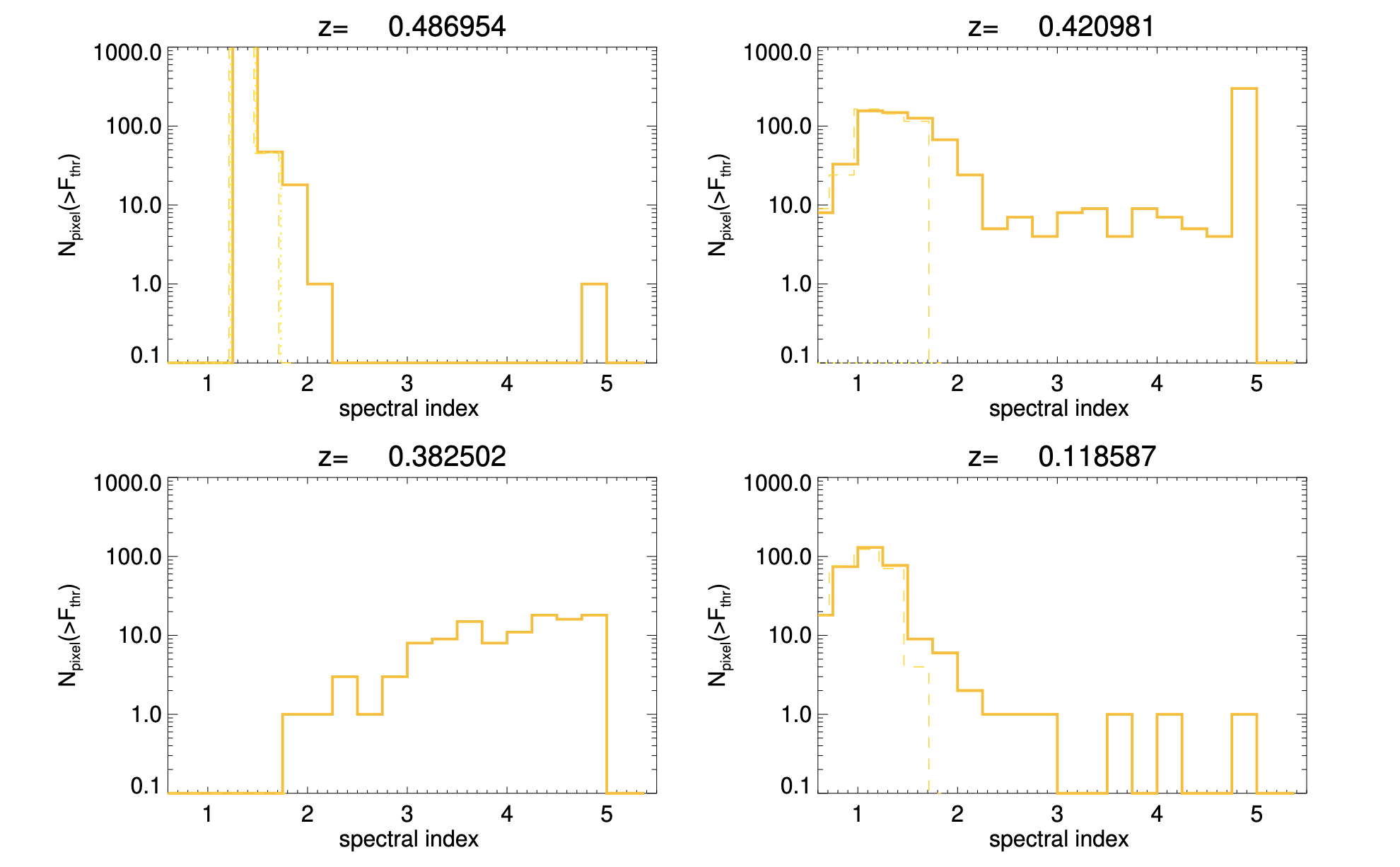}
    \caption{Distributions of radio spectral index between $140$ and $50$ MHz, only for $\geq 1 \sigma_{\rm rms}$ detectable pixels in radio maps of all runs at four different epochs, only for  the electron evolution model including all loss and re-acceleration terms.} 
    \label{fig:pdf_spectra_all}
\end{figure*}

For all runs, the distributions of detectable spectral indices, measured at the same epochs, confirm these trends:  Figure ~\ref{fig:pdf_spectra_all} shows that at most epochs the detectable emission is dominated by steep spectra and that, in general, runs with a higher power lead to a larger number of detectable and steep emission patches.  

The detection of such steep spectra emission is challenging and it demands for sensitive low-frequency radio observations. However, the detection of such structures has become increasingly more common with LOFAR and MWA, and indeed several extremely steep spectrum sources, often associated with remnant plasma from radio galaxy activity, and possibly interacting with the ICM, have been reported in the latest years: \citet{fdg17} ($\alpha=4.5$ in Abell 1033), \citet{2021MNRAS.508.3995B} ($\alpha=3.2$ in RX J1720.1+2638), \citet{Hodgson21a} ($\alpha=5.97$ in Abell 2877). In particular, some of the simulations analysed in Paper I were indeed already used by \citet{Hodgson21a} for the interpretation of real MWA data. They argued that the late evolution of fossil radio plasma, which can produce shortly ($\leq 100 \rm Myr$) detectable structures if a few remnant lobes mix and get compressed by weak shocks triggered by ICM activity, can represent a viable scenario for the formation of such diffuse steep-spectrum sources.

Recent work has investigated the relation between the radio power of central radio galaxies and the X-ray luminosity of the host cluster \citep[][]{pasini20}. In particular, \citet{pasini21} observed 227 radiogalaxies with LOFAR-HBA (140MHz) and correlated their radio emission with the X-ray emission in the 0.5-2~keV band, in the the eROSITA Final Equatorial Depths Survey (eFEDS), reporting  a positive correlation between the two quantities, as well as large scatter in radio power for any given X-ray luminosity. 

We can produce a similar statistics in our runs, by computing the total radio emission from our electrons at 140 MHz, compared to the X-ray emission computed from the simulated gas distribution, in which we assumed a single temperature value for each simulated cell, a constant composition (with metallicity $Z/Z_{\odot}=0.3$), and collisional equilibrium. This allowed us to  use the emissivity,  $\Lambda$, from the B-Astrophysical Plasma Emission Code (B-APEC)  {\footnote{https://heasarc.gsfc.nasa.gov/xanadu/xspec/manual/Models.html.}}, and to compute the total X-ray (continuum and line) emission within the same eROSITA energy band. 

\begin{figure}
    \centering
    \includegraphics[width=0.499\textwidth]{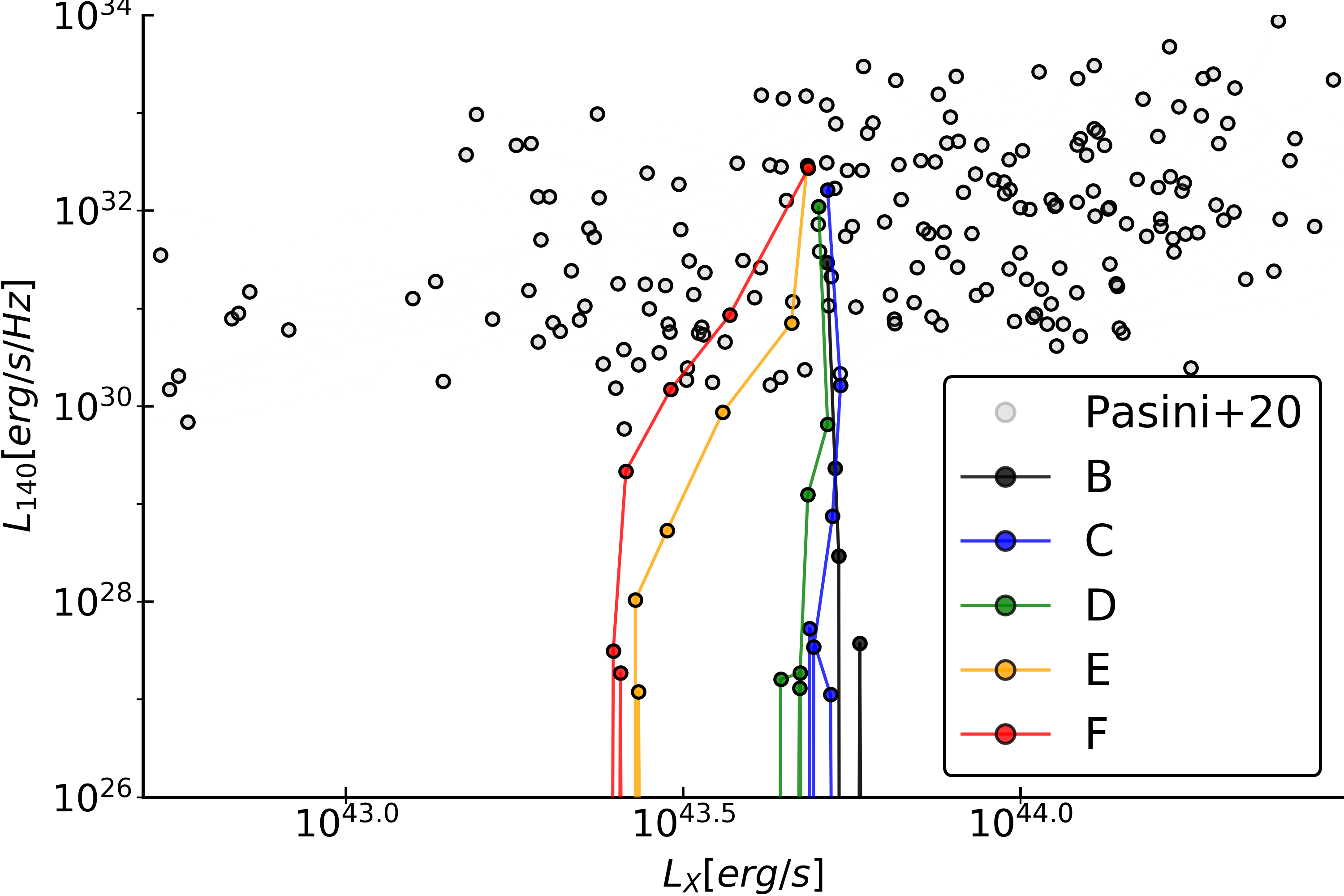}
    \includegraphics[width=0.499\textwidth]{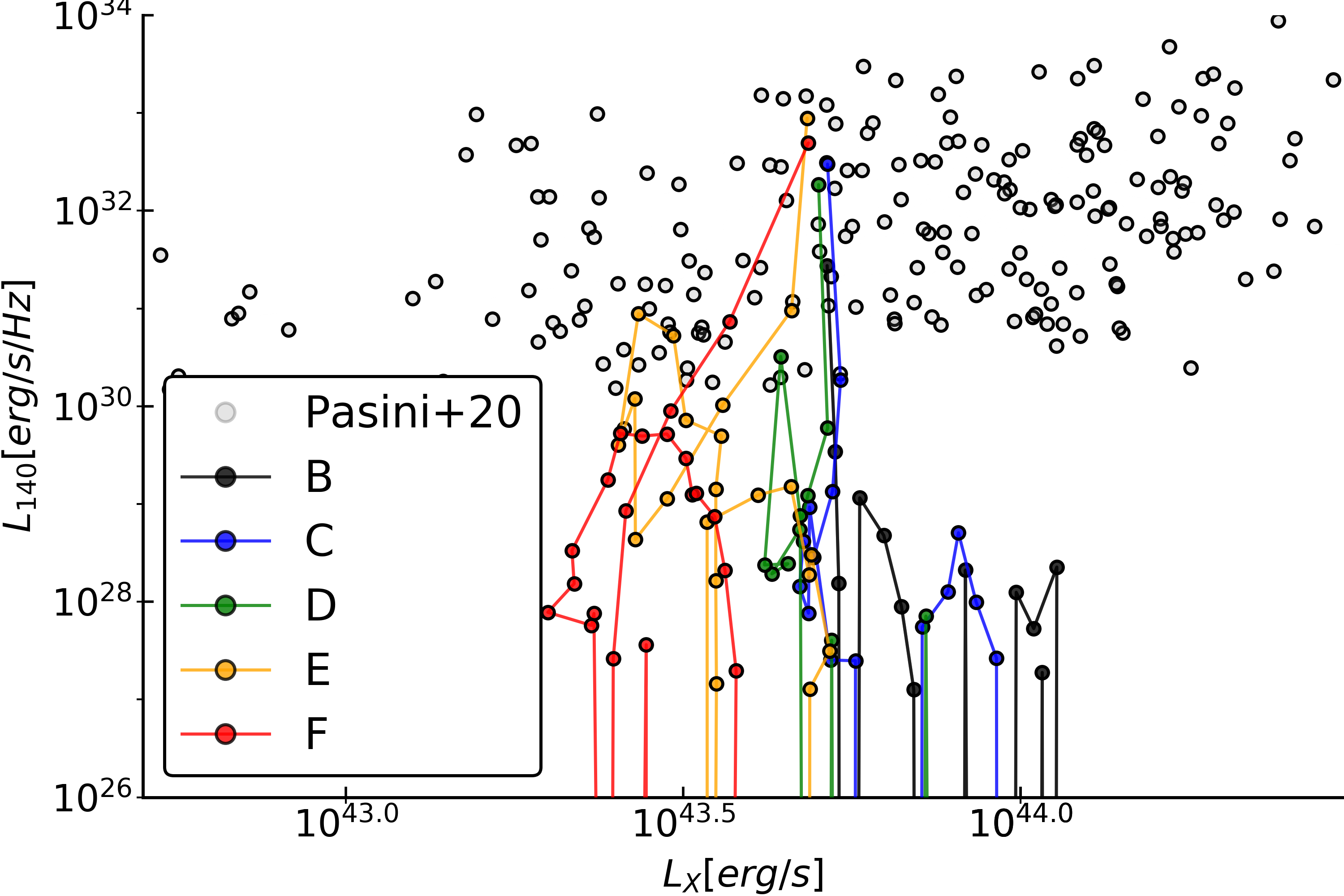}    
    \caption{Evolution of the total radio emission within $\leq 500 \rm ~kpc$ from the group centre at 140MHz, as a function of the X-ray emission in the [0.5-2]keV range within the same region, from $z=0.5$ to $z=0.4$ and for our five runs. The top panel shows our prediction for a model in which electrons are injected by jets and are then only subject to radiative and adibatic changes, while in the lower panel all re-acceleration mechanisms are considered. The grey points show the real LOFAR-HBA and eRosita observations from \citet{pasini21}.}  
    \label{fig:radio_vs_X}
\end{figure}

This is shown in Figure~\ref{fig:radio_vs_X}, where we plot the evolution of the total radio emission from our electrons (in the two extreme cases in which only injection and cooling are considered, or when also all reacceleration processes are used), within the same fixed area ($\leq 500 \rm ~kpc$) from the group centre used by \citet{pasini21}. 

 While it is impossible with our single simulated group to establish of a correlation between X-ray and radio power, we can study the impact of the jet power in the simulated X-ray and radio emission as a function of time. 

In the simulated case, the X-ray luminosity is little affected by the jet bursts (the X-ray luminosity from the innermost region only varies by a factor $\sim 2$ at most in the time interval under consideration) but the radio emission changes by many orders of magnitude in the same period of time. If we compare the same absolute epoch the relative drop of the radio emission is much less pronounced for a higher jet power: for example, at the reference epoch of $z=0.4$ (as shown in the plots) the radio emission in run F has dropped by a factor $\sim 10^3$ compared to the initial value, while this is $\sim 4 \cdot 10^4$ in run E while this is  $\geq 10^6$ in the lowest power runs (albeit with in a non-monotonic way, due to re-acceleration mechanisms). If no-reacceleration mechanism is considered, all simulated evolution show a quick crossing of the range spanned by real observations, followed by a longer stage of undetectable radio emission. On the opposite, when Fermi I and Fermi II re-acceleration are allowed, the radio emission drops more slowly with time, and at least in the two highest power runs the remnant radio emission remain for a longer period of time within the range of real observations. 

Interestingly, we find that our jets have a different impact on the X-ray emission of the group, i.e. depending on their initial power, the total X-ray luminosity of the group is found to alternatively increase (in low power jets) or decrease (in the highest power cases) over time, with the medium power run (D, $P_j = 3 \cdot 10^{44} \rm erg/s$ roughly separating the two regimes. 
Jets with  $P_j \geq 9 \cdot 10^{43} \rm ~erg/s$ (run E and D) are powerful enough to smooth the gas core density, carve prominent X-ray cavities and reduce (by a factor $\sim 50 \%$) the core X-ray emission until $z=0.4$.

Conversely, jets with $P_j \leq 10^{44} \rm ~erg/s$ reduce the X-ray emission from the core only in the very first stage of activity, while later on the group increases its X-ray emission owing to the gradual increase in its density, due to the effect of further matter accretions, and passing by substructures. 
The above trends makes run E and F able to produce two overall positively correlated trajectories in the ($L_{140,}$,$L_X$) plane, as in observations, while the other produce a negative and steep correlation between radio and X-ray powers, for the time in which (some) of their the radio emission is detectable.

Although it is impossible to reliably use these trends to compare to the observed positive correlation reported by \citet{pasini20}, it is interesting to notice that the large amount of vertical evolution in all our runs does not seem in conflict with the large scatter in real observations, even in the very homogeneous selection presented by \citet{pasini21}. This suggests that the observed scatter might not be caused by the host cluster, but rather that it may trace the complex evolution of real sources through the ($L_{140}$,$L_X$) plane on a time scale that depends on the power of the underlying radio jet.

\section{Discussion: how does jet power influence the evolution of radio emission?}

Our simulations helped us to identify several relations between the input jet power (of a single episode of AGN feedback) and the late-time ($\geq 100 \rm Myr$) evolution of injected relativistic particles and of their associated radio signal.

First, the magnetisation of jets (followed by the magnetisation of remnant lobes) depends on the input jet power for $\sim 1 \ \rm Gyr$ since the jet release (Figures~\ref{fig:scatter}-\ref{fig:pdf_evol}).

Depending on the initial power of jets, the injected particles can exit the cluster core and continue to be dispersed up to the cluster virial radius by large-scale turbulent motions, or instead remain confined around the cluster core (Figure~\ref{fig:distance}). For the $M_{\rm 100}\approx 2 \cdot 10^{14} M_{\odot}$ host system considered in this work, the critical jet power appears to be in the $P_j \geq 3-9 \cdot 10^{44} \rm erg/s$ range. 

Third, the typical distance reached by electrons at fixed times is a regular function of the jet power (Figure~\ref{fig:power_distance}), which can be well described by a $\langle D \rangle \propto P_j^{w_p}$ dependence up to $\sim 1-2 \rm ~Gyr$ since the injection of jets. 
Fourth, the budget of relativistic electrons a few Gyr after the injection is found to increased with $P_j$ (Figure \ref{fig:pdf_Eratio}-\ref{fig:pdf_Eratio_1e3}), albeit typically with a complex relation that depends on the time-integrated effects of energy losses  and re-acceleration terms. 

Consequently, the radio emission at a given frequency is also typically higher at any given time for higher jet powers, even if in this case spectral effects at different frequencies further complicate the search for simple trends. In general, we find that the stronger the initial jet power, the longer do radio lobes remain inside the range of typical radio luminosities, always with very steep radio spectra (Figure \ref{fig:pdf_spectra_all}-\ref{fig:radio_vs_X}).

Our conclusions are based on a single and impulsive ($\leq 32 \rm ~Myr$) AGN burst for each resimulation. Only with more sophisticated simulations it will be possible to assess the amplitude of such differences, in the presence of multiple AGN feedback events and with bursts of different durations. 

\section{Caveats \& discussion}

In this Section, we briefly comment what we believe are the most important weaknesses and the range of validity of the theoretical work presented here. 

First, our simulations are uniquely concerned with the single fluid ideal MHD  approach, i.e. resistive and multi-fluid effects are entirely neglected. 
Therefore, since the viscous and resistive scales are essentially the same, our simulations refer to an ICM which has a magnetic Prandtl number $P_M=\nu/\eta \approx 1$ (where $\nu$ is the kinematic viscosity and $\eta$ is the resistivity), which  surely is an unrealistic assumption  \citep[e.g.][]{2004ApJ...612..276S,bl11b,2016ApJ...817..127B}. To probe the small-scale dynamo, the entirely different and computationally expensive type of kinetic simulations are needed \citep[e.g.][]{2016PNAS..113.3950R}. In this work, we are not concerned with the effects of dynamo amplification, which is bound to be underestimated at this resolution anyway. 
To bracket the likely value of magnetic field amplification produced by plasma processes below our resolution scale, we indeed used the post-processing recalibration of the magnetic field on tracers (Section 
\ref{subsec:Bfield}). 

Another important limitation of this suite of simulation is that, for the sake of simplicity, we did not even attempt to self-consistently link the cooling-heating cycle of baryons induced by AGN feedback, which would introduce a number of numerical and physical complications which are far from being solved in cosmological simulations \citep[e.g.][and references therein]{2017MNRAS.467.3475N,2022MNRAS.511.3751H}. 

Having established with this first work (jointly with Paper I) the role between jet energetics and the circulation of non-thermal electrons, we defer to future work to design efficient numerical modules in {\enzo} also to couple jets to the mass growth of SMBH, with a further distinction between hot and cold gas accretion modalities  \citep[e.g.][]{2021MNRAS.503.3492T}.

Moreover, our treatment of jets is not suitable to follow the actual interaction between the relativistic (either electron or proton dominated) content of jets and the thermal ICM. Our jets are entirely made of hot gas and magnetic fields. 
This choice is very commonly adopted in the literature, and is not expected to significantly affect the long term ($\geq 100 \rm ~Myr$) evolution of the jets, or of the particles they carry. However, on small evolutionary stages the effective equation of state of a truly cosmic-ray dominated jet may produce somewhat different dynamics  \citep[][]{2018MNRAS.481.2878E}, also related the very first, relativistic bulk motion of jets \citep[][]{2017MNRAS.471L.120P}.

In summary, the simulated large-scale ($\geq 100 \rm~ kpc$) and long-term ($\geq 100 \rm ~Myr$) circulation of our jets and of the injected lobe remnants is reasonably well captured by our model. Yet, the small-scale interplay between gas, magnetic fields and cosmic rays, as well as the coupling with the mass growth process of SMBH deserve more extended, ad-hoc simulations.

\section{Conclusions}

We  have presented new simulations of the evolution of relativistic electrons injected by radio galaxies in the ICM, studying how the spatial and energy evolution of electrons is changed by the increase of the jet power. 
We modelled the evolution of electrons assuming they are {\it passively} advected by flows in the ICM and we studied the evolution of their energy spectra by including realistic loss and gain terms.

Our main results can be summarized as follows:

\begin{itemize}

\item The propagation of tracers in the ICM is the combination of an initial stage ($\leq 200 \rm ~Myr$ since the the release of jets) mostly dominated by jet dynamics, followed by a longer ($\geq 2 \rm ~Gyr$ since the release of jets) second stage  dominated by the diffusion by ICM turbulent motions. The final average distance covered by electrons, at fixed epochs, is found to strongly depend on the input jet power, through a well-constrained $\langle D \rangle =A_p P_j^{w_p} $ relation.

\item Our simulated jets do not cause significant thermodynamical differences in the evolution of the ICM that can be detected after about $\sim 500 \rm ~Myr$ since their burst. On the other hand, magnetic fields in the ICM are affected by jets even $\sim 1 \rm ~Gyr$ after the burst.

\item  Depending on the input jet power and on the efficiency of Fermi I and Fermi II re-acceleration on the injected electrons, the innermost ICM volume can have a cosmic ray (electron) to thermal gas energy ratio in the range  $\sim 10^{-6}$ to $\sim 10^{-4}$ even $\sim 3 \rm ~Gyr$ since the jet release (with the latter being produced by higher power jets, and with efficient Fermi II re-acceleration from the ICM).  This budget comes from the vast majority of fossil $\gamma \leq 10^3$ electrons, originally injected by the single central radio galaxy simulated in this works, Hence, this constitutes a lower limit on the real electron budget in the ICM.

\item  The spectrum of radio-emitting electrons, already after $\sim 200 \rm ~Myr$ since the jet injection, crucially depends on the ICM dynamics capable of re-accelerating fossil electrons injected by radio galaxies. Only if Fermi I and Fermi II processes are considered, the remnant electrons can become detectable again later in the cluster evolution, typically at low frequencies and with very steep radio spectra ($\alpha \geq 2$). 

\end{itemize}

This work confirms that the ICM can be enriched at low redshifts with a volume-filling population of fossil relativistic electrons, seeded by radio galaxies. ICM dynamics induced by mergers can re-energise this population leading to diffuse radio emission \citep[e.g.][]{sa99,bj14,2019SSRv..215...16V}. 

While the existence of such a reservoir of fossil electrons is a mere hypothesis, and whether or not radio galaxies could produce them in their lifetime remains unproven, our simulations supports that this mechanism is realistic, for a range of jet powers and coupling between ICM dynamics and particle re-acceleration. 

Recent low-frequency observations of nearby radio sources in clusters and groups of galaxies have started scratching the surface of the low-frequency emission from remnant electrons 
\citep[e.g.][]{fdg17,wilber18,2020A&A...634A...4M,Hodgson21a,brienza21,brienza22} and new numerical simulations of these complex processes represent a tool to constrain non-thermal physics in clusters. Whether the single release of relativistic electrons by a radiogalaxy in the ICM is enough to contribute all fossil relativistic electron required to reproduce the morphology and observed power of most radio relics or radio halos, or whether instead multiple radio galaxies and a more extended activity period is required, will be subject of follow-up investigations. 

Moreover, the evidence that already for $\geq 100 \rm ~Myr$ the distribution of spectral index in lobes mixing with the ICM has different shapes depending on the underlying re-acceleration mechanisms suggests that, in general, age estimates based on the modelling of radio spectra would increasingly overestimate the age of sources, if only loss and adiabatic terms are included in the modelling (see also Paper I). Further analysis of this aspect will be the subject of future work.

\section*{Acknowledgements}
 We thank our anonymous reviewer for their very detailed and helpful comments to the first version of this draft. 

  The authors gratefully acknowledge the Gauss Centre for Supercomputing e.V. (www.gauss-centre.eu) for supporting this project by providing computing time through the John von Neumann Institute for Computing (NIC) on the GCS Supercomputer JUWELS at J\"ulich Supercomputing Centre (JSC), under projects "radgalicm" and "radgalicm2". \\ 
  
 F.V. acknowledges financial support from the European Union's Horizon 2020 program under the ERC Starting Grant "MAGCOW", no. 714196. D.W. is funded by the Deutsche Forschungsgemeinschaft (DFG, German Research Foundation) - 441694982. MB acknowledges support from the Deutsche Forschungsgemeinschaft under Germany's Excellence Strategy - EXC 2121 "Quantum Universe" - 390833306. TP is supported by the DLR Verbundforschung under grant number 50OR1906.  Finally, we made use of the Cosmological Calculator by E. Wright  (http://www.astro.ucla.edu/~wright/CosmoCalc.html). 
 \bibliographystyle{aa}
 \bibliography{franco3}
\appendix
\section{Tests of the Cosmic Ray Solver}
In order to ensure that our procedure to evolve electron spectra (Section ~\ref{subsec:fokker}) and to compute their radio synchrotron spectra (Section~\ref{subsec:sync}), we tested our prediction against the independent results of the BRATS code \citep[][]{brats} which is a code of reference for the fitting of synchrotron ageing models to the observation of real radio sources (e.g. \citealp{2020brienza, 2021biava, 2022kukreti}). 

Here we show the reference test of a radio source which has been continuously injected for 25 Myr, and then has passively cooled for the following 200 Myr (CIOFF model). In BRATS, the energy spectrum was simulated using a minimum energy of $\gamma_{\rm min}=10$,  a maximum energy of $\gamma_{\rm max}=10^6$,  an initial energy spectrum of $\delta=2$ and a constant magnetic field of $B=10~ \rm \mu G$. To compare with our model prediction, we imposed the same constant magnetic field and input spectrum on electrons in a test uniform population of electrons, and we also switched off the adiabatic loss terms in our model.  This allowed us to verify a very good  correspondence between our simulated spectra and BRATS predictions until an epoch of $\sim 200 \rm ~Myr$ since the injection, which confirms that our implementation of electrons ageing and synchrotron emission works well, and moreover that the input spectra imposed on our sources, which assumed a CIOFF model, represent a realistic enough model of radio galaxies, with an initial radio spectrum compatible with observations.

\begin{figure}
    \centering
    \includegraphics[width=0.45\textwidth]{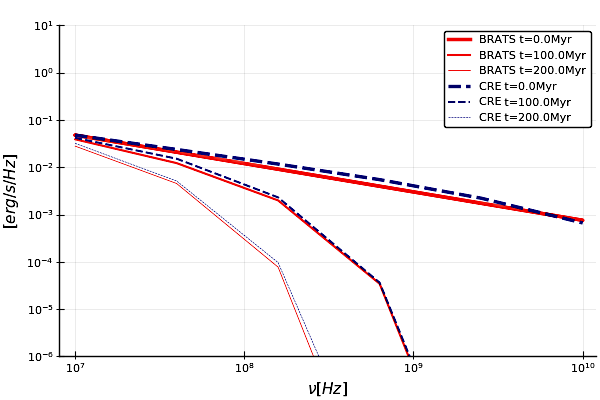}
    \caption{Comparison of the simulated radio emission spectra for a source with an input  $\alpha = 0.6$ radio spectrum and with a constant $B=10~ \rm \mu G$ uniform magnetic field, evolved using our cosmic ray electron solver, or using  using BRATS \citep[][]{brats}.}
    \label{fig:distance}
\end{figure}

\begin{figure*}
    \centering
    \includegraphics[width=0.8\textwidth]{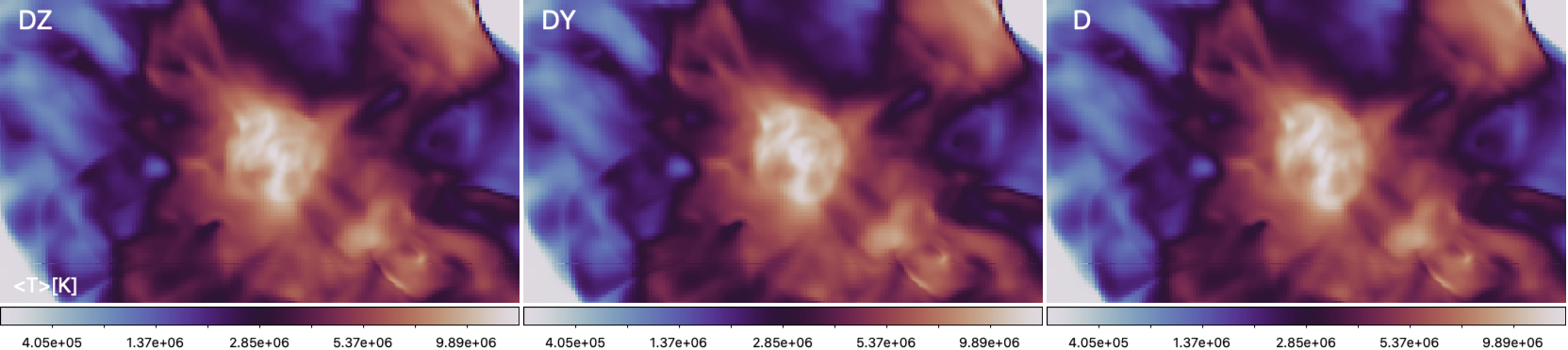}
    \includegraphics[width=0.8\textwidth]{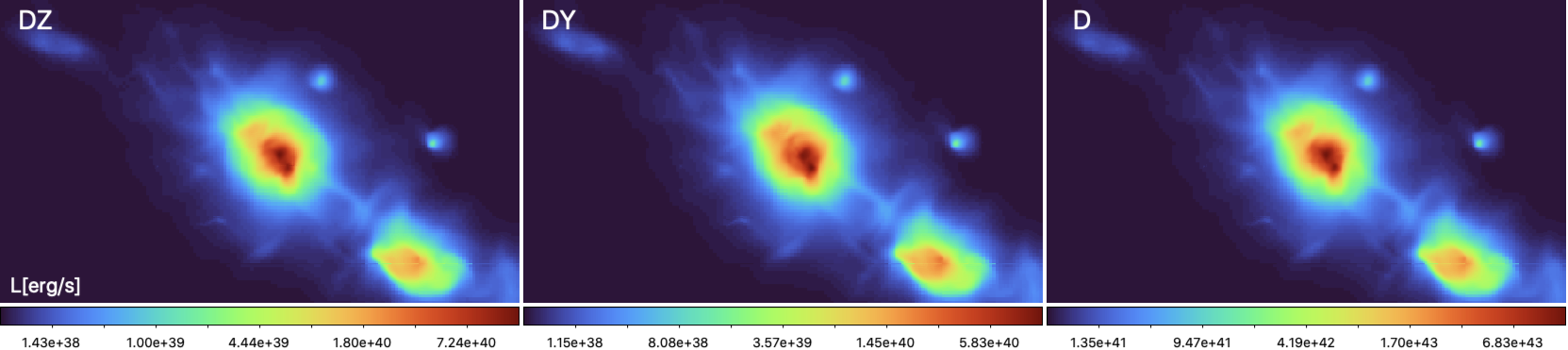}
    \includegraphics[width=0.8\textwidth]{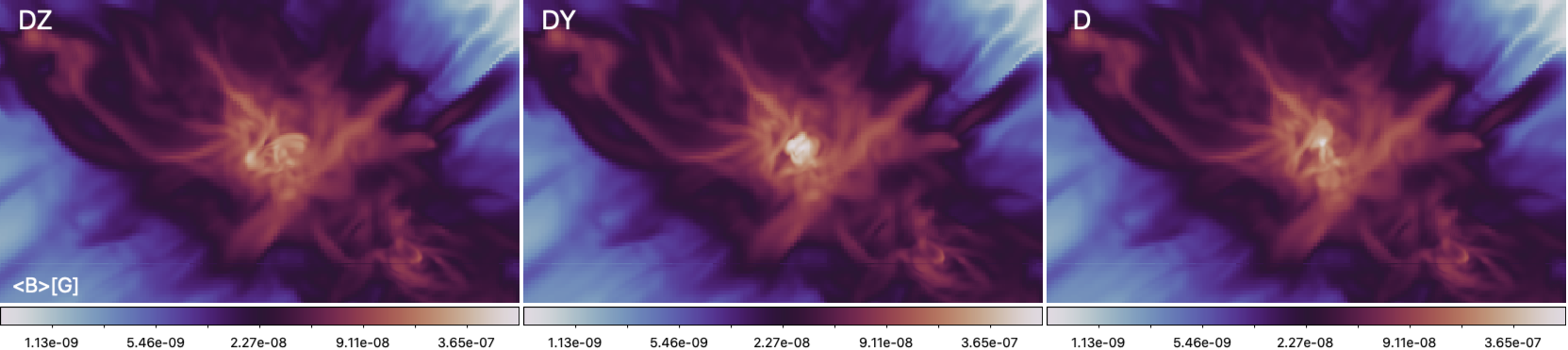}
    \caption{Maps of projected mass weighted mean gas temperature (top row), total X-ray emission in the [0.5-2~keV band (middle row), and projected mass weighted mean magnetic field strength (bottom row) at z=0.391 for our three resimulations of jets with $P_j=4 \cdot 10^{44} ~\rm erg/s$ power, with jets expanding either along the $x$, $y$ or $z$ axis of the simulation. Each panel has size 4.2 $\times$3.2 Mpc.} 
    \label{fig:map_test}
\end{figure*}

\begin{figure}
    \centering
    \includegraphics[width=0.49\textwidth]{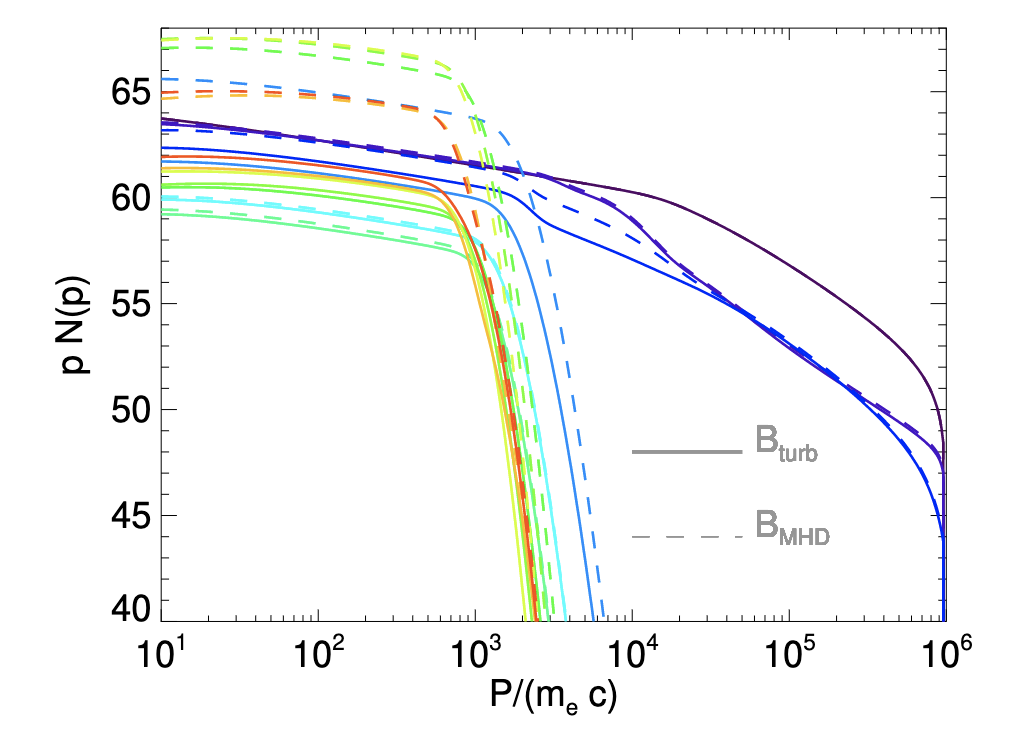}
    \caption{Evolution of the relativistic electron spectra in the innermost $\leq 300 \rm ~kpc$ of our cluster in the run E model and including all loss and gain terms, using the rescaled magnetic field as in the main text (solid lines) or the original magnetic field produced by the MHD simulation (dashed lines). The different colors mark 10 different epochs, roughly equally spaced in time, from $z=0.5$ (black) to $z=0.1$ (red).} 
    \label{fig:bturb_test}
\end{figure}

\begin{figure*}
    \centering
    \includegraphics[width=0.95\textwidth]{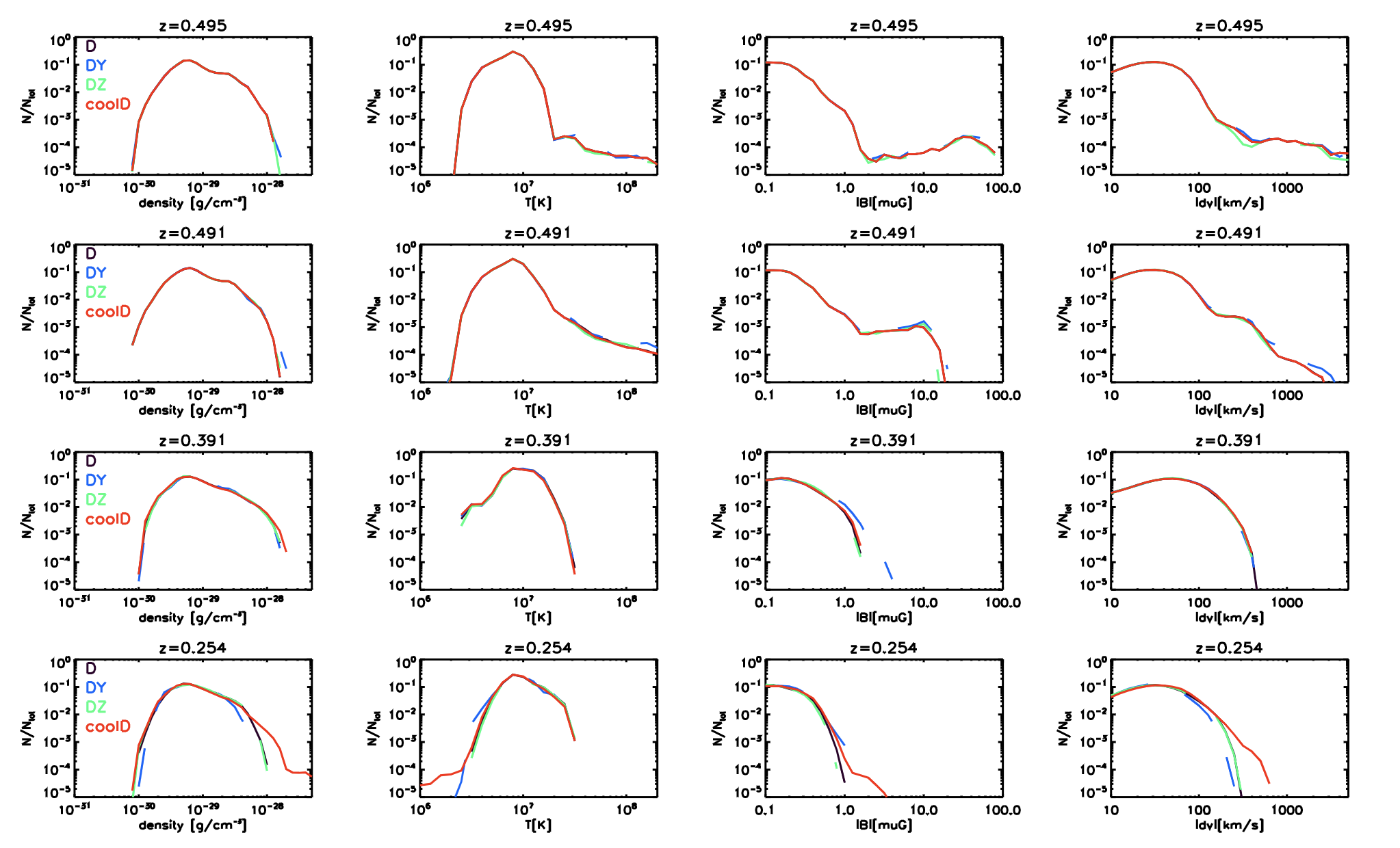}
    \caption{Distributions of gas density, gas temperature, velocity dispersion and magnetic field strength for four epochs, comparing different prescriptions for the jet model in run D (see text for explanations). The distributions are taken within a reference comoving $1^3 \rm ~Mpc^3$ around the (moving) cluster center of mass at four different redshifts.} 
    \label{fig:pdf_evol_test}
\end{figure*}

\section{Tests with the original magnetic field of the simulation}

As explained in Section~\ref{subsec:Bfield}, the spatial resolution obtained with these runs is not enough to ensure a realistically large Reynolds number across the simulated ICM volume, which prevents the  the amplification of the cluster magnetic field by the small-scale dynamo \citep[e.g.][]{review_dynamo}. We therefore evolved our relativistic electrons under the effect of a re-normalised magnetic field, extrapolated based on the local measured value of the solenoidal turbulent energy flux, and assuming a fixed conversion of this energy flux ($2\%$) into the creation of new magnetic field \citep[e.g.][]{bm16,va18mhd}.

In Figure \ref{fig:bturb_test}, we show the simulated electrons spectra for the innermost ($\leq 300 \rm ~kpc$) region of our group in the run E model and using all loss and gain terms, obtained either with the rescaled magnetic field as in the main paper ($B_{\rm turb}$), or with the original magnetic field directly produced by the MHD simulation ($B_{\rm MHD}$). 
As commented already in the main paper, the average difference in the magnetic field across the tracer distribution is not dramatic, i.e. $ \langle B_{\rm turb} \rangle \sim 2-3 \langle B_{\rm MHD} \rangle$. However, the difference can be more significant for those tracers sampling spectra in very turbulent regions, and these differences can be amplified in simulated spectra, as the can crucially affect the balance between synchrotron losses (which scales as $\propto B^{-2}$) and the turbulent re-acceleration term (which scales as $\propto B$, see Equation~\ref{eq:ASA}).

For this reason, if turbulent re-acceleration is considered in combination with an unrealistically low magnetic field, an unrealistically large amount of low energy electrons can be re-accelerated. This is clearly shown by Figure \ref{fig:bturb_test}, where we can see the progressive buildup of the electron distribution at $p \leq 10^4$, leading to a final excess of order $\sim 10^2-10^3$ in the total energy of fossil relativistic electrons, compared to our more realistic choice of the rescaled magnetic field, $B_{\rm turb}$. Since this also leads to the overproduction of low-frequency radio emission in our galaxy group, we consider the rescaled magnetic field a better option to obtain a realistic view of the evolving population of fossil electrons in the ICM.

\section{Tests with additional physical variations}

While our main paper focuses on the influence of the initial power of jets on the long term evolution of gas and electrons in a reference group of galaxies, additional differences can be expected for different choices in  the initial direction of jets. Therefore, we produced two additional resimulations of the intermediate case of run D in which we released exactly the same power ($P_j=4 \cdot 10^{44} \rm erg/s$) along the other 
 other two possible perpendicular directions: run Dy and Dx. These are meant to explore whether the large scale circulation of injected electrons can show different properties, depending on the different sectors in the ICM the jets expand into, and in the possible different amount of frustration that jets can experience depending on the ICM flow they encounter.
 Figure \ref{fig:map_test} gives the projected view of the three simulations after $1 \rm ~Gyr$ since the jet injection  (showing minimal differences between runs, with the obvious exception of the cluster core), while
 figure \ref{fig:pdf_evol_test} shows that no significant differences in the statistics of the thermodynamical properties of the ICM are seen in the three runs, meaning that our main results are not affected by jet orientations.
 With run coolD, we also tested whether the evolution of the cluster after the jet injection can be affected by radiative cooling, which is switched on in this case: again, with the exception of small tail of low temperature gas ($T \leq 10^6 ~\rm K$) forming at low redshift in the latter case, we report that the main finding of our paper are unchanged.
 Future work will investigate the more realistic case of having cooling active since the start of the simulation, which can potentially leading to multi-phase ICM and to a higher level of gas clumping, interacting with jets.

\end{document}